\documentclass[a4paper,pagesize,DIV=12]{scrartcl}
\pdfoutput=1

\usepackage[utf8]{inputenc}
\usepackage[T1]{fontenc}
\usepackage[english,american]{babel}
\usepackage{lmodern}
\usepackage{amsmath, amssymb, amsfonts}
\usepackage{csquotes}
\usepackage{graphicx}
\usepackage{epstopdf, epsfig}
\usepackage{mathtools}
\usepackage{color}
\usepackage{xcolor}
\usepackage{siunitx}
\usepackage{braket}
\usepackage{pdfpages}

\usepackage[footnotesize]{caption}[2008/08/24]

\usepackage[font=footnotesize]{subcaption}

\usepackage[numbers,sort&compress]{natbib}

\usepackage[activate={true,nocompatibility}]{microtype}

\usepackage[colorlinks=true, linkcolor=blue, citecolor=blue, urlcolor=blue]{hyperref}
\hypersetup{
	pdftitle={A boundary integral method with volume-changing objects for ultrasound-triggered margination of microbubbles},
	pdfauthor={Achim Guckenberger, Stephan Gekle},
	pdfkeywords={mathematical theory of boundary integral methods; Fredholm theory; periodic domain; volume-changing objects; blood flow; microbubbles; margination; targeted drug delivery; ultrasound}
}

\newcommand{\dd}{\mathrm{d}}
\newcommand{\vect}[1]{\boldsymbol{#1}}
\renewcommand{\vec}[1]{\boldsymbol{#1}}

\newcommand{\vx}{\vec{x}}
\newcommand{\vxz}{\vx_0}
\newcommand{\Obj}{O}
\newcommand{\Bub}{\mathcal{B}}
\newcommand{\Caps}{\mathcal{C}}
\newcommand{\Wall}{\mathcal{W}}
\newcommand{\uv}{\vec{u}}
\newcommand{\fv}{\vec{f}}
\newcommand{\tfv}{\triangle\fv}

\newcommand{\Rey}{\mathrm{Re}}

\makeatletter
\newcommand{\lambdabar}{{\mathchoice
		{\smash@bar\textfont\displaystyle{0.25}{1.2}\lambda}
		{\smash@bar\textfont\textstyle{0.25}{1.2}\lambda}
		{\smash@bar\scriptfont\scriptstyle{0.25}{1.2}\lambda}
		{\smash@bar\scriptscriptfont\scriptscriptstyle{0.25}{1.2}\lambda}
	}}
	\newcommand{\smash@bar}[4]{%
		\smash{\rlap{\raisebox{-#3\fontdimen5#10}{$\m@th#2\mkern#4mu\mathchar'26$}}}%
	}
	\makeatother

\newtheorem{theorem}{Theorem}

\setcapindent{0em}

\begin{document}
	\title{A boundary integral method with volume-changing objects for ultrasound-triggered margination of microbubbles}
	\author{%
		\normalsize{}Achim Guckenberger\thanks{Corresponding author: \href{mailto:achim.guckenberger@uni-bayreuth.de}{achim.guckenberger@uni-bayreuth.de}}\,\,\,\thanks{Biofluid Simulation and Modeling, Fachbereich Physik, Universit\"at Bayreuth, Bayreuth}\, ,
		Stephan Gekle\footnotemark[1]}
	\date{\normalsize{}Dated: \today}

	\maketitle

	\begin{abstract}
A variety of numerical methods exist for the study of deformable particles in dense suspensions. 
None of the standard tools, however, currently include volume-changing objects such as oscillating microbubbles in three-dimensional periodic domains.
In the first part of this work, we develop a novel method to include such entities based on the boundary integral method.
We show that the well-known boundary integral equation must be amended with two additional terms containing the volume flux through the bubble surface.
We rigorously prove the existence and uniqueness of the solution.
Our proof contains as a subset the simpler boundary integral equation without volume-changing objects (such as red blood cell or capsule suspensions) which is widely used but for which a formal proof in periodic domains has not been published to date.

In the second part, we apply our method to study microbubbles for targeted drug delivery.
The ideal drug delivery agent should stay away from the biochemically active vessel walls during circulation. 
However, upon reaching its target it should attain a near-wall position for efficient drug uptake.
Though seemingly contradictory, we show that lipid-coated microbubbles in conjunction with a localized ultrasound pulse possess precisely these two properties.
This ultrasound-triggered margination is due to hydrodynamic interactions between the red blood cells and the oscillating lipid-coated microbubbles which alternate between a soft and a stiff state. We find that the effect is very robust, existing even if the duration in the stiff state is more than three times lower than the opposing time in the soft state.

	\end{abstract}

	\clearpage
	\tableofcontents

\section{Introduction}

Margination refers to the effect that stiff objects such as platelets, leukocytes or stiff synthetic microparticles preferentially travel along the walls in suspension flows, e.g.\ in the vascular system \citep{Freund2007, Freund2012, Kumar2014, Namdee2013, Fedosov2014b, Fitzgibbon2015, Vahidkhah2015, DApolito2015, Muller2015, Muller2016, Mehrabadi2016, Gekle2016, Bacher2017}. 
This is a result of softer particles such as red blood cells (RBCs), drops or capsules migrating towards the center away from the boundaries \citep{Aarts1988, Fedosov2011, Farutin2013, Mukherjee2014, Singh2014, Katanov2015}.
A similar effect occurs not only in Poiseuille but also in bounded shear flow due to the image stresslet generated by the presence of the wall \citep{Smart1991,Mukherjee2013, Singh2014}.
Migration behavior has implications for targeted drug delivery as efficient drug uptake is only possible if the drug delivery agents are positioned close to the walls of the blood vessels near the target organ \citep{Rychak2007,Kooiman2014, Lammertink2015}.
Accordingly, it advocates the use of stiff particles as drug delivery agents.
Yet, during the transport phase towards the target the agents should remain buried in the vessel interior to avoid high shear stress and premature biochemical interaction with the endothelial wall.
This would speak in favor of using soft particles.
Being able to combine both seemingly contradicting properties might lead to a very effective drug administration protocol.

One of the most promising approaches for targeted drug delivery is the use of coated microbubbles \citetext{ultrasound contrast agents, see e.g.\ \citealp{Klibanov2002,Lindner2004,Faez2013, Unnikrishnan2012}} to actively and selectively enhance drug uptake \citep{Ferrara2007,Couture2014,Unger2014, Owen2014, Lammertink2015,Kooiman2014,Kotopoulis2016}. 
In the simplest scenario, microbubbles are injected together with the actual drug suspension and an ultrasound pulse is applied at the target organ which makes the bubbles oscillate periodically. 
This strongly enhances drug uptake due to the forces that the oscillating microbubbles exert on nearby endothelial cells~\citep{Kooiman2014}. 
More recently, there have also been numerous attempts to use the bubbles themselves as drug carriers by biochemically attaching active drug substances e.g.\ on the bubble surface which are released due to an ultrasound pulse at the target organ \citep{Couture2014, Unger2014, Kooiman2014}. 

Microbubbles coated with a phospholipid layer \citep{Ferrara2007, Faez2013} are usually rather soft deformable objects in the absence of ultrasound \citep{Marmottant2005,Overvelde2010}.
They would therefore be expected to be buried inside the blood stream akin to RBCs, as also concluded from \textit{in-vivo} experiments \citep{Lindner2002}.
This allows for their safe transport, but makes them at first sight unsuitable candidates for drug delivery.
Yet, during ultrasound exposure bubble expansion beyond a critical radius~$R_\mathrm{soft}$ in the low pressure phase causes stiffening of the bubble shell \citep{Marmottant2005, Rychak2006, Overvelde2010} which might induce margination.
On the other hand, shrinking in the high pressure phase leads to buckling (softening) of the phospholipid shell \citep{Sijl2011}.
Since the bubbles thus rapidly oscillate between a soft and a stiff state, it is \textit{a priori} unclear if and to what extent such objects would show margination.

We study this question by means of 3D numerical simulations.
Many methodologies are available for computing flows with hard or deformable particles, provided that the volume of each particle remains unchanged.
These methods are able to accurately reproduce the margination of stiff particles which originates from their hydrodynamic interaction with the surrounding RBCs.
Here however, we require a method that can handle deformable volume-changing microbubbles together with RBCs in a periodic channel.
Such a method is currently not available.
The first part of our paper therefore deals with the development of our \emph{volume-changing object boundary integral method} (VCO-BIM) in periodic domains.
Compared to existing boundary integral formulations we find that additional terms occur which account for the volume fluxes across the bubble surfaces.
We prove mathematically that the resulting Fredholm integral equation has exactly one solution.
The proof and the method hold for an arbitrary amount of volume-changing objects and capsule-like entities (RBCs, vesicles, drops, etc.)\ with arbitrary viscosity ratios.
We give the proof in some detail and generality since a number of recent publications \citep[e.g.\@][]{Loewenberg1996,Zinchenko2000,Zhao2010,Lindbo2010} derive and use boundary integral equations in periodic domains (without bubbles), but a proof of uniqueness and existence of their solution has not been established to date.
We also note that very occasionally boundary integral methods have been used with expanding bubbles \citep{Power1992,Power1996,Power1992a,Nie2002}, but these attempts have been restricted to infinite domains, making them unsuitable for blood flow simulations. 

In the second part we use our VCO-BIM to find that microbubbles indeed show what we call \emph{ultrasound-triggered margination} (UTM): Ultrasound exposure causes rapid and reliable margination of otherwise soft microbubbles.
UTM is caused by the special properties of the lipid bubble shell and their interaction with the red blood cells.
The effect is robust and rapidly drives microbubbles towards the endothelial wall even if the \enquote{stiff time} (i.e.\ the time during which the bubble size is larger than the critical radius $R_\mathrm{soft}$) is more than three times smaller than the opposing \enquote{soft time}.
Phospholipid coated microbubbles are thus shown to simultaneously possess two highly desirable, but seemingly contradicting properties: safe passage in the low-shear zones of the vessel interior and near-endothelial position at the target organ, the latter being easily controllable by ultrasound exposure.

\section{The volume-changing object boundary integral method}\label{sec:Theory}

Obtaining numerical solutions of the Stokes equation via boundary integral methods has a long history starting with the publication by \citet{Youngren1975}.
Well established is the \emph{direct} method that is suitable for the simulation of incompressible deformable particles with viscosity ratios $\lambda \neq 0,\infty$ in an infinite domain \citep{Pozrikidis2001a}.
Rigorous proofs of existence and uniqueness of the solution exist \citep[e.g.\@][]{Odqvist1930,Ladyzhenskaya1969,PozrikidisBook92,Kohr2004}.
They are enabled by the fact that the equation is a Fredholm integral equation of the second kind, allowing for the application of the Fredholm theory \citep[e.g.\@][]{Kress2014}.
If deformable bubbles ($\lambda = 0$) with volume changes are included, only the method in an infinite domain but no complete proof exists \citep{Nie2002}.

\emph{Indirect} boundary integral methods solve a (typically second kind) equation for an auxiliary field, from which the physical velocity can be computed afterwards.
Such a formulation has been used to model expanding bubbles in an infinite domain with established existence and uniqueness results for the solution \citep{Power1992,Power1992a,Power1996}.
Another indirect variant is the completed double-layer boundary integral method (CDLBIEM) tailored for simulating rigid objects ($\lambda = \infty$), with proofs in infinite domains being well-established \citep[e.g.\@][]{Power1987, Karrila1989, Sangtae2005, Kohr2004}.

Without bubbles, equations in \emph{periodic} domains for direct 
\citep[e.g.\@][]{Hasimoto1959, Zick1982, Loewenberg1996, Zinchenko2000, Zhao2010, Freund2011a, Freund2013, Freund2014a}
and indirect methods 
\citep[e.g.\@][]{Zhao2011a, Zhao2012a, Fan1998, Lindbo2010, Freund2012, Wang2013a, Fitzgibbon2015, Spann2016, afKlinteberg2014, afKlinteberg2016}
are well known.
The general geometry Ewald-like method (GGEM) also uses an indirect formulation to make the equations amenable to an accelerated computation.
This was mostly used for problems where two of the three spacial directions are periodic
\citep{Hernandez-Ortiz2007, Pranay2010, Kumar2011, Kumar2012a, Kumar2014, Zhu2015a, Sinha2015, Sinha2016, Zhao2017}.
Yet, statements regarding existence and uniqueness of the solution are lacking so far.
This may be of some concern since some well-known proofs for the infinite domain \citep[e.g.\@][ch.~4.5]{PozrikidisBook92} require an auxiliary field that would violate the conservation of the ambient fluid if applied to periodic domains, even if all individual objects are volume-conserving.

The purpose of the present section is thus two-fold: First, we show that the presence of volume-changing objects in periodic domains leads to new non-trivial terms in the equation for the direct boundary integral method.
Second, we rigorously proof the existence and uniqueness of the solution of this periodic boundary integral equation employed in the present work and in other publications as listed above.

For this, we start by deriving the Fredholm boundary integral (FBI) equation for 3D periodic domains with deformable capsule-like ($\lambda \neq 0,\infty$) volume-conserving particles and deformable volume-changing objects such as bubbles.
The final result for $N_\Obj$ objects $\Obj_k$ in a periodic domain with unit cell $\Gamma$ of volume $V_\Gamma$ is
\begin{align}\label{eq:FBI:Intro}
	\begin{aligned}
		u_j(\vxz) = {}&\frac{2}{1 + \lambdabar_{\Obj_k}} 
		\Bigg[
			 \braket{u_j}_\Gamma
			- \frac{1}{8 \pi \mu} \sum_{q=1}^{N_\Obj} (\mathcal{N}_{\partial \Obj_q} \vec{F})_j(\vxz) \\
			&\qquad\qquad\quad + \frac{1}{8\pi} \sum_{q=1}^{N_\Obj} (1 - \lambdabar_{\Obj_q}) (\mathcal{K}_{\partial \Obj_q} \vec{u})_j(\vxz)
			+ \frac{1}{V_\Gamma} \sum_{q=1}^{N_\Bub} Q_{\Bub_q} \chi^{(\Bub_q)}_j
		\Bigg] \\
		&- \frac{1 - \lambdabar_{\Obj_k}}{1 + \lambdabar_{\Obj_k}} z^{(k)}_j(\vxz) \left[ \oint_{\partial \Obj_k} u_l(\vx) n_l(\vx) \, \dd S(\vx) - Q_{\Obj_k} \right] \, ,
	\end{aligned}
	\\
	\vxz \in \partial\Obj_k \, , \quad k=1,\ldots,N_{\Obj} \, , \quad j = 1,2,3 \, .
	\nonumber
\end{align}
This equation forms the basis of our VCO-BIM.
Here, $k$ is the index of the object on whose surface the evaluation point $\vxz$ is located.
Moreover, $\uv$ on the left-hand side is the velocity on the surface $\partial \Obj_k$ of the $k$'th object, $\braket{u_j}_\Gamma$ the prescribed average flow through~$\Gamma$ and $\mu$ the dynamic viscosity.
$\lambdabar_{\Obj_k}$ is an effective viscosity ratio for the $k$'th object defined in equation \eqref{eq:DefLambdabar} below.
Furthermore, $\vec{F}$ is the outer traction in case of bubbles and otherwise the jump of the traction across the interfaces.
$\mathcal{N}_{\partial\Obj_q} \vec{F}$ and $\mathcal{K}_{\partial\Obj_q}\vec{u}$ are the usual single- and double-layer integrals, respectively, evaluated with the Green's functions for a 3D periodic domain (given by equations~\eqref{eq:SL} and \eqref{eq:DL}).
The second term on the second line is the first novel contribution from the $N_\Bub$ volume-changing bubbles and contains the centroid $\vec{\chi}^{(\Bub_k)}$ as well as the volume flux $Q_{\Bub_k}$ into or out of the bubble. The latter is a, possibly time-dependent, prescribed quantity chosen such that the sum of all fluxes is zero.
Finally, the last line is essentially a part of the so-called Wielandt deflation \citep{Sangtae2005} where $\vec{z}^{k}$ is a known function.
Again, for bubbles a new term due to the flux $Q_{\Obj_k}$ appears.
As we will show, the last line is imperative for bubbles ($\lambdabar_{\Obj_k} = 0$) as it ensures uniqueness, contrary to objects with $\lambdabar_{\Obj_k} \neq 0$ where it is merely an optional accelerator for the numerical procedure.

After introducing the system components in section \ref{sec:SystemBIM}, we use section \ref{sec:BI} to derive the general boundary integral equation including volume-changing bubbles in periodic domains.
Section \ref{sec:Fredholm} then turns the boundary integral equation into the numerically solvable Fredholm boundary integral equation \eqref{eq:FBI:Intro} and, most importantly, rigorously proves existence and uniqueness of the solution.
This proof includes the periodic BIM equation without bubbles which is solved numerically by a number of existing codes \citep[e.g.\@][]{Loewenberg1996,Zinchenko2000,Zhao2010,Lindbo2010}.
Finally, section \ref{sec:BubbleDetails} gives some details about how we model oscillating bubbles and \ref{sec:BIM:Numerics} outlines the numerical implementation of our method.
Symbols are defined and explained on their first use, but are also listed in appendix~\ref{sec:ListOfSymbols} as a quick reference.

\subsection{System description}\label{sec:SystemBIM}

\subsubsection{Periodicity and the unit cell}\label{sec:DefPeriod}
\begin{figure}
	\centering
	\includegraphics{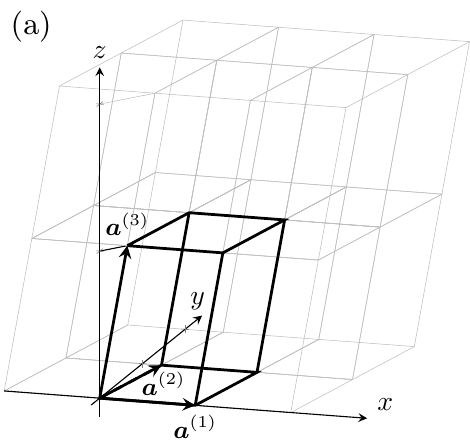}
	\hfill
	\includegraphics{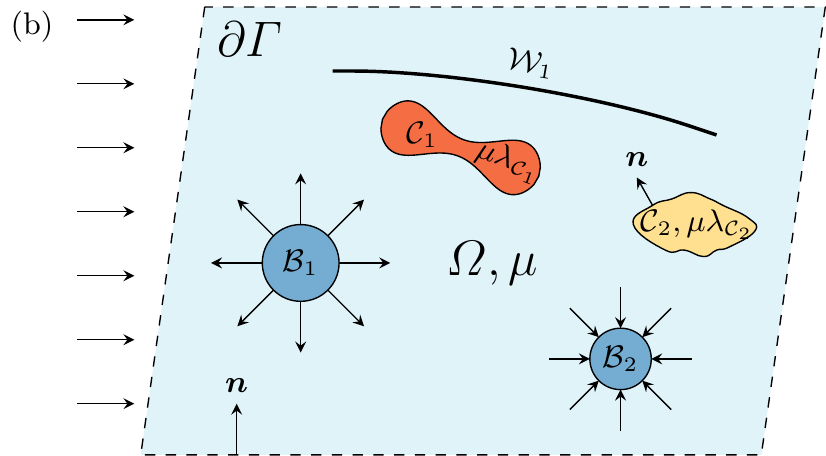}
	\caption{%
	(a) Example of a triclinic unit cell $\Gamma$ (thick lines), together with 11 replicas (thin, gray lines). In this example, the two basis vectors $\vec{a}^{(1)}$ and $\vec{a}^{(2)}$ lie in the $x$-$y$-plane, but $\vec{a}^{(2)}$ is not parallel to the $y$-axis.
	(b) Two-dimensional sketch of the general three-dimensional problem. $\Gamma \subset \mathbb{R}^3$ contains the whole unit cell (everything within the dashed border $\partial \Gamma$). 
	$\Omega \subset \Gamma$ is the ambient fluid with dynamic viscosity~$\mu$ (light blue).
	As an example, two capsule-like entities ($\Caps_1$, $\Caps_2$), two volume-changing bubbles ($\Bub_1$, $\Bub_2$) and a wall $\Wall_1$ are shown.
	The normal vectors always point into $\Omega$.
	The arrows on the left represent an imposed flow.
	}
	\label{fig:UnitCell:AbstractSetup}
\end{figure}

We mostly consider flows in 3D periodic systems.
To this end, we introduce a triclinic unit cell $\Gamma \subset \mathbb{R}^3$ that is spanned by the basis $\{\vec{a}^{(1)}, \vec{a}^{(2)}, \vec{a}^{(3)} \}$ as shown in figure~\ref{fig:UnitCell:AbstractSetup}~(a).
The three basis vectors $\vec{a}^{(i)} \in \mathbb{R}^3$, $i=1,2,3$, define a right-handed coordinate system.
In the most general case they are not required to be of unit length or orthogonal and might depend on time, although the latter will not be explicitly considered here.
Nevertheless, even static but skewed bases can be useful in practice, for example to model so-called deterministic lateral displacement devices \citep{Huang2004,Kruger2014,Zhang2015b}.
The surface $\partial \Gamma$ of the unit cell is not included in the open set~$\Gamma$.
We will denote the volume by $V_\Gamma$.

To cast the concept of periodicity into mathematical terms, we introduce by
\begin{equation}\label{eq:PerGridVec}
	\vec{X}^{(\vec{\alpha})} := \alpha_1 \vec{a}^{(1)} + \alpha_2 \vec{a}^{(2)} + \alpha_3 \vec{a}^{(3)}
\end{equation}
some grid vector with some grid index $\vec{\alpha} \in \mathbb{Z}^3$.
The unit cell $\Gamma$ is assumed to be replicated infinitely throughout space by displacing it with all possible grid vectors.
We will show in appendix~\ref{sec:overlap} that actually any of these boxes with any origin can be chosen as $\Gamma$.
Furthermore, a general function $f: \mathbb{R}^3 \rightarrow \mathbb{R}$ is said to periodic if it satisfies
\begin{equation}
	f(\vx + \vec{X}^{(\vec{\alpha})}) = f(\vx) \, \quad \forall \vx \in \mathbb{R}^3 \quad \text{and} \quad \forall \vec{\alpha} \in \mathbb{Z}^3 \, .
\end{equation}
The reciprocal (Fourier) unit cell is spanned by the reciprocal basis vectors $\vec{b}^{(j)} \in \mathbb{R}^3$, $j=1,2,3$, defined by $\vec{a}^{(i)} \cdot \vec{b}^{(j)} = 2 \pi \delta_{ij}$.
Some general Fourier grid vector is then denoted by
\begin{equation}\label{eq:FourierDef}
	\vec{k}^{(\vec{\kappa})} := \kappa_1 \vec{b}^{(1)} + \kappa_2 \vec{b}^{(2)} + \kappa_3 \vec{b}^{(3)} \, ,
\end{equation}
with $\vec{\kappa} \in \mathbb{Z}^3$.

\subsubsection{System components}\label{sec:SysComp}
The general system considered in this section is depicted in figure~\ref{fig:UnitCell:AbstractSetup}~(b).
We introduce the ambient fluid $\Omega \subset \Gamma$ which represents the open set of the space within the unit cell~$\Gamma$ but outside of any immersed object.
The fluid is assumed to have a constant dynamic viscosity $\mu$ and is governed by the usual Stokes equation and incompressibility condition.
Furthermore, we have $N_\Obj$ objects $\Obj_k \subset \Gamma$, $k=1,\ldots,N_\Obj$ in the unit cell as further detailed below.
For now we assume that all of them are completely located within $\Gamma$ and relax this requirement later on (which is required for dynamic simulations of dense suspensions, see appendix~\ref{sec:overlap}).
No object shall overlap with or contain any other object.
The 2D surfaces of the objects embedded in 3D space will be denoted by the symbol \enquote{$\partial$}, e.g.\ $\partial \Obj_k$.
We also assign a velocity $\uv(\vx)$ to each point $\vx \in \bar{\Gamma}$, where the bar represents the closure of the set.

Regarding the immersed objects, we consider three different types.
First, deformable particles that can be used to mimic \enquote{capsule-like} entities (such as vesicles, drops or red blood cells).
The $i$'th capsule will be denoted by the open set $\Caps_k \subset \Gamma$ with $k=1,\ldots,N_\Caps$, where $N_\Caps$ are the number of capsules.
Their inside is filled with some Stokesian fluid that has a dynamic viscosity of $\mu \lambda_{\Caps_k}$, where $\lambda_{\Caps_k} > 0$ is the viscosity ratio between the inner and outer fluids.
Their volume is conserved and as such the net flux $Q_{\Caps_k}$ into or out of the capsules is zero.
Deformations are governed by the jump of the traction $\triangle \vec{f}$ across their surfaces $\partial \Caps_k$.
This traction jump is calculated from an appropriate constitutive law that determines the actual object properties.
The velocity across the interfaces is assumed to be continuous.

Second, $N_\Wall$ non-closed objects $\Wall_k$, $k = 1,\ldots,N_\Wall$, may exist that can be used to model deformable (fixed traction jump) or rigid (prescribed velocity) walls.
Note that we set $\Wall_k = \partial \Wall_k$, i.e.\ these objects do not have some \enquote{inside}.
The velocity across the walls is assumed to be continuous.

Third, we introduce our novel ingredient, $N_\Bub$ bubble-like objects $\Bub_k$, $k=1,\ldots,N_\Bub$ whose volumes are allowed to change.
Contrary to capsules their inside is filled with some compressible fluid such as air that has a much lower viscosity than the ambient fluid.
Therefore, the detailed flow field inside the bubbles will not be considered and, instead of the Stokes equations, the essential model assumption for this inner fluid is simply
\begin{equation}\label{eq:Bubble:Div}
	\vec{\nabla} \cdot \vec{u}(\vx) = c_k \, , \quad \vx \in \Bub_k \, , \quad k = 1,\ldots,N_\Bub \, .
\end{equation}
Here, $c_k$ are some prescribed constants, meaning that the particles expand or contract homogeneously over their whole interior and that any inhomogeneities such as eddies are considered to vanish instantaneously.
But note that $c_k$ may depend on the time allowing for oscillating bubbles.
As shown below (eq.~\eqref{eq:Bub:Flux:c}), $c_k$ is related to the net flux $Q_{\Bub_k}$ into or out of the bubble by $c_k = Q_{\Bub_k} / V_{\Bub_k}$, where $V_{\Bub_k}$ is the $k$'th bubble's volume.
Moreover, the surface deformation of the bubbles is governed by the prescribed traction at the outside of the surfaces, as further elaborated in section~\ref{sec:BubbleDetails}.

For convenience, some arbitrary object of any type will be denoted by $\Obj_k$.
We further define $N_\Obj := N_\Caps + N_\Bub + N_\Wall$ as well as $\Obj := \Caps \cup \Bub \cup \Wall$ where $\Caps$/$\Bub$/$\Wall$ mean the unions of all capsules/bubbles/walls.
All in all, we have $\Gamma = \Omega \cup \bar{\Obj}$.
Furthermore, unit normal vectors are denoted by the symbol $\vec{n}$.
We choose the convention that the normal vector of every capsule and bubble points into $\Omega$.
The normal vector of walls may point in any of the two possible directions.
The normal vector on $\partial \Gamma$ shall point into $\Omega$.

\subsection{Deriving the boundary integral equation with volume-changing objects}\label{sec:BI}

\subsubsection{The general boundary integral equation}\label{sec:GenBI}

In order to introduce the required notation and equations for later on, we start from the standard Stokes and continuity equations for flows at low Reynolds numbers which corresponds to the typical situation encountered in the microcirculation \citep{Misbah2012,Freund2014}:
\begin{subequations}\label{eq:Stokes}
	\begin{align}
		- \vec{\nabla} P(\vx) + \mu \nabla^2 \vec{u}(\vx) &= 0 \, , \\
		\vec{\nabla} \cdot \vec{u}(\vx) &= 0 \, , \label{eq:Incompress}
	\end{align}
\end{subequations}
for $\vx \in \Omega$.
Here, $P$ is the pressure, $\nabla^2$ the usual 3D Laplace operator and $\vec{u}$ the fluid velocity.
The inside of capsules is governed by analogous expressions, but with a different viscosity in general.
Body forces such as gravity will be neglected, but can be easily incorporated via an effective pressure \citep[eq.~(1.2.9)]{PozrikidisBook92}.

Next, we introduce the traction $\fv$ acting on the surface $\partial \Obj_k$ of some object $\Obj_k$ via
\begin{equation}\label{eq:Trac}
	f_i(\vx) := \sigma_{ij}(\vx) n_j(\vx) \, , \quad \vx \in \partial \Obj_k \, , \quad i = 1,2,3 \, ,
\end{equation}
where the fluid stress tensor is defined by
\begin{equation}\label{eq:Sigma}
	\sigma_{ij} := -P \delta_{ij} + \mu \left(\frac{\partial u_i}{\partial x_j} + \frac{\partial u_j}{\partial x_i}\right) \, , \quad \vx \in \Omega \, , \quad i,j=1,2,3 \, .
\end{equation}
Summation over repeated Cartesian components is implied throughout this work.
One needs to take the limit of the stress tensor onto the surface for the evaluation of equation~\eqref{eq:Trac}.
This limit can be taken from both sides, leading to the traction on the outside ($\fv^+$, limit from $\Omega$ onto the surface) and the inside ($\fv^-$, defined with the outer normal vector)
of a closed object's surface.
The difference is the traction jump
\begin{equation}\label{eq:TracJump}
	\tfv(\vx) := \fv^+(\vx) - \fv^-(\vx) \, , \quad \vx \in \partial \Obj_k \, ,
\end{equation}
which is the major quantity coupling the flow with the surface mechanics. 
For a wall, $\tfv$ is the difference of the tractions on its two sides.

Equations~\eqref{eq:Stokes} can be efficiently and accurately solved via the boundary integral method \citep{Freund2014,PozrikidisBook92,Pozrikidis2001a}.
Assuming that all objects are located within the unit cell $\Gamma$ (for objects crossing the unit cell boundary, see appendix~\ref{sec:overlap}), one can derive the boundary integral (BI) equation \citep[compare][ch.~2.3, pages~37 and 143]{PozrikidisBook92}
\begin{align}\label{eq:BI:GeneralWithGamma}
	\begin{aligned}
		u_j(\vxz) = &- \frac{1}{8 \pi \mu} \sum_{q=1}^{N_\Obj} (\mathcal{N}_{\partial \Obj_q} \vec{F})_j(\vxz) + \frac{1}{8\pi} \sum_{q=1}^{N_\Obj} (1 - \lambdabar_{\Obj_q}) (\mathcal{K}_{\partial \Obj_q} \vec{u})_j(\vxz) \\
			&- \frac{1}{8 \pi \mu} (\mathcal{N}_{\partial \Gamma} \vec{f})_j(\vxz) + \frac{1}{8\pi} (\mathcal{K}_{\partial\Gamma} \uv)_j(\vxz) \, ,
	\end{aligned}
	\\
	\vxz \in \Omega \, , \quad j = 1,2,3 \, . \nonumber
\end{align}
Note that the evaluation point $\vxz$ is in the ambient fluid and not within any object or on any surface.
The single-layer integral (or single-layer potential) over some surface~$S$ is defined as
\begin{equation}\label{eq:SL}
	(\mathcal{N}_{S} \vec{f})_j(\vxz) := \int_{S} f_i(\vx) G_{ij}(\vx,\vxz) \, \dd S(\vx) \, , \quad j=1,2,3 \, ,
\end{equation}
and the double-layer integral is
\begin{equation}\label{eq:DL}
	(\mathcal{K}_{S} \vec{u})_j(\vxz) := \int_{S} u_i(\vx) T_{ijl}(\vx,\vxz) n_l(\vx) \, \dd S(\vx) \, , \quad j=1,2,3 \, .
\end{equation}
These integrals exist for $\vxz \in \mathbb{R}^3$, notably in the improper sense if $\vxz \in S$ or if $\vxz$ is on some periodic image of $S$ \citep[ch.~6]{Kress2014}.
The improper double-layer integral is in this case sometimes also denoted as a principal value integral \citep[p.~27]{PozrikidisBook92}.

$\vec{F}$ is a \enquote{unified traction} which represents either the traction $\fv^+$ on the outside in case of the bubbles, or the traction jump $\tfv$ in case of the capsules and the walls.
For notational convenience we have abbreviated $\vec{f} := \vec{f}^+$ in the last line of equation~\eqref{eq:BI:GeneralWithGamma}.
Moreover, an \enquote{effective} viscosity ratio is defined as
\begin{equation}\label{eq:DefLambdabar}
	\lambdabar_{\Obj_q} :=
	\begin{cases}
		\lambda_{\Caps_q} & \text{if $\Obj_q$ is a capsule $\Caps_q$,} \\
		1 & \text{if $\Obj_q$ is a wall $\Wall_q$,} \\
		0 & \text{if $\Obj_q$ is a bubble $\Bub_q$,}
	\end{cases}
	\quad
	\text{for} \quad q = 1,\ldots,N_\Obj \, .
\end{equation}
Thus, if $\lambdabar_{\Obj_q} = 1$, the corresponding terms in the second sum are always absent (regardless of the value of $\mathcal{K}_{\partial \Obj_q} \uv$).

The wall contributions in equation~\eqref{eq:BI:GeneralWithGamma} can be derived by considering as a starting point an object with a finite thickness.
Taking the limit as the thickness tends to zero, integrals that previously went over the distinct opposing sides of the wall now go over essentially the same surface, except that the integrands are still evaluated on their respective sides.
The normal vectors on these two sides are perfectly antiparallel.
Additionally using the continuity of the velocity, one finds that the double-layer integral drops out (represented by $\lambdabar_{\Wall} = 1$).
Furthermore, for the single-layer integral we have 
$\int_{\partial\Wall^+ \cup \partial\Wall^-} f_i^+ G_{ij} \dd S = \int_{\partial\Wall^+} (f_i^+ - f_i^-) G_{ij} \dd S \equiv \int_{\partial\Wall} \triangle f_i G_{ij} \dd S$
where $\partial \Wall^\pm$ denotes the two sides of the zero-thickness wall, and equation~\eqref{eq:TracJump} was used to introduce the traction jump \citep[p.~37]{PozrikidisBook92}. The integration $\int_{\partial\Wall}$ goes over only one side of the wall.
Walls spanning through the whole domain $\Gamma$ can be introduced by a similar limiting procedure.

Note that the BI equation~\eqref{eq:BI:GeneralWithGamma} is actually valid for \emph{any} Green's functions, no matter if periodic or not.
However, in order to be able to compute anything in practice, they should follow the general symmetries of the system.
As we want to implement a periodic domain, we will next introduce the appropriate expressions.

\subsubsection{Appropriate Green's functions for a 3D periodic domain}\label{ch:Green}
Green's functions $G_{ij}$ and $p_j$ for the velocity and pressure, respectively, are obtained by solving the singularly forced Stokes equation.
It is possible to derive specialized Green's functions such as for singly \citep{Pozrikidis1996} or doubly \citep{Pozrikidis1996,Greengard2004,Cortez2014a} periodic domains or with incorporated stationary walls \citep{Blake1971,Janssen2008,Liron1976,Staben2003}.
However, we want to be able to apply our method for general channel geometries, and therefore employ the standard Green's functions for a 3D periodic domain.
According to \cite{Hasimoto1959} and \cite{Pozrikidis1996}, they are given by
\begin{align}
	G_{ij}(\vx,\vxz) &= \braket{G_{ij}}_\Gamma + \frac{8\pi}{V_\Gamma} \sum_{\vec{\kappa} \neq 0} \frac{1}{|\vec{k}^{(\vec{\kappa})}|^2} \left(\delta_{ij} - \frac{k^{(\vec{\kappa})}_i k^{(\vec{\kappa})}_j}{|\vec{k}^{(\vec{\kappa})}|^2}\right) e^{-\mathrm{i} \vec{k}^{(\vec{\kappa})} \cdot (\vx - \vxz)} \, , \label{eq:Green:G} \\
	p_j(\vx,\vxz) &= \frac{8 \pi}{V_\Gamma} x_j + \frac{8 \pi}{V_\Gamma} \mathrm{i} \sum_{\vec{\kappa} \neq 0} \frac{k^{(\vec{\kappa})}_j}{|\vec{k}^{(\vec{\kappa})}|^2} e^{-\mathrm{i} \vec{k}^{(\vec{\kappa})} \cdot (\vx - \vxz)} \, , \label{eq:Green:P} \\
	&\text{with} \quad \vx \in \mathbb{R}^3 \setminus \{ \vxz \} \, , \quad i,j=1,2,3  \nonumber
\end{align}
for the velocity ($G_{ij}$) and the pressure ($p_j$), respectively.
$\vxz \in \bar{\Gamma}$ is arbitrary but fixed.
The sums go over all possible Fourier grid vectors $\vec{k}^{(\vec{\kappa})}$ as defined in eq.~\eqref{eq:FourierDef} with $\vec{\kappa} \in \mathbb{Z}^3 \setminus \{ \vec{0} \}$, and $\mathrm{i}$ is the imaginary unit.
$G_{ij}$ satisfies the incompressibility condition,
\begin{equation}\label{eq:GreenIncompress:u}
	\frac{\partial G_{ij}(\vx,\vxz)}{\partial x_i} = 0 \, , \quad \vx \in \mathbb{R}^3 \, , \quad i,j=1,2,3 \, .
\end{equation}
Furthermore, $\braket{G_{ij}}_\Gamma := \frac{1}{V_\Gamma} \int_\Gamma G_{ij}(\vx,\vxz) \, \dd x^3$ is independent of $\vxz$ and constitutes a free parameter representing an imposed average net flow for the Green's function system.
We choose the coordinate system such that it is zero \citep{Zinchenko2000}:
\begin{equation}\label{eq:AvgG:Zero}
	\braket{G_{ij}}_\Gamma = 0 \, , \quad i,j=1,2,3 \, .
\end{equation}
Note that the full system can nevertheless have an imposed flow which will be introduced in section~\ref{ch:UCellSurf:DL} via double-layer integrals over the unit cell boundary.

Combining both of the above equations \citep[via][eq.~(2.1.8)]{PozrikidisBook92}, the Stresslet is found to be
\begin{equation}\label{eq:Green:T}
	T_{ijl}(\vx,\vxz) = -\frac{8\pi}{V_\Gamma} x_j \delta_{il} + \breve{T}_{ijl}(\vx,\vxz) \, , \quad \vx \in \mathbb{R}^3 \setminus \{ \vxz \} \, , \quad i,j,l=1,2,3 \, .
\end{equation}
$T_{ijl}$ contains a linear (non-periodic) part, whereas the periodic part is
\begin{equation*}
	\breve{T}_{ijl}(\vx,\vxz) := - \frac{8\pi}{V_\Gamma} \mathrm{i} \sum_{\vec{\kappa} \neq 0} \left(\frac{k^{(\vec{\kappa})}_j \delta_{il} + k^{(\vec{\kappa})}_l \delta_{ij} + k^{(\vec{\kappa})}_i \delta_{lj}}{|\vec{k}^{(\vec{\kappa})}|^2} - 2 \frac{k^{(\vec{\kappa})}_i k^{(\vec{\kappa})}_j k^{(\vec{\kappa})}_l}{|\vec{k}^{(\vec{\kappa})}|^4}\right) e^{-\mathrm{i} \vec{k}^{(\vec{\kappa})} \cdot (\vx - \vxz)} \, .
\end{equation*}
The periodicity holds for both arguments, as well as for the Stokeslet:
\begin{subequations}
	\begin{align}
		G_{ij}(\vx,\vxz) &= G_{ij}(\vx + \vec{X}^{(\vec{\alpha})}, \vxz + \vec{X}^{(\vec{\alpha}'})) \, , 
		\label{eq:G:Periodic}
		\\
		\breve{T}_{ijl}(\vx,\vxz) &= \breve{T}_{ijl}(\vx + \vec{X}^{(\vec{\alpha})}, \vxz + \vec{X}^{(\vec{\alpha}')}) 
		\quad \forall \vec{\alpha}, \vec{\alpha}' \in \mathbb{Z}^3 \, , \quad i,j,l=1,2,3 \, . \label{eq:T:Periodic}
	\end{align}
\end{subequations}
This also implies that $\vxz$ can actually be located within almost the whole space $\mathbb{R}^3 \setminus \{\vx\}$.

Equations~\eqref{eq:Green:G} and \eqref{eq:Green:T} are the Green's functions used below.
The above given forms, however, are not computable in practice due to their slow convergence.
Dramatic speedups are achieved by using the Ewald decomposition \citep{Hasimoto1959}.
The final formulas are given by \citet{Zhao2010}.

We will additionally need two general relations. 
Hence, let $\Obj$ be some object with a closed surface $\partial \Obj$, and let $\{ \Obj^{(\vec{\alpha})} \}$ be the set of $\Obj$ that is offset with all possible periodic grid vectors $\vec{X}^{(\vec{\alpha})}$ from equation~\eqref{eq:PerGridVec}.
Then, for some proper Green's functions:
\begin{align}
	\oint_{\partial \Obj} T_{ijl}(\vxz,\vx) n_j(\vx) \, \dd S(\vx) &= \delta_{il}
	\begin{cases}
		8\pi & \text{if $\vxz \in \{ \Obj^{(\vec{\alpha})} \}$} \\
		4\pi & \text{if $\vxz \in \{ \partial \Obj^{(\vec{\alpha})} \}$} \\
		0 & \text{otherwise}
	\end{cases}
	\, , \quad
	\vxz \in \mathbb{R}^3
	\, ,
	\label{eq:TnInt1}
	\\
	&\text{\citep[eq.~(3.2.7)]{PozrikidisBook92},}
	\nonumber
	\\
	\oint_{\partial\Obj} n_i(\vx) G_{ij}(\vx,\vxz) \, \dd S(\vx) &= \oint_{\partial\Obj} G_{ji}(\vx,\vxz) n_i(\vx) \, \dd S(\vx) = 0 \, , \quad \vxz \in \mathbb{R}^3
	\, , 
	\label{eq:GnInt}
	\\
	&\text{\citep[eqs.~(2.1.4) and (3.1.3)]{PozrikidisBook92}.}
	\nonumber
\end{align}

\subsubsection{Boundary conditions for a 3D periodic domain}
As we wish to simulate a 3D periodic domain with the BI equation~\eqref{eq:BI:GeneralWithGamma}, we stipulate as a boundary condition that the velocity $\uv$ shall be periodic, i.e.\
\begin{equation}\label{eq:uFull:Periodic}
	\uv(\vx + \vec{X}^{(\vec{\alpha})}) = \uv(\vx) \quad \forall \vx \in \mathbb{R}^3 \quad \text{and} \quad \forall \vec{\alpha} \in \mathbb{Z}^3 \, .
\end{equation}
Note that in principle $\uv$ may contain a linear component, as long as the arrangement of the objects and their images retain a triclinic periodicity.
This could be used to simulate an overall linear shear flow \citep{Phan-Thien1991,Pozrikidis1993,Pozrikidis1999,Li1995,Loewenberg1996}.
However, this also usually implies that the basis vectors $\vec{a}^{(i)}$ of the unit cell have to be time dependent, leading to additional problems for longer simulation times \citep[p.~221]{Blawzdziewicz2007}.

As for the pressure Green's function from equation~\eqref{eq:Green:P}, it is possible to have a non-zero pressure gradient over the unit cell that drives a certain average flow $\braket{\uv}_\Gamma$. 
Hence, we can decompose the pressure as $P(\vx) = \braket{\vec{\nabla} P}_\Gamma \cdot \vx + \breve{P}(\vx)$ for $\vx \in \mathbb{R}^3$ where $\breve{P}$ denotes the periodic part, and
\begin{equation}\label{eq:DefAverage}
	\braket{\bullet}_\Gamma := \frac{1}{V_\Gamma} \int_{\Gamma} \bullet \,\, \dd x^3
\end{equation}
is the average over the whole unit cell.
Using equations~\eqref{eq:Sigma} and \eqref{eq:Trac}, we find for the traction \citep[also compare][eq.~(2.23)]{Zick1982}
\begin{equation}\label{eq:f:Decomp}
	f_i(\vx) = - \braket{\vec{\nabla} P}_\Gamma \cdot \vec{x} \, n_i(\vx) + \breve{f}_i(\vx) \, , \quad i = 1,2,3 \, ,
\end{equation}
where $\vec{\breve{f}}$ denotes the periodic part.

We finally remark that the BI equation~\eqref{eq:BI:GeneralWithGamma} can be used with \emph{any} Green's function to simulate a periodic domain, provided that the periodicity condition \eqref{eq:uFull:Periodic} is enforced.
Using the periodic Green's functions from equations~\eqref{eq:Green:G} and \eqref{eq:Green:T} is merely a convenient choice since the integrals over $\partial \Gamma$ then yield simple expressions, as shown next.

\subsubsection{Computing the integrals over the unit cell's surface for a periodic domain}

\paragraph{\textit{Single-layer integral:}}
\label{ch:UCellSurf:SL}
With the help of the decomposition~\eqref{eq:f:Decomp}, the periodicity of the involved quantities (due to eqs.~\eqref{eq:G:Periodic} and \eqref{eq:uFull:Periodic}) and the divergence theorem together with eqs.~\eqref{eq:AvgG:Zero} an \eqref{eq:GreenIncompress:u}, one can show \citep{Zick1982}
\begin{equation}
	(\mathcal{N}_{\partial \Gamma} \vec{f})_j(\vxz) = 0 \, , \quad \vxz \in \Omega \, , \quad j=1,2,3 \, .
\end{equation}

\paragraph{\textit{Double-layer integral:}}
\label{ch:UCellSurf:DL}
It is tempting to assume that the double-layer integral $\mathcal{K}_{\partial\Gamma} \uv$ over the unit cell surface $\partial \Gamma$ is also zero.
This, however, will turn out to be wrong if a net flow is imposed and/or if volume-changing objects are included.
The latter will lead to an additional novel contribution to the equations.
Following \citet{Zick1982}, the first step for the evaluation is to use the decomposition~\eqref{eq:Green:T}.
The integral term containing $\breve{T}_{ijl}$ vanishes due to periodicity (compare equations~\eqref{eq:uFull:Periodic} and \eqref{eq:T:Periodic}, and note that normal vectors on opposing sides of $\partial \Gamma$ are antiparallel).
The other term is treated by adding and subtracting integrals over the objects, resulting in
\begin{equation}
	(\mathcal{K}_{\partial\Gamma} \uv)_j(\vxz) = -\frac{8\pi}{V_\Gamma}
	\Bigg[\>
		\oint\limits_{\partial \Omega} x_j u_i(\vx) n_i(\vx) \, \dd S(\vx)
		- \oint\limits_{\mathclap{\partial \Caps \cup \partial \Bub}} x_j u_i(\vx) n_i(\vx) \, \dd S(\vx)
	\Bigg]
	\, .
\end{equation}
Integrals over walls give zero contributions since the velocity $\uv$ is continuous across their surface, but the normal vectors on opposite sides have different signs.
Using the continuity of the normal velocity across the interfaces and applying the divergence theorem while watching out for the normal vector convention (always into the ambient fluid $\Omega$), one obtains
\begin{equation}
	\begin{split}
		(\mathcal{K}_{\partial\Gamma} \uv)_j(\vxz) = -\frac{8\pi}{V_\Gamma}
		\Bigg[
			&-\int_\Omega u_j(\vx) \, \dd x^3 - \int_{\mathcal{\Caps \cup \Bub}} u_j(\vx) \, \dd x^3 \\
			&- \int\limits_{\mathclap{\Omega \cup \Caps }} x_j \vec{\nabla} \cdot \uv(\vx) \, \dd x^3
			- \int\limits_{\Bub} x_j \vec{\nabla} \cdot \uv(\vx) \, \dd x^3
		\Bigg]
		\, .
	\end{split}
\end{equation}
The integrals in the first line can be combined to $\int_\Gamma \uv \, \dd x^3 = V_\Gamma \braket{\uv}_\Gamma$, with the average defined in equation~\eqref{eq:DefAverage} (walls are nullsets).
Moreover, the divergence of the velocity vanishes in $\Omega$ and the capsules $\Caps$ because of eq.~\eqref{eq:Incompress} (Stokesian fluids).
Furthermore, the last term is absent in existing formulations without volume-changing objects, but here it is non-zero in general and therefore requires special attention.

A more usable form of this last term may be obtained by using the model from equation~\eqref{eq:Bubble:Div}.
For a particular bubble $\Bub_k$, $k=1,\ldots,N_\Bub$, we immediately find
\begin{equation}\label{eq:xDivU}
	\int_{\Bub_k} x_j \vec{\nabla} \cdot \uv(\vx) \, \dd x^3 = c_k V_{\Bub_k} \chi^{(\Bub_k)}_j \, , \quad j=1,2,3 \, ,
\end{equation}
where we have defined the geometric centroid
\begin{equation}\label{eq:Centroid}
	\vec{\chi}^{(\Bub_k)} := \frac{1}{V_{\Bub_k}} \int_{\Bub_k} \vx \, \dd x^3 \, .
\end{equation}
$V_{\Bub_k}$ is the bubble volume.
We obtain a connection between $c_k$ to the more intuitive flux~$Q_{\Bub_k}$ out of or into a bubble by computing
\begin{equation}\label{eq:Bub:Flux:c}
	Q_{\Bub_k} 
	:= \oint\limits_{\partial \Bub_k} u_i n_i \, \dd S 
	= \int\limits_{\Bub_k} \vec{\nabla} \cdot \uv \, \dd x^3
	= c_k V_{\Bub_k} \, .
\end{equation}

Putting it all together, the double-layer integral over $\partial \Gamma$ hence becomes
\begin{equation}\label{eq:Bub:Flux:allTogether}
	(\mathcal{K}_{\partial\Gamma} \uv)_j(\vxz) = 8 \pi \braket{u_j}_\Gamma + \frac{8 \pi}{V_\Gamma} \sum_{k=1}^{N_\Bub} Q_{\Bub_k} \chi^{(\Bub_k)}_j \, , \quad \vxz \in \Omega \, , \quad j=1,2,3 \, .
\end{equation}
Similar to $\braket{G_{ij}}_\Gamma$ for the Green's function from equation~\eqref{eq:Green:G}, the average velocity $\braket{u_j}_\Gamma$ is a free parameter that can be used to drive a flow through the system \citep{Zhao2010}.
Since the flux $Q_{\Bub_k}$ is also a prescribed input parameter, and the centroid of an object can be easily computed \citep[see e.g.\@][]{Zhang2001a}, we have therefore obtained an expression of the BI equation that is actually usable in practice.

\subsubsection{The full boundary integral equation and some remarks}\label{sec:FinalBIEq}
The BI equation~\eqref{eq:BI:GeneralWithGamma} thus becomes
\begin{equation}\label{eq:BI:Periodic}
	\begin{split}
		u_j(\vxz) = {}&{} \braket{u_j}_\Gamma - \frac{1}{8 \pi \mu} \sum_{q=1}^{N_\Obj} (\mathcal{N}_{\partial \Obj_q} \vec{F})_j(\vxz) + \frac{1}{8\pi} \sum_{q=1}^{N_\Obj} (1 - \lambdabar_{\Obj_q}) (\mathcal{K}_{\partial \Obj_q} \vec{u})_j(\vxz) \\
		&+ \frac{1}{V_\Gamma} \sum_{k=1}^{N_\Bub} Q_{\Bub_k} \chi^{(\Bub_k)}_j 
		\, , \qquad \vxz \in \Omega \, , \quad j = 1,2,3
	\end{split}
\end{equation}
with the novel bubble term in the last line.
The single- as well as the double-layer integrals must be evaluated with the appropriate Green's functions from equations~\eqref{eq:Green:G} and \eqref{eq:Green:T}, respectively.

We first remark that equation~\eqref{eq:BI:Periodic} reduces to the case of the infinite system for $V_\Gamma \rightarrow \infty$ (i.e.\ $\frac{1}{V_\Gamma} = 0$), as the flux terms vanish and the Green's functions converge to the well-known expressions for an infinite system, i.e.\ \citep{PozrikidisBook92}
\begin{subequations}\label{eq:Green:Inf}
	\begin{align}
		G_{ij}(\vx,\vxz) = \frac{\delta_{ij}}{|\hat{\vec{x}}|} + \frac{\hat{x}_i \hat{x}_j}{|\hat{\vec{x}}|^3} \, , \quad \vx,\vxz \in \mathbb{R}^3 \setminus \{ \vx = \vxz \} \, , \quad i,j=1,2,3 \label{eq:G:Inf} \\
\intertext{for the Stokeslet, and}
		T_{ijl}(\vx,\vxz) = -6 \frac{\hat{x}_i \hat{x}_j \hat{x}_l}{|\hat{\vec{x}}|^5}  \, , \quad \vx,\vxz \in \mathbb{R}^3 \setminus \{ \vx = \vxz \} \, , \quad i,j,l=1,2,3 \label{eq:T:Inf}
	\end{align}
\end{subequations}
for the Stresslet, where $\hat{\vec{x}} := \vx - \vx_0$.

Second, the imposed average flow $\braket{u_j}_\Gamma$ can be interpreted as the flow that would prevail in the absence of any objects, and is the most convenient quantity to prescribe a certain flow.
A relationship to the corresponding pressure gradient is easily established \citep[eq.~(8)]{Zhao2010}. 

Third, the prescribed fluxes for the bubbles cannot be chosen arbitrarily.
To see this, consider on the one hand
\begin{equation}
	\oint_{\partial\Gamma} u_i n_i \, \dd S = 0 \, ,
\end{equation}
where we used once again the periodicity of the velocity $\uv$ and the opposite signs of the normal vector $\vec{n}$ on opposing sides of the unit cell surface $\partial\Gamma$.
On the other hand,
\begin{equation}
	\oint_{\partial\Gamma} u_i n_i \, \dd S
	= \oint\limits_{\partial \Omega} u_i n_i \, \dd S - \oint\limits_{\mathclap{\partial\Caps \cup \partial\Bub}} u_i n_i \, \dd S
	= - \sum_{k=1}^{N_\Bub} Q_{\Bub_k} \, ,
\end{equation}
where the divergence theorem and the incompressibility of the velocity in $\Omega$ and $\Caps$ together with the definition of the flux have been employed.
Combining these two equations, we find
\begin{equation}\label{eq:SumFluxZero}
	\sum_{k=1}^{N_\Bub} Q_{\Bub_k} = 0 \, .
\end{equation}
Hence, the fluxes must be chosen such that the total flux is zero, i.e.\ that the outer fluid volume is conserved.
This implies that at least two bubbles are required that oscillate out-of-phase for the periodic system.
Furthermore, if the ambient fluid domain $\Omega$ is not simply connected (imagine a tube separating the unit cell $\Gamma$ into an inner and an outer domain), the fluxes must be chosen such that the volume within the respective domains is conserved.

Fourth, the initial assumption that all objects are completely located within the unit cell can be relaxed.
Surface integrals can be evaluated continuously over the objects' surfaces even if these surfaces cross the boundary of the unit cell, as is common in simulations of dense suspensions. 
This property is usually silently assumed in the literature, although it is \textit{a priori} unclear if it holds due to the linear part in the stresslet~\eqref{eq:Green:T} and the non-periodic centroid term in equation~\eqref{eq:BI:Periodic}.
With volume-changing objects it actually follows from the non-trivial interplay between the integrals over the unit cell's surface $\partial \Gamma$ and the centroid term.
We proof this result explicitly in appendix~\ref{sec:overlap}.

\subsection{Fredholm integral equation} \label{sec:Fredholm}

The BI equation~\eqref{eq:BI:Periodic} can be used to compute the flow velocity everywhere within the ambient fluid $\Omega$ if the tractions/traction-jumps, the velocities and the fluxes are known.
However, we usually prescribe either the tractions/traction-jumps \emph{or} the velocities, while the other quantity is unknown.
The basic idea to obtain a determining equation is to use eq.~\eqref{eq:BI:Periodic} and move the evaluation point $\vxz$ onto the surface of the objects.
We thereby obtain a so-called Fredholm integral equation which can be solved for the unknown variables.
Section~\ref{sec:FBI:Basic} summarizes the result of this standard procedure. The subsequent sections are devoted to ensuring and proving the uniqueness of the solution. 
This cannot be taken for granted if bubbles are included. But even without bubbles it has so far not yet been rigorously proven in periodic systems.

\subsubsection{Basic equation}\label{sec:FBI:Basic}
We now assume that all objects have surfaces of Lyapunov type \citetext{i.e.\ are \enquote{smooth}, see \citealp[p.~96]{Kress2014}, for more details}.
If corners or edges within the surfaces existed, the results would change, see e.g.\ \citet[ch.~2.5]{Kress2014} and \citet[p.~37]{PozrikidisBook92}.
For smooth objects the single-layer potential is continuous \citep[ch.~3.4.4]{Kohr2004} if $\vxz$ is moved across the surface and the double-layer potential makes a jump \citep[eq.~(2.3.12)]{PozrikidisBook92}.
Following these two references, we obtain the Fredholm boundary integral (FBI) equation as
\begin{align}\label{eq:FBI:NonUnique}
	\begin{aligned}
		u_j(\vxz) = \frac{2}{1 + \lambdabar_{\Obj_k}} 
		\Bigg[
			& \braket{u_j}_\Gamma
			- \frac{1}{8 \pi \mu} \sum_{q=1}^{N_\Obj} (\mathcal{N}_{\partial \Obj_q} \vec{F})_j(\vxz) \\
			&+ \frac{1}{8\pi} \sum_{q=1}^{N_\Obj} (1 - \lambdabar_{\Obj_q}) (\mathcal{K}_{\partial \Obj_q} \vec{u})_j(\vxz)
			+ \frac{1}{V_\Gamma} \sum_{q=1}^{N_\Bub} Q_{\Bub_q} \chi^{(\Bub_q)}_j
		\Bigg] \, ,
	\end{aligned}
	\\
	\vxz \in \partial\Obj_k \, , \quad k=1,\ldots,N_{\Obj} \, , \quad j = 1,2,3 \, .
	\nonumber
\end{align}
Note that the evaluation point $\vxz$ is located directly on the surfaces of the objects.
The single- as well as the double-layer integrals exist as improper integrals \citep[ch.~6]{Kress2014}.
Equation~\eqref{eq:FBI:NonUnique} corresponds to the first two lines in equation~\eqref{eq:FBI:Intro}. 
The missing two terms will be introduced in section~\ref{sec:PatchedFBI} to ensure uniqueness of the solution.

The above FBI equation can in principle be used to find the unknown quantity\,--\,if the solution were unique in all cases.
In our application presented in section~\ref{sec:Application}, we prescribe the traction/traction jump $\vec{F}$ on all objects (also on walls for efficiency reasons).
This leads to a so-called Fredholm equation of the second kind that is amenable to the Fredholm theory.
As will be analyzed and fixed below, equation~\eqref{eq:FBI:NonUnique} has multiple solutions if bubbles are included.
Without bubbles, the solution is unique, as will also be shown below.

On the other hand, prescribing the velocities on all objects leads to a Fredholm equation of the first kind which has various unfavorable properties:
The solution is in general not unique, the condition number grows with resolution \citetext{compare \citealp[p.~127]{Karrila1989} and \cite{Marin2012a}}, and no general mathematical theory exists.
These are the reasons why alternative approaches for such problem statements have been invented, e.g.\ the completed double-layer boundary integral method \citep{Kohr2004,Karrila1989,Zhao2012a}.

Finally, prescribing the velocities on some objects and the tractions on others yields a mixed type equation.
Similar to the first kind type, no general theory exists and at least parts of the system have \enquote{difficult} properties.

\subsubsection{Fredholm theory and the non-uniqueness of solution}\label{ch:FredholmTheory}
Henceforth, we consider the case when equation~\eqref{eq:FBI:NonUnique} is a Fredholm integral equation of the second kind, i.e.\ when all velocities are unknown.
In order to apply the Fredholm theory, we need to introduce the homogeneous version of equation~\eqref{eq:FBI:NonUnique},
\begin{align}\label{eq:FBI:NonUnique:Hom}
	h_j(\vxz) = 
	\frac{1}{4\pi} \frac{1}{1 + \lambdabar_{\Obj_k}} \sum_{q=1}^{N_\Obj} (1 - \lambdabar_{\Obj_q}) \oint_{\partial \Obj_q} h_i(\vx) T_{ijl}(\vx,\vxz) n_l(\vx) \, \dd S(\vx)
	\, ,
	\\
	\vxz \in \partial\Obj_k \, , \quad k=1,\ldots,N_{\Obj} \, , \quad j = 1,2,3 \, ,
	\nonumber
\end{align}
where $\vec{h}$ denotes an eigensolution to the eigenvalue $1$.
Note again that the double-layer integral is meant to be absent for $\lambdabar_{\Obj_q} = 1$ objects, especially walls.
The corresponding adjoint equation \citetext{\citealp[p.~106]{PozrikidisBook92} and \citealp{Kress2014}} is given by
\begin{equation}\label{eq:FBI:NonUnique:Adj}
	a_j(\vxz) = 
		\frac{1 - \lambdabar_{\Obj_k}}{4 \pi} M_j[\vec{a}](\vxz)
	\, ,
	\quad
	\vxz \in \partial\Obj_k \, , \quad k=1,\ldots,N_{\Obj} \, , \quad j = 1,2,3 \, ,
\end{equation}
with the eigensolution $\vec{a}$ and the abbreviation
\begin{equation}\label{eq:A:M}
	M_j[\vec{a}](\vxz) := \sum_{q=1}^{N_\Obj} \frac{1}{1 + \lambdabar_{\Obj_q}} n_l(\vxz) \oint_{\partial \Obj_q} a_i(\vx) T_{jil}(\vxz,\vx) \, \dd S(\vx) \, .
\end{equation}

The integral kernels and their adjoints are weakly singular \citetext{see \citealp[p.~31 and theorem 4.12]{Kress2014}, \citealp[pages~36 and 113]{PozrikidisBook92}, as well as \citealp[p.~137]{Karrila1989}}.
This means that all occurring integral operators are compact \citep[theorem 2.30]{Kress2014}, and that the eigensolutions of the homogeneous and adjoint equations are continuous \citep[see][p.~58]{Kress2014}.

For walls (i.e.\ open objects) we adopt the convention that closed surface integrals~$\oint$ go over both sides.
Due to the continuity of the eigensolutions, however, they provide no contribution.
Alternatively, as in the derivation of the BI equation, one can also revert back to walls with finite thickness and take the limit afterwards.
The formulas in the following have to be interpreted in the same way.
Notice that in the adjoint equation integrals over objects appear which actually have $\lambdabar_{\Obj_q} = 1$.

The compactness of the integral operators also implies that the Fredholm theory can be used to make precise statements about uniqueness and existence of solutions \citetext{see \citealp[p.~114]{PozrikidisBook92} and \citealp[p.~55~f.]{Kress2014}}.
For the present purpose the major theorem can be written as follows:
\begin{theorem}[Fredholm alternative]\label{theo:Fredholm}
	\begin{enumerate}
		\item The homogeneous and adjoint equations \eqref{eq:FBI:NonUnique:Hom} and \eqref{eq:FBI:NonUnique:Adj} have the same finite number of eigensolutions. 
		
		\item If the homogeneous equation \eqref{eq:FBI:NonUnique:Hom} has only the trivial solution $\vec{h} \equiv 0$, then the full equation~\eqref{eq:FBI:NonUnique} has exactly one solution (existence and uniqueness).
		
		\item If the homogeneous equation \eqref{eq:FBI:NonUnique:Hom} has a nontrivial solution, then the full equation~\eqref{eq:FBI:NonUnique} has solutions if and only if all eigensolutions $\vec{a}$ of the adjoint equation~\eqref{eq:FBI:NonUnique:Adj} satisfy
		\begin{equation}\label{eq:CondR}
			\sum_{k=1}^{N_\Obj} \oint_{\partial\Obj_k} R_j(\vx) a_j(\vx) \, \dd S(\vx) = 0 \, , \quad k = 1,\ldots,N_\Obj \, .
		\end{equation}
		Here, $\vec{R}$ contains all fully known terms (i.e.\ terms that are missing in the homogeneous equation).
	\end{enumerate}
\end{theorem}

To arrive at uniqueness and existence statements therefore requires to know \emph{all} solutions of the adjoint equation.
In case of equation~\eqref{eq:FBI:Unique:Adj}, the solutions are
\begin{equation}\label{eq:AdjSolutions:NonUnique}
	\vec{a}^{(m)}(\vxz) =
	\begin{cases}
		\vec{n}(\vxz) & \text{if $\vxz \in \partial \Bub_m$} \\
		0 & \text{otherwise}
	\end{cases}
	\, , \quad
	m = 1,\ldots,N_\Bub \, , \quad \vxz \in \partial \Obj \, .
\end{equation}
That these are indeed solutions can be easily shown with the help of equation~\eqref{eq:TnInt1}.
To show that they are the \emph{only} solutions requires a somewhat longer procedure, similar to section~\ref{sec:Proof} (we skip it as it is not of any major interest here).
Thus, the homogeneous equation~\eqref{eq:FBI:NonUnique:Hom} also as $N_\Bub$ solutions.
Furthermore, we have 
$R_j(\vxz) = \frac{2}{1 + \lambdabar_{\Obj_k}} 
		\big[
			 \braket{u_j}_\Gamma
			- \frac{1}{8 \pi \mu} \sum_{q=1}^{N_\Obj} (\mathcal{N}_{\partial \Obj_q} \vec{F})_j(\vxz)
			+ \frac{1}{V_\Gamma} \sum_{q=1}^{N_\Bub} Q_{\Bub_q} \chi^{(\Bub_q)}_j
		\big]$,
and all solutions $\vec{a}^{(m)}$ satisfy condition~\eqref{eq:CondR} due to equation~\eqref{eq:GnInt}.
Hence, by virtue of the Fredholm alternative, the FBI equation~\eqref{eq:FBI:NonUnique} has more than one solution if bubbles are included.

\subsubsection{Ensuring uniqueness: The full equation}
\label{sec:PatchedFBI}

Equation~\eqref{eq:FBI:NonUnique} does not have a unique solution because the flux of the bubbles is not determined by the equation.
To remove this ambiguity, we introduce additional terms into equation~\eqref{eq:FBI:NonUnique} in such a way that the solution of the new equation is unique and simultaneously also a solution of the old equation~\eqref{eq:FBI:NonUnique}.
In analogy to \citet{Nie2002}, the modified equation is then given by eq.~\eqref{eq:FBI:Intro}, which is repeated here for convenience:
\begin{align}\label{eq:FBI:Unique}
	\begin{aligned}
		u_j(\vxz) = {}&\frac{2}{1 + \lambdabar_{\Obj_k}} 
		\Bigg[
			 \braket{u_j}_\Gamma
			- \frac{1}{8 \pi \mu} \sum_{q=1}^{N_\Obj} (\mathcal{N}_{\partial \Obj_q} \vec{F})_j(\vxz) \\
			&\qquad\qquad\quad + \frac{1}{8\pi} \sum_{q=1}^{N_\Obj} (1 - \lambdabar_{\Obj_q}) (\mathcal{K}_{\partial \Obj_q} \vec{u})_j(\vxz)
			+ \frac{1}{V_\Gamma} \sum_{q=1}^{N_\Bub} Q_{\Bub_q} \chi^{(\Bub_q)}_j
		\Bigg] \\
		&- \frac{1 - \lambdabar_{\Obj_k}}{1 + \lambdabar_{\Obj_k}} z^{(k)}_j(\vxz) \left[ \oint_{\partial \Obj_k} u_l(\vx) n_l(\vx) \, \dd S(\vx) - Q_{\Obj_k} \right] \, ,
	\end{aligned}
	\\
	\vxz \in \partial\Obj_k \, , \quad k=1,\ldots,N_{\Obj} \, , \quad j = 1,2,3 \, .
	\nonumber
\end{align}
This is the central equation that is solved in our VCO-BIM.
It is a direct method as the solution $\uv$ is the physical velocity rather than an auxiliary field.
The fact that eq.~\eqref{eq:FBI:Unique} has exactly one solution will be proven below and constitutes a central result of the present work.
$\vec{z}^{(k)}$ can be chosen arbitrarily, as long as the restriction
\begin{equation}\label{eq:z:n}
	\oint_{\partial\Obj_k} z^{(k)}_j(\vx) n_j(\vx) \, \dd S(\vx) = 1 \, , \quad k = 1,\ldots,N_{\Obj}
\end{equation}
is satisfied.
A convenient choice is
\begin{equation}\label{eq:z:WithSurf}
	\vec{z}^{(k)}(\vx) = \vec{n}(\vx) / S_{\Obj_k} \, , \quad \vx \in \partial \Obj_k \, , \quad  k = 1,\ldots,N_{\Obj} \, ,
\end{equation}
where $S_{\Obj_k}$ is the surface area of the $k$'th object.
Furthermore, $Q_{\Obj_k}$ is the prescribed flux of object $\Obj_k$, which must be zero for all entities except the bubbles.

The integral term in the last line of eq.~\eqref{eq:FBI:Unique} can be interpreted as part of the so-called Wielandt deflation procedure \citep{Sangtae2005} for objects with viscosity ratios $\lambdabar_{\Obj_k} > 0$.
This method is sometimes used to accelerate the convergence rate of iterative solution algorithms \citep{Zinchenko1997,Zinchenko2000}, but is 
otherwise optional for $\lambdabar_{\Obj_k} > 0$.
Choosing not to use it amounts to setting $\vec{z}^{(k)} = 0$ (in which case condition~\eqref{eq:z:n} must be disregarded).
For bubbles ($\lambdabar_{\Obj_k} = 0$) that oscillate ($Q_{\Obj_k}\neq 0$), however, an additional new term including the surface flux $Q_{\Obj_k}$ needs to be taken into account. 
Note that the last line is an essential ingredient to ensure uniqueness of the solution for bubbles (with and without volume changes), contrary to the usual situation found in the literature.
We also remark that the FBI equation remains valid in an infinite system ($\Gamma \rightarrow \mathbb{R}^3$) similar to the BI equation from section~\ref{sec:FinalBIEq}.

The solution of the patched equation~\eqref{eq:FBI:Unique} is still a solution of the old equation~\eqref{eq:FBI:NonUnique}.
This can be shown by multiplying eq.~\eqref{eq:FBI:Unique} with the normal vector $n_j$, summing over $j$ and integrating over some object's surface $\partial \Obj_k$.
Using relations~\eqref{eq:TnInt1}, \eqref{eq:GnInt} and \eqref{eq:z:n} gives $\oint_{\partial \Obj_k} u_i n_i \, \dd S = Q_{\Obj_k}$, meaning that the flux out of or into the object matches with the prescribed value $Q_{\Obj_k}$, as expected.
Moreover, substituting it back into eq.~\eqref{eq:FBI:Unique} recovers the original equation~\eqref{eq:FBI:NonUnique}.

Despite the patch, equation~\eqref{eq:FBI:Unique} is still a Fredholm integral equation of the second kind for the velocities on all objects if $\vec{F}$ is given.
In order to apply the Fredholm theory, we need to introduce again the adjoint of the homogeneous equation, namely
\begin{align}\label{eq:FBI:Unique:Adj}
	a_j(\vxz) = 
		\frac{1 - \lambdabar_{\Obj_k}}{4 \pi} M_j[\vec{a}](\vxz)
		- \frac{1 - \lambdabar_{\Obj_k}}{1 + \lambdabar_{\Obj_k}} n_j(\vxz) \oint_{\partial\Obj_k} z^{(k)}_l(\vx) a_l(\vx) \, \dd S(\vx)
	\, ,
	\\
	\vxz \in \partial\Obj_k \, , \quad k=1,\ldots,N_{\Obj} \, , \quad j = 1,2,3 \, ,
	\nonumber
\end{align}
with the abbreviation $M_j$ from equation~\eqref{eq:A:M}.
$\vec{a}$ denotes again the eigensolutions to the eigenvalue $1$.

The goal now is to prove that eq.~\eqref{eq:FBI:Unique:Adj} has only the obvious solution $\vec{a} = 0$.
Theorem~\ref{theo:Fredholm} then implies that the actual FBI equation~\eqref{eq:FBI:Unique} has exactly one solution.
Unfortunately, the procedure used by \citet[p.~116~f.]{PozrikidisBook92} cannot be adapted for the proof in the periodic system because the artificial flow that he introduces is a source field.
This works in infinite domains where the fluid can escape to infinity, but violates the conservation of the outer fluid volume in periodic domains (even if all objects were volume conserving).
We keep the proof rather general, as none has been published before for the periodic system to the best of our knowledge.

\subsubsection{Proof of existence and uniqueness of the solution}\label{sec:Proof}

As a start, we assume that there is at least one non-trivial solution $\vec{a}$ of the adjoint equation~\eqref{eq:FBI:Unique:Adj}.
We then define an artificial velocity field similar to \citet[\S 4]{Odqvist1930} by
\begin{equation}\label{eq:DefA}
	A_j(\vxz) := \sum_{q=1}^{N_\Obj} \frac{1}{1 + \lambdabar_{\Obj_q}} \oint_{\partial\Obj_q} a_i(\vx) G_{ij}(\vx,\vxz) \, \dd S(\vx) \, , \quad \vxz \in \mathbb{R}^3 \, , \quad j=1,2,3 \, .
\end{equation}
This field has a few special properties.
First of all, $\vec{A}$ is defined within the whole space $\mathbb{R}^3$ because it inherits the domain from the periodic Stokeslet and because such a single-layer integral exists in the improper sense if $\vxz$ is located on any surface.
Moreover, because the eigensolutions $\vec{a}$ of the adjoint equation are continuous as explained above, $\vec{A}$ is continuous as $\vxz$ crosses any object surface $\partial\Obj$ \citep[ch.~3.4.4]{Kohr2004}.
The field is also periodic due to eq.~\eqref{eq:G:Periodic}, and we have
\begin{equation}\label{eq:A:Incompress}
	\vec{\nabla} \cdot \vec{A}(\vxz) = 0 \, , \quad \vxz \in \mathbb{R}^3
\end{equation}
due to equation~\eqref{eq:GreenIncompress:u}
and
\begin{equation}\label{eq:A:ZeroFlow}
	\braket{\vec{A}}_\Gamma = 0
\end{equation}
because of equation~\eqref{eq:AvgG:Zero}.

Furthermore, if we define the associated pressure as
\begin{equation}\label{eq:A:Press}
	P^{\vec{A}}(\vxz) := \mu \sum_{q=1}^{N_\Obj} \frac{1}{1 + \lambdabar_{\Obj_q}} \oint_{\partial \Obj_q} a_i(\vx) p_i(\vxz,\vx) \, \dd S(\vx)
\end{equation}
with the Green's function $\vec{p}$ for the pressure from eq.~\eqref{eq:Green:P}, one can show with the help of the singular Stokes equation \citep[eq.~(2.2)]{Pozrikidis1996} as well as equations~\eqref{eq:Green:G} and \eqref{eq:Green:P} that the Stokes equation with $\vec{A}$ is satisfied everywhere but on the surfaces, i.e.\
\begin{equation}\label{eq:A:Stokes}
	-\vec{\nabla} P^{\vec{A}}(\vx) + \mu \nabla^2 \vec{A}(\vx) = 0 \, , \quad \vx \in \mathbb{R}^3 \setminus \partial \{ \Obj^{(\vec{\alpha})} \} \, .
\end{equation}
$\{ \Obj^{(\vec{\alpha})} \}$ denotes the objects and all of their periodic images.
The Stokes equation can alternatively be written as
\begin{equation}\label{eq:A:Stokes:Sigma}
	\frac{\partial \sigma^{\vec{A}}_{ij}(\vx)}{\partial x_i} = 0 \, , \quad \vx \in \mathbb{R}^3 \setminus \partial \{ \Obj^{(\vec{\alpha})} \} \, , \quad j=1,2,3 \, ,
\end{equation}
where the stress tensor is given by $\sigma^{\vec{A}}_{ij} := - P^{\vec{A}} \delta_{ij} + \mu \big( \frac{\partial A_i}{\partial x_j} + \frac{\partial A_j}{\partial x_i} \big) $.
Continuing, the traction (cf.\ eq.~\eqref{eq:Trac}) associated with $\vec{A}$ at the outside ($\vec{f}^{\vec{A},+}$) and inside surface ($\vec{f}^{\vec{A},-}$, normal vector to the outside) of some object $\Obj_k$ can be expressed as \citetext{compare \citealp[eq.~(3.4.61)]{Kohr2004} and \citealp[eq.~(2.15)]{Odqvist1930}}
\begin{subequations}\label{eq:A:TracInsideOutside}
	\begin{align}
		f^{\vec{A},+}_j(\vxz) &= - \frac{4\pi\mu}{1 + \lambdabar_{\Obj_k}} a_j(\vxz) + \mu M_j[\vec{a}](\vxz) \, , \\
		f^{\vec{A},-}_j(\vxz) &= + \frac{4\pi\mu}{1 + \lambdabar_{\Obj_k}} a_j(\vxz) + \mu M_j[\vec{a}](\vxz) \, ,
		\\ & \qquad \qquad \vxz \in \partial \Obj_k \, , \quad k = 1,\ldots,N_\Obj \, , \quad j=1,2,3 \, , \nonumber
	\end{align}
\end{subequations}
where $M_j$ was defined in equation~\eqref{eq:A:M}.
Solving equations~\eqref{eq:A:TracInsideOutside} for $\vec{a}$ and $\vec{M}$ leads to
\begin{subequations}\label{eq:A:aM}
	\begin{align}
		a_j(\vxz) &= - \frac{1 + \lambdabar_{\Obj_k}}{8 \pi \mu} \left[f^{\vec{A},+}_j(\vxz) - f^{\vec{A},-}_j(\vxz)\right] \, , \label{eq:A:aM:a} \\
		M_j[\vec{a}](\vxz) &= \frac{1}{2 \mu} \left[f^{\vec{A},+}_j(\vxz) + f^{\vec{A},-}_j(\vxz)\right] \, , 
		\\
		&\qquad\qquad \vxz \in \partial \Obj_k \, , \quad k = 1,\ldots,N_\Obj \, , \quad j=1,2,3 \, ,  \nonumber
	\end{align}
\end{subequations}

The last required property of the artificial field $\vec{A}$ is the energy conservation.
Following \citet[ch.~1.5]{PozrikidisBook92} and using equations~\eqref{eq:A:Incompress} and \eqref{eq:A:Stokes}, we can derive
\begin{equation*}
	\sum_{k=1}^{N_\Obj} \, \oint\limits_{\partial \Obj_k} f^{\vec{A},+}_j(\vx) A_j(\vx) \, \dd S(\vx) 
	+  \oint\limits_{\partial \Gamma} f^{\vec{A},+}_j(\vx) A_j(\vx) \, \dd S(\vx) 
	= -2 \mu \int\limits_{\Omega} \sum_{i,j=1}^{3} \left[E^{\vec{A}}_{ij}(\vx)\right]^2 \, \dd x^3 \, .
\end{equation*}
The strain rate tensor is defined as
\begin{equation}\label{eq:A:StrainRateTens}
	E^{\vec{A}}_{ij}(\vx) := \frac{1}{2} \left(\frac{\partial A_i(\vx)}{\partial x_j} + \frac{\partial A_j(\vx)}{\partial x_i}\right) \, , \quad \vx \in \mathbb{R}^3 \setminus \partial \{ \Obj^{(\vec{\alpha})} \} \, , \quad i,j=1,2,3 \, .
\end{equation}
The integral over the unit cell's surface $\partial\Gamma$ is simply zero.
This follows similar to the derivation of the double-layer integral in section~\ref{ch:UCellSurf:DL} because $\vec{A}$ is periodic, $\vec{f}^{\vec{A},+}$ contains at most a linear component (due to the definition of the traction, eq.~\eqref{eq:Trac}, the pressure, eq.~\eqref{eq:A:Press}, and the linear term in the pressure Green's function, eq.~\eqref{eq:Green:P}), $\vec{A}$ is incompressible according to eq.~\eqref{eq:A:Incompress} and because the average flow is zero as given by equation~\eqref{eq:A:ZeroFlow}.
Furthermore, a similar equation can be derived for the inside of the objects since $\vec{A}$ is defined everywhere.
In the end, we obtain
\begin{subequations}\label{eq:A:Ineq}
	\begin{align}
		\sum_{k=1}^{N_\Obj} \, \oint\limits_{\partial \Obj_k} f^{\vec{A},+}_j(\vx) A_j(\vx) \, \dd S(\vx) 
			= -2 \mu \int\limits_{\Omega} \sum_{i,j=1}^{3} \left[E^{\vec{A}}_{ij}(\vx)\right]^2 \, \dd x^3 \leqslant 0 \, , \label{eq:A:Ineq:1}
\intertext{and}
		\oint\limits_{\partial \Obj_k} f^{\vec{A},-}_j(\vx) A_j(\vx) \, \dd S(\vx) 
			= 2 \mu \int\limits_{\Obj_k} \sum_{i,j=1}^{3} \left[E^{\vec{A}}_{ij}(\vx)\right]^2 \, \dd x^3 \geqslant 0
			\, , \label{eq:A:Ineq:2} \\
			\quad k=1,\ldots,N_\Obj \, . \nonumber
	\end{align}
\end{subequations}
The inequalities follow because the viscosity $\mu$ is $> 0$ and the integrals contain kernels that are obviously greater or equal to zero.

With all required properties of the artificial field $\vec{A}$ established, we now proceed to show that the adjoint equation~\eqref{eq:FBI:Unique:Adj} does not have any non-trivial solution $\vec{a}$.
We will do this by a \textit{reductio ad absurdum} argument.
Hence, assume that there is at least one non-trivial solution denoted by $\vec{a}$.
Following \cite[\S 4]{Odqvist1930}, we begin by substituting eqs.~\eqref{eq:A:aM} into the adjoint~\eqref{eq:FBI:Unique:Adj}, leading to
\begin{align}
	f^{\vec{A},+}_j(\vxz) = \lambdabar_{\Obj_k} \left[ f^{\vec{A},-}_j(\vxz) + \frac{8 \pi \mu}{1 + \lambdabar_{\Obj_k}} n_j(\vxz) \oint_{\partial \Obj_k} z^{(k)}_l(\vx) a_l(\vx) \, \dd S(\vx) \right]
	\, , \\
	\vxz \in \partial \Obj_k \, , \quad k = 1,\ldots,N_\Obj \, , \quad j=1,2,3 \, . \nonumber
\end{align}
We now multiply with $A_j$, sum over $j$ and integrate over the surface of all objects.
The contribution from the second term is simply zero because we can use eq.~\eqref{eq:DefA} and write
\begin{equation}
	\begin{split}
		&\oint\limits_{\partial \Obj_k} A_j(\vxz) n_j(\vxz) \, \dd S(\vxz) = \\
		&= \sum_{q=1}^{N_\Obj} \frac{1}{1 + \lambdabar_{\Obj_q}} 
		\oint\limits_{\partial \Obj_q} a_i(\vx) 
		\underbrace{\Bigg[ \, \oint\limits_{\partial \Obj_k} G_{ij}(\vx,\vxz) n_j(\vxz) \, \dd S(\vxz) \Bigg]}_{= 0 \text{ because of eq.~\eqref{eq:GnInt}}}
		\dd S(\vx)
		= 0 \, .
	\end{split}
\end{equation}
Thus we find
\begin{equation}
	0 \geqslant \sum_{k=1}^{N_\Obj} \oint_{\partial \Obj_k} f^{\vec{A},+}_j(\vx) A_j(\vx) \, \dd S(\vx) = \sum_{k=1}^{N_\Obj} \lambdabar_{\Obj_k} \oint_{\partial \Obj_k} f^{\vec{A},-}_j(\vx) A_j(\vx) \, \dd S(\vx) \geqslant 0 \, .
\end{equation}
The inequality signs follow from the energy conservation~\eqref{eq:A:Ineq} and equation~\eqref{eq:DefLambdabar}.
Both inequality signs together imply
\begin{equation}
	\sum_{k=1}^{N_\Obj} \oint_{\partial \Obj_k} f^{\vec{A},+}_j(\vx) A_j(\vx) \, \dd S(\vx) = 0 \, ,
\end{equation}
and due to eq.~\eqref{eq:A:Ineq:1} we thus have $E^{\vec{A}}_{ij}(\vx) = 0$ for all $\vx \in \Omega$ and $i,j=1,2,3$.
This in turn means that $\vec{A}$ can only represent rigid-body motion within $\Omega$ \citep[ch.~1.5]{PozrikidisBook92}, i.e.\
\begin{equation}
	\vec{A}(\vx) = \vec{U}^\Omega + \vec{\omega}^\Omega \times \vec{x} \, , \quad \vec{x} \in \Omega \, .
\end{equation}
$\vec{U}^\Omega$ and $\vec{\omega}^\Omega$ are constants that do not depend on $\vec{x}$.
The symbol \enquote{$\times$} denotes the cross product.
Furthermore, using the periodicity of $\vec{A}$, we immediately find $\vec{\omega}^\Omega = 0$.

Next, we recall that $\vec{A}$ is continuous across the objects' surfaces, i.e.\ $\vec{A}|_{\partial \Obj} = \vec{A}|_{\Omega} = \vec{U}^\Omega$,
to derive
\begin{equation}\label{eq:A:fjMinusInt}
	\sum_{k=1}^{N_\Obj} \oint_{\partial \Obj_k} f^{\vec{A},-}_j(\vx) A_j(\vx) \, \dd S(\vx)
	= U^\Omega_i \sum_{k=1}^{N_\Obj} \int_{\Obj_k} \frac{\partial \sigma_{ji}^{\vec{A}}(\vx) }{\partial x_j} \, \dd x^3 
	= 0 \, .
\end{equation}
Here we used the definition of the traction from eq.~\eqref{eq:Trac}, the divergence theorem, the symmetry of the stress tensor and finally the Stokes equation~\eqref{eq:A:Stokes:Sigma}.
Summing expression~\eqref{eq:A:Ineq:2} over all objects, using eq.~\eqref{eq:A:fjMinusInt} and that the integral arguments on the right-hand side of eq.~\eqref{eq:A:Ineq:2} are positive, we find $E^{\vec{A}}_{ij}(\vx) = 0$ for all $\vx \in \Obj_k$, $k=1,\ldots,N_\Obj$ and $i,j=1,2,3$.
Hence, $\vec{A}$ must also represent rigid-body motion within every object:
\begin{equation}
	\vec{A}(\vx) = \vec{U}^{(k)} + \vec{\omega}^{(k)} \times \vec{x} \, , \quad \vx \in \Obj_k \, , \quad k=1,\ldots,N_\Obj \, .
\end{equation}
The $2 N_\Obj$ constants $\vec{U}^{(k)}$ and $\vec{\omega}^{(k)}$ could in principle be different for each $k$.
But because of the continuity of $\vec{A}$ across $\partial \Obj_k$ we have $\vec{U}^{(k)} = \vec{U}^\Omega$ and $\vec{\omega}^{(k)} = 0$ for all $k=1,\ldots,N_\Obj$.
All in all, we derived the following explicit expression
\begin{equation}
	\vec{A}(\vx) = \vec{U}^\Omega = \mathrm{const} \, , \quad \vec{x} \in \mathbb{R}^3
\end{equation}
for the artificial field $\vec{A}$.

Next, we exploit the Stokes equation~\eqref{eq:A:Stokes} which immediately leads to $\vec{\nabla} P^{\vec{A}}(\vx) = 0$ for $\vx \in \mathbb{R}^3 \setminus \partial \{ \Obj^{(\vec{\alpha})} \}$.
The pressure associated with $\vec{A}$ is thus a simple constant in every connected set, which we will write as
\begin{equation}\label{eq:A:PressSol}
	P^{\vec{A}}(\vx) =
	\begin{cases}
		-C_\Omega & \text{if $\vec{x} \in \Omega$,} \\
		-C_k & \text{if $\vec{x} \in \Obj_k , \>\> k=1,\ldots,N_\Obj$,}
	\end{cases}
	\quad \text{with} \quad \vec{x} \in \Gamma \setminus \partial \Obj \, .
\end{equation}
Only the values within the unit cell will be needed (the pressures within the periodic images might be different at first).
The stress tensor is thus
\begin{equation}
	\sigma_{ij}^{\vec{A}}(\vx) =
	\begin{cases}
		C_\Omega \delta_{ij} & \text{if $\vec{x} \in \Omega$,} \\
		C_k \delta_{ij} & \text{if $\vec{x} \in \Obj_k , \>\> k=1,\ldots,N_\Obj$,}
	\end{cases}
	\quad \text{with} \quad \vec{x} \in \Gamma \setminus \partial \Obj \, .
\end{equation}
Taking the limit onto the surfaces from the outside and inside and multiplying them with the outer normal vector gives the tractions $\vec{f}^{\vec{A},+}$ and $\vec{f}^{\vec{A},-}$.
By substituting them into eq.~\eqref{eq:A:aM:a} we obtain
\begin{equation}\label{eq:SolAdjoint:Explicit}
	a_j(\vx) = \widetilde{C}_k n_j(\vx) \, , \quad \vec{x} \in \partial \Obj_k \, , \quad k=1,\ldots,N_\Obj \, , \quad j=1,2,3 \, ,
\end{equation}
with the constants $\widetilde{C}_k := -\frac{1 + \lambdabar_{\Obj_k}}{8 \pi \mu} (C_\Omega - C_k)$ for $k=1,\ldots,N_\Obj$.
This result is somewhat remarkable: \emph{Every} solution to the adjoint equation~\eqref{eq:FBI:Unique:Adj} must be of the form given by equation~\eqref{eq:SolAdjoint:Explicit}.
It also means that the global linear dependency of the pressure that appears in eq.~\eqref{eq:A:Press} via the Green's function drops out, which is consistent with expression~\eqref{eq:A:PressSol}.

Now, the initial assumption was that there is a non-trivial solution to the adjoint equation.
Since all solutions are of the above form~\eqref{eq:SolAdjoint:Explicit}, there must exist one $k' \in \{ 1,\ldots,N_\Obj \}$ with $\widetilde{C}_{k'} \neq 0$.
We thus substitute it into eq.~\eqref{eq:FBI:Unique:Adj} for $\vxz \in \partial \Obj_{k'}$, and with the help of equation~\eqref{eq:TnInt1} arrive at
\begin{equation}\label{eq:a:contradict}
	1 = \frac{1 - \lambdabar_{\Obj_{k'}}}{1 + \lambdabar_{\Obj_{k'}}} \left(1 - \oint_{\partial \Obj_{k'}} z^{(k')}_l(\vx) n_l(\vx) \, \dd S(\vx)\right) .
\end{equation}
The Wielandt deflation term can be active ($\vec{z}^{(k')} \neq 0$) or inactive ($\vec{z}^{(k')} = 0$) for a particular object.
If it is active, condition~\eqref{eq:z:n} and $\lambdabar_{\Obj_{k'}} \geqslant 0$ immediately lead to the contradiction $1 = 0$.
On the other hand, if the term is inactive, equation~\eqref{eq:a:contradict} can only be satisfied for $\lambdabar_{\Obj_{k'}} = 0$. But this means that by definition $\Obj_{k'}$ is a bubble, where we demanded that the Wielandt term is always active. 
Thus we also get a contradiction.
This means that our initial assumption (that there is a non-trivial solution to the adjoint equation) must have been wrong, i.e.\ equation~\eqref{eq:FBI:Unique:Adj} only has the solution $\vec{a} \equiv 0$.

To complete the proof, we use the Fredholm alternative from theorem~\ref{theo:Fredholm}.
The homogeneous equation therefore also has only the trivial solution, and consequently the full FBI equation~\eqref{eq:FBI:NonUnique} has exactly one solution (existence and uniqueness).
This holds as long as the Wielandt term exists for objects $\Obj_k$ with $\lambdabar_{\Obj_k} = 0$.
Note that for $\lambdabar_{\Obj_{k}} > 0$ the Wielandt term does not matter concerning uniqueness of the solution, but may be used to accelerate the numerical convergence as remarked before.
We also mention that the above procedure carries over to other systems and Green's functions such as for an infinite domain without any essential changes.

\subsection{Bubble model details}\label{sec:BubbleDetails}

\subsubsection{The traction and the constitutive law}

As stated in section~\ref{sec:BI} we prescribe a certain outer traction $\vec{f}^+$ on the surface of the bubbles.
This is necessary because the introduction of the traction jump $\triangle \vec{f}$ (as is done for capsules) in the BI equation would require the application of the Stokes equation at the inside \citep[compare sec.~\ref{sec:GenBI} and][pages~37 and 143]{PozrikidisBook92} which is not possible because the inside is a compressible fluid with very low viscosity.
Due to this very low viscosity, however, we can neglect the shear stress acting on the inside surface of the bubbles and only the inner pressure $P_{\Bub_k}$ will be of relevance.
Hence, the outer traction is expressible as \citep{Youngren1976,Power1992}
\begin{equation}\label{eq:Bub:TracWithPress}
	\vec{f}^+(\vx) \approx \triangle \vec{f}(\vx) - P_{\Bub_k} \vec{n}(\vx) \, , \quad \vx \in \partial \Bub_k \, , \quad k=1,\ldots,N_\Bub \, .
\end{equation}
Note that the minus before the pressure comes from the fact that $\vec{f}^+$ represents the force exerted by the fluid on the membrane, and not vice versa.
The traction jump $\triangle \vec{f}$ must be determined by some constitutive law for the interface, such as the ordinary Young-Laplace equation
\begin{equation}\label{eq:YoungLapl}
	\triangle \vec{f}(\vx) = 2 \gamma_{\Bub_k} H(\vx) \vec{n}(\vx) \, , \quad \vx \in \partial \Bub_k \, , \quad k=1,\ldots,N_\Bub \, .
\end{equation}
$H$ is the mean curvature, taken to be positive for a sphere.
This equation is valid for a spatially constant surface tension $\gamma_{\Bub_k}$, i.e.\ for interfaces between two immiscible substances.
Additional surfactants can lead to a position dependency and non-zero tangential components \citep{Pozrikidis2001a}.

\subsubsection{Imposing bubble volume changes}
We can now prescribe a certain traction jump and an (in general time dependent) internal pressure to model an oscillating bubble.
The traction can then be computed via equation~\eqref{eq:Bub:TracWithPress} and substituted into the FBI equation~\eqref{eq:FBI:Intro}.
This should work fine in principle.
However, after the substitution we observe that the $P_{\Bub_k} \vec{n}$ term simply drops out due to equation~\eqref{eq:GnInt}, leaving us unable to enforce a certain pressure and thus any volume changes.
This deficiency of the FBI equation is because of the fact that the single-layer potential is incapable of producing any flow with sinks or sources \citep[ch.~4.1]{PozrikidisBook92}.
This in turn originates from the incompressibility~\eqref{eq:GreenIncompress:u} of the flow produced by the Stokeslet.

Rather than prescribing a certain internal pressure we therefore prescribe a certain flux $Q_{\Bub_k}$. 
This is easily implemented as the flux appears explicitly in the patched FBI equation~\eqref{eq:FBI:Intro}.
For the purpose of solving eq.~\eqref{eq:FBI:Intro} we then set $\vec{f}^+ = \triangle \vec{f}$.

\subsection{Numerical implementation}\label{sec:BIM:Numerics}
Our volume-changing object boundary integral method (VCO-BIM) solves the FBI equation~\eqref{eq:FBI:Intro} numerically.
For this, we discretize the surfaces of all objects with flat triangles.
Dynamic refinement and coarsening via Rivara's longest-edge bisection algorithm \citep{Rivara1984} is employed (see the supplementary information for some examples).
Object centroids and volumes are calculated as explained by \citet{Zhang2001a}.
Integrals are computed with a standard Gaussian quadrature \citep{Cowper1973}, where quantities at points within the triangles are obtained from their nodal values via linear interpolation \citep{PozrikidisBook92}.
Surface integrals where the Green's functions are singular are treated in polar coordinates in case of the single-layer integrals \citep{Pozrikidis1995}, and by adapting the usual surface subtraction scheme for the double-layer integrals \citep[eq.~(8.8)]{Pozrikidis2001a}. 
In both cases the Green's functions of the infinite domain from eq.~\eqref{eq:Green:Inf} are used to eliminate the singularities \citep{Loewenberg1996,Zinchenko2000} since they are faster to calculate than their periodic counterparts and coincide with them for $\vect{x}\approx\vect{x}_0$.
Nearly singular behavior (occurring when objects come near to each other) is additionally removed for the double-layer potentials \citep{Loewenberg1996}.
After discretization, equation~\eqref{eq:FBI:Intro} becomes a linear system that we solve with \mbox{GMRES} \citep{Saad1986}, bypassing the need to explicitly construct the system's matrix.
We remark that BiCGSTAB \citep{vanderVorst1992} was found to be slower in most cases. 

The computation of the discretized integral equation with the periodic Green's functions from equations~\eqref{eq:Green:G} and \eqref{eq:Green:T} is accelerated by two different means.
First of all, the Ewald decomposition by \citet{Hasimoto1959} is used to split the expressions into fast converging real and Fourier space parts, also see \cite{Lindbo2010}.
The final expressions are given by \citet[ch.~5.1]{Zhao2010}.
Second, we employ the smooth particle mesh Ewald method (SPME) to further accelerate the computation of the Fourier parts via fast Fourier transforms \citep{Saintillan2005}.

The time evolution of the objects is obtained by solving the kinematic condition 
\begin{equation}\label{eq:TimeEvol}
	\frac{\dd \vec{x}}{\dd t} = \uv(\vx) \, , \quad \vx \in \partial \Obj
\end{equation}
for each mesh node by some standard explicit ODE integrator, such as Runge-Kutta or the adaptive Bogacki-Shampine \citep{Bogacki1989} and Cash-Karp methods \citep{Cash1990}.
Unfortunately, the average volume of the objects would slowly shrink with time due to unavoidable discretization errors.
To counter this, we employ two different strategies.
First, we use the discretized version of the no-flux condition $\oint \vec{u} \cdot \vec{n} \, \dd S = 0$ for objects with zero flux.
This equation effectively represents a hyperplane. 
We then rotate the solution vector onto this hyperplane.
This procedure is similar to \citet[eq.~(43)]{Farutin2014}.
Second, to eliminate the volume drift completely, we additionally employ the rescaling method as explained by \citet[eq.~(63)]{Farutin2014}.

Bending forces for capsule-like objects follow the Canham-Helfrich model \citep{Canham1970, Helfrich1973}. Various numerical implementations are explained in the article by \citet{Guckenberger2016} and reviewed by \citet{Guckenberger2017}.
Shear and area dilatation elasticity of cells and capsules is implemented as detailed by \citet{Kruger2012Springer} and \citet{Guckenberger2016}.
Large distortions of the mesh are prevented automatically in this case as the forces depend explicitly on the triangle deformations.
Bubble surfaces, on the other hand, do not feature in-plane tensions.
This results in their mesh becoming inhomogeneous very quickly, leading to numerical instabilities.
To prevent this, we observe that the nodes need to follow the fluid velocity only in the normal vector direction since any tangential displacement leaves the bubble shape unchanged. Thus, an artificial tangential displacement of 
\begin{equation}\label{eq:MeshStab}
	\delta x_{i,\alpha+1}^{(a)} = \zeta \sum_{j=1}^{3} (\delta_{ij} - n_{i,\alpha} n_{j,\alpha}) \frac{\sum_b (x_{j,\alpha}^{(b)} - x_{j,\alpha}^{(a)}) w^{ab}_\alpha}{\sum_b w^{ab}_\alpha} \, , \quad i=1,2,3
\end{equation}
can be applied after every time step without modifying the physical behavior.
We apply this formula in an iterative process, indicated by the Greek subscript $\alpha$. 
The superscripts $a$ and $b$ denote different nodes, the Latin subscript indicates a certain Cartesian component and the sums go over the first ring of neighbors $b$ of node $a$.
The parameter $\zeta=0.3$ controls the stiffness of the scheme. 
The iteration stops once the maximal displacement falls below a predefined threshold.
Finally, the weights are chosen as $w^{ab}_\alpha = A^{ab}_\alpha / |\vec{x}_{\alpha}^{(b)} - \vec{x}_{\alpha}^{(a)}|$, where the sum of the areas $A^{ab}_\alpha$ of the two triangles containing nodes $a$ and $b$ tends to homogenize the triangle areas, and the denominator tries to keep possibly applied refinement local to where it had been applied.
A similar approach has been used by \citet[eq.~(59)]{Farutin2014}.

We tested our code extensively by comparing the integrals with analytically known values \citep[compare][sec.~8.3]{Farutin2014}, as well as by studying usual benchmark systems such as the deformation of a capsule in an infinite shear flow \citep{Guckenberger2016}. 
The code was also successfully applied to the diffusion of particles near elastic membranes \citep{Daddi-Moussa-Ider2016, Daddi-Moussa-Ider2016a, Daddi-Moussa-Ider2016b, Daddi-Moussa-Ider2017a, Daddi-Moussa-Ider2017b, Daddi-Moussa-Ider2017} and was used to compare with experimental obtained shapes of red blood cells in microchannel flows \citep{Quint2017Accepted}.
Further verifications can be found in the supplementary information (SI).
We parallelized our code with OpenMP and MPI, and we use explicit SIMD vectorization via the \texttt{Vc} library \citep{Kretz2012} in some core parts.

\section{Ultrasound-triggered margination of microbubbles}\label{sec:Application}

\begin{figure}
	\centering
	\includegraphics[width=\linewidth]{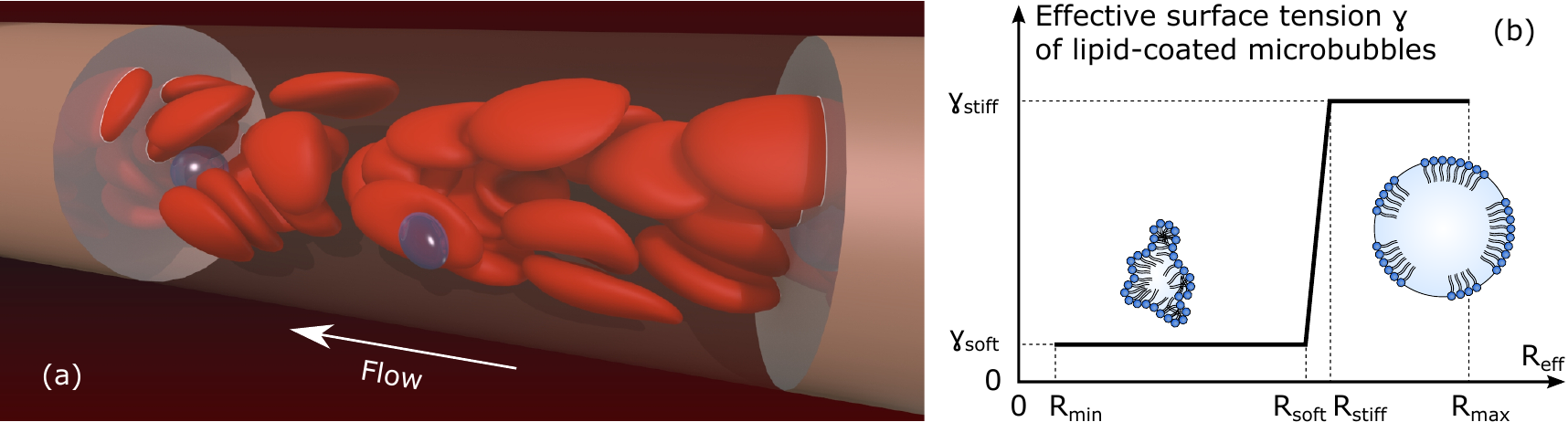}
	\caption{%
	Simulation setup.
	(a)~Snapshot of the large simulation containing 30 red blood cells and two microbubbles. 
	Periodic boundary conditions are used, i.e.\ the left and right non-translucent cylinders are periodic images of the center one which has a length of $\SI{48}{\micro\meter}$ in this case. 
	(b)~Sketch of Marmottant \textit{et al.}'s~model for lipid coated microbubbles \citep{Marmottant2005,Overvelde2010}. 
	The effective surface tension $\gamma$ is a function of the effective radius which varies between $R_\mathrm{min}$ and $R_\mathrm{max}$ during an ultrasound period.
	The bubble is in the soft buckled state for $R_\mathrm{eff} \leqslant R_\mathrm{soft}$ and in the stiff ruptured state for $R_\mathrm{eff} > R_\mathrm{stiff}$, as indicated by the two inset sketches. 
	}
	\label{fig:SetupAndMarmottant}
\end{figure}

We now use our VCO-BIM as introduced in the previous section to investigate the behavior of ultrasound contrast agents (lipid-coated microbubbles) in microcapillary blood flow. 
Our numerical simulations consist of two ultrasound contrast agents and several red blood cells within a cylindrical blood vessel as depicted in figure~\ref{fig:SetupAndMarmottant}~(a).
The lipid coating of the microbubbles leads to a radius-dependent effective surface tension which will be modeled as detailed in section~\ref{sec:lipidBubbles}.
Red blood cells and the remaining ingredients are described in section~\ref{sec:bloodFlow}.
Our central result, namely the occurrence of ultrasound-triggered margination (UTM) is given in section~\ref{sec:Results}.

\subsection{Lipid coated microbubbles}
\label{sec:lipidBubbles}

\subsubsection{Surface forces}
The coating of microbubbles leads to deviations from the simple coupling between bubble volume and external pressure expected from the expansion/compression of an ideal gas \citep{Marmottant2005,deJong2007,Overvelde2010,Frinking2010,Doinikov2011,Sijl2011}.
There are various models available for varying types of surface coatings that produce such nonlinear responses \citep[e.g.\@][]{Sarkar2005, Faez2013}.
One that incorporates the special properties of phospholipid coated bubbles (i.e.\ the size-dependent state of the coating) was presented by \citet{Marmottant2005}, providing a suitable description for their behavior \citep{Overvelde2010,Kooiman2014,Frinking2010,Doinikov2011,Sijl2011,Faez2013}.
The major ingredient is the introduction of an \emph{effective} surface tension that depends nonlinearly on the bubble's size. 
Such a size dependency is the most important aspect of the model for the purpose of margination.
Hence, a more elaborate surface model including surface viscosity \citep[e.g.\@][]{Paul2010} is not required here.
The relation which we employ can be divided into three major regions \citep{Marmottant2005,Overvelde2010} as illustrated in figure~\ref{fig:SetupAndMarmottant}~(b):
\begin{enumerate}
	\item In the high compression regime, the area available per lipid molecule is smaller than its extent, leading to pronounced buckling as observed by ultrahigh-speed imaging \citep{Sijl2011}. This is modeled by an effective surface tension of $\gamma_\mathrm{soft} \approx 0$ below an effective radius $R_\mathrm{soft}$ \citep{Overvelde2010}. The bubble is highly deformable in this state \citep{Marmottant2005,Rychak2006}.
	\item With increasing radius, a very narrow elastic regime occurs, extending up to a maximal radius $R_\mathrm{stiff}$.
	\item Above $R_\mathrm{stiff}$ the shell ruptures, leaving floating rafts of lipids on the surface \citep{Marmottant2005,Borden2005}.
	This leads to the very high surface tension $\gamma_\mathrm{stiff}$ of a direct air-water interface.
\end{enumerate}

Due to the smallness of the second regime \citep{Overvelde2010}, we set in the following $R_\mathrm{soft} = R_\mathrm{stiff}$ such that the effective surface tension in our case can be written as
\begin{equation}\label{eq:EffTens}
	\gamma(R_\mathrm{eff}) = 
		\begin{cases}
			\gamma_\mathrm{soft} & \text{if } R_\mathrm{eff} \leqslant R_\mathrm{soft}, \\
			\gamma_\mathrm{stiff} & \text{if } R_\mathrm{eff} > R_\mathrm{soft},
		\end{cases}
\end{equation}
where $R_\mathrm{eff} := \sqrt[3]{3 V / (4 \pi)}$ is the instantaneous effective radius and $V$ the bubble volume.
The traction jump then follows from the Young-Laplace equation~\eqref{eq:YoungLapl}.

The equilibrium radius $R_0$, i.e.\ the radius assumed when no ultrasound is present, can be located at different positions relative to the transition radius $R_\mathrm{soft}$ depending on the process of bubble preparation. 
Most importantly, it was shown that they can be created in the buckled state ($R_0 \leqslant R_\mathrm{soft}$) \citep{Borden2005,Rychak2006,Frinking2010} as desired for safe transport by default within the blood vessels \citep{Lindner2002}.

\subsubsection{Modeling the effect of an acoustic source}

Because we are interested in the margination behavior, the exact form of the oscillations is expected to be irrelevant. 
We therefore prescribe the flux of the $i$'th bubble as $Q_i(t) = A_i \sin(2\pi f t)$ to model the effect of an external acoustic source. Here, $t$ is the time, $A_i$ the flux amplitude and $f$ the frequency.
This results in a volume oscillation of 
\begin{equation}\label{eq:VolOsciLaw}
	V_i(t) = V^{(0)}_i + A_i/(2 \pi f) [1 - \cos(2 \pi f t)]
\end{equation}
for the $i$'th bubble, with $V^{(0)}_i$ being the initial volume at $t=0$.
To conserve the total outer fluid volume as required by eq.~\eqref{eq:SumFluxZero}, both bubbles are set to oscillate out-of-phase, i.e.\ $A_1 = - A_2$.
Although not being entirely realistic, it is mandated on a very fundamental level by the periodic boundary conditions and the incompressibility of the ambient fluid.
We do not expect that this small restriction affects the validity of the presented conclusions since (i) bubble-bubble interactions are strongly shielded by the RBCs and (ii) margination hinges upon the stiffness variations of the individual bubbles and is therefore independent of the phase of the oscillations.

Continuing, we emphasize that we impose only the instantaneous bubble volume and not a spherical shape. Hence, the bubbles are still deformable, a property which is crucial for the hydrodynamic interaction with the RBCs. 

The most important quantity in the present context is the ratio of the stiff to soft duration which we denominate as $\delta = T_+ / T_-$. Here, $T_+$ is the time spent in the stiff state (i.e.\ $R_\mathrm{eff} > R_\mathrm{soft}$) and $T_-$ the time in the soft state ($R_\mathrm{eff} \leqslant R_\mathrm{soft}$). 
Since margination would trivially be expected for $\delta \gg 1$, we concentrate on $0 \leqslant \delta \leqslant 1$ in the present study, in agreement with experiments \citep{Marmottant2005,deJong2007,Overvelde2010,Sijl2011}.
We remark that $\delta$ does not depend on the frequency $f$.

\subsubsection{Bubble parameters}
In our simulations we set the surface tensions in the soft and stiff state to $\gamma_\mathrm{soft} = 0.5 \kappa_\mathrm{S}$ and $\gamma_\mathrm{stiff} = 10 \kappa_\mathrm{S}$, respectively. $\kappa_\mathrm{S}$ is the shear modulus of the red blood cells (see below). 
These choices sensibly describe the stiffness of the bubbles relative to RBCs regarding margination while at the same time ensuring numerical stability. 
Realistic values of $\gamma_\mathrm{stiff}=\SI{7e-2}{N/m}$ and $\gamma_\mathrm{soft} \approx 0$ \citep{Overvelde2010,Sijl2011} would lead to a numerically very stiff problem and consequently require extremely small time steps. 
The supplementary information shows that $\gamma_\mathrm{soft} = 0.1 \kappa_\mathrm{S}$ and $\gamma_\mathrm{stiff} = 25 \kappa_\mathrm{S}$ do not change the results significantly.
Furthermore, we fix $R_\mathrm{min} = \SI{1.7}{\micro\meter}$ and $R_0 = R_\mathrm{soft} = \SI{2}{\micro \meter}$ which are typical values for microbubbles \citep{Borden2005,Overvelde2010,Kooiman2014} (using $R_0 = \SI{1}{\micro\meter}$ leaves the results qualitatively unchanged, see SI).
Taking $\delta$ as the major control parameter, $R_\mathrm{max}$ and the amplitudes $A_i$ are uniquely determined via the prescribed volume oscillation law~\eqref{eq:VolOsciLaw}.
Assuming an ideal gas within the bubbles and an atmospheric pressure of $\SI{100}{kPa}$, a value of $\delta = 1$ then corresponds to an acoustic pressure amplitude of around $P_\mathrm{A} \approx \SI{45}{kPa}$, in agreement with experimentally used values \citep{Overvelde2010}.

In most current applications, ultrasound pressure amplitudes and frequencies are in the kilo-pascal and mega-hertz range, respectively \citep{Kooiman2014,Lammertink2015}.
Such values lead to strong primary and secondary radiation forces \citep{Rychak2005, Johnson2016} making the bubbles agglomerate in a small spot on the vessel wall opposite of the ultrasound transducer \citep{Dayton1999, Rychak2007, Kilroy2014}.
This strong localization is highly undesirable for drug delivery applications where a uniform bubble distribution over the entire vessel wall is required.
In contrast, we will show below that ultrasound-triggered margination is able to reliably achieve an isotropic distribution if the ultrasound parameters are chosen such that radiation forces become subdominant.
For $P_\mathrm{A} \approx \SI{45}{kPa}$ we therefore keep the acoustic frequency at $f = \SI{1}{kHz}$ in the following.
The magnitude of the primary radiation force is then typically of the order of $|\vect{F}_\mathrm{rad}| \approx 10^{-15}\,\si{N}$, meaning that it plays only a secondary role as shown explicitly in the SI.
We consequently neglect it in what follows.
In order to exploit UTM also at higher frequencies, one can reduce $P_\mathrm{A}$ as exemplified in the SI where we consider $f \leqslant \SI{10}{kHz}$ for $P_\mathrm{A} \approx \SI{6}{kPa}$.

\subsection{Blood flow in capillaries}
\label{sec:bloodFlow}

\subsubsection{Blood flow constituents}\label{sec:bloodFlowConst}
We model the blood flow by explicitly resolving the red blood cells and treating the surrounding blood plasma as a Newtonian fluid \citep{Chien1966}.
For our simulations we use mostly $15$ RBCs that are distributed randomly within the blood vessel if not noted otherwise. 
Each RBC has an initial large radius of $R_\mathrm{RBC} = \SI{4}{\micro \meter}$ \citep{Evans1972, Freund2014}.
Continuing, the RBC shear elasticity is modeled via Skalak's constitutive energy \citep{Skalak1973} with a shear modulus of $\kappa_\mathrm{S} = \SI{5e-6}{N/m}$ \citep{Yoon2008, Freund2014} and the typical discocyte shape as the reference geometry.
This model also includes an area dilatation modulus that is set to $\kappa_\mathrm{A} = 10 \kappa_\mathrm{S}$.
Furthermore, we additionally introduce an extra surface dilatation energy $E_\mathrm{a} = (\kappa_\mathrm{a} / 2) (S - S_0)^2 / S_0$ \citep{Kruger2012Springer} with the corresponding area dilatation modulus $\kappa_\mathrm{a} = 10 \kappa_\mathrm{S}$, the instantaneous surface area $S$ and the reference surface area $S_0$.
This leads to area deviations of typically $\lesssim \SI{4}{\percent}$.
Moreover, bending forces are modeled according to the Canham-Helfrich law \citep{Canham1970, Helfrich1973, Guckenberger2017} with a bending modulus of $\kappa_\mathrm{B} = \SI{2e-19}{N.m}$ \citep{Park2010, Freund2014} and the spontaneous curvature set to zero.
For numerical efficiency we employ the usual approximation that inner and outer viscosities are equal \citep{Zhao2012a,Kumar2014,Kruger2012Springer, Freund2014}, i.e.\ the viscosity ratio is $\lambda_\mathrm{RBC} = 1$.
As a result, any double-layer integrals over the RBC surfaces vanish, and Wielandt deflation terms cannot appear for RBCs (compare section~\ref{sec:PatchedFBI}).
Nevertheless, both are still present for the bubbles.

The periodic vessel has a length of usually $\SI{24}{\micro \meter}$ and a radius of $R_\mathrm{Vessel} = \SI{11}{\micro \meter}$. 
Together with the $15$ RBCs this results in a hematocrit of around $\SI{16}{\percent}$, a typical value encountered in capillaries \citep{Klitzman1979,House1987}.  
The larger simulation from figure~\ref{fig:SetupAndMarmottant}~(a) as well as higher hematocrit values lead to the same results which are presented below and in the supplementary information.
Furthermore, one possibility for the boundary condition of the vessel wall would be to set its velocity to zero.
This, however, leads to a mixed kind Fredholm integral equation.
As explained in sec.~\ref{sec:FBI:Basic}, no general mathematical theory exists and this type can be rather performance-intensive although it might work in practice.
We therefore follow \citet{Freund2007} and fix the wall's nodes $\vec{x}_i$ via springs to their original position $\vec{x}_i^{(0)}$, leading to a traction jump of $\triangle \vec{f} = \kappa_\mathrm{W} (\vec{x}_i - \vec{x}_i^{(0)})$, where $\kappa_\mathrm{W} = \SI{6.25e6}{N/m^3}$ is the spring constant.
Increasing $\kappa_\mathrm{W}$ by a factor of $5$ does not change results qualitatively as shown in the SI.
Thus we end up with a Fredholm integral equation of the second kind having exactly one solution as proven in section~\ref{sec:Proof}.

\subsubsection{Hydrodynamics}
We use our VCO-BIM for 3D periodic domains as presented in section~\ref{sec:Theory} to solve the Stokes equation.
The core of this method is equation~\eqref{eq:FBI:Intro} which we solve for an imposed average flow chosen such that the maximal flow velocity in the middle of the vessel is roughly $u_\mathrm{max} \approx \SI{4.7}{mm/s}$, if not noted otherwise.
This value matches with physiological flow velocities in capillaries and arterioles~\citep{Popel2005}.

The Stokes equation is a good approximation if the Reynolds numbers are much smaller than unity.
For the translational motion we find for our system $\Rey_\mathrm{T} = 2 R_\mathrm{RBC} \, u_\mathrm{max} \rho \, / \mu \approx 0.03 \ll 1$, where $\mu = \SI{1.2e-3}{kg/(s.m)}$ is the dynamic viscosity of blood plasma \citep{Skalak1989} and $\rho \approx 10^3\,\si{kg/m^3}$ its density.
A different Reynolds number can be defined based on the radial oscillations as $\Rey_\mathrm{R} = (2 R_0)^2 \rho f / \mu$. For $f \leqslant \SI{10}{kHz}$ (as used in the SI) this results in $\Rey_\mathrm{R} < 0.07 \ll 1$.
We thus conclude that the Stokes equation can faithfully capture the considered RBC and bubble interactions.

\subsubsection{Numerical procedure}
The general methodology of our numerical implementation was already explained in section~\ref{sec:BIM:Numerics}.
Here we only mention the remaining aspects that are specific to the present application.
The triangle count for the blood vessel is 630 for the $\SI{24}{\micro\meter}$ long channel.
Rivara's longest-edge bisection algorithm \citep{Rivara1984} is used to refine high curvature and close contact regions for the dynamic objects.
Hence, the number of triangles varies over time with typical averages of around 1500 for the bubbles and 780 for the RBCs.
See the SI for some illustrations.
Artificial overlapping between the objects within the channel is further suppressed by the introduction of a short-range repulsive potential $E_\mathrm{Rep}(r_{ij}) = \left[b/(r_{ij} - l_\mathrm{m})\right] \exp\left[l_\mathrm{c} / (r_{ij} - l_\mathrm{c})\right]$ \citep{Noguchi2005a,McWhirter2009} with $r_{ij}$ denoting the distance between two nodes (vertices), $l_\mathrm{m}$ being the minimal possible distance and $l_\mathrm{c}$ the distance where the potential smoothly drops to zero. 
We choose $l_\mathrm{m} = 0.01 R_\mathrm{RBC}$ and $l_\mathrm{c} = 0.125 R_\mathrm{RBC}$, the latter being of the order of the typical edge length of the initial bubble meshes.

The traction jump on the RBCs for the elasticity and dilatation contributions is computed by differentiating the energies with respect to the mesh vertices, as explained by \citet{Kruger2012Springer} and \citet[sec.~4.2]{Guckenberger2016}.
The repulsive potential is handled in the same way.
Bending forces (for the RBCs) and the mean curvature (required for the bubbles, cf.\ eq.~\eqref{eq:YoungLapl}) are obtained via Method~C as given by \citet{Guckenberger2016}.
Despite being less precise than some alternative methods described in that work, it proved to be more stable than the others.

SPME errors for the computation of the Green's functions are kept below $\lesssim \SI{0.01}{\percent}$.
Increasing the precision by one order of magnitude did not change the results (see SI).
The Gaussian quadrature rules for the integrals use $7$ Gauss points per triangle for the bubbles and the blood vessel and $4$ points for the RBCs.
We solve the integral equation via GMRES with a residuum of max.\ $10^{-4}$.
Furthermore, the time evolution from eq.~\eqref{eq:TimeEvol} is obtained by the adaptive Bogacki-Shampine method \citep{Bogacki1989} with the relative tolerance fixed to $10^{-5}$ and the absolute tolerance set to $10^{-4} R_\mathrm{RBC}$ \citep{Press2007a}.
We use the volume rescaling approach for bubbles and RBCs and additionally the hyperplane method for RBCs to handle any artificial volume drift as explained in section~\ref{sec:BIM:Numerics}.
No special mesh control scheme was necessary for the vessel and the RBCs due to the nature of the prescribed forces, but for bubbles we use equation~\eqref{eq:MeshStab} where the iteration stops once the maximal displacement falls below $10^{-4} R_\mathrm{RBC}$.
Typical simulations times are in the 1\,--\,2 weeks regime on a recent 20 core Intel system.

\subsection{Results and discussion}
\label{sec:Results}

\subsubsection{Microbubbles with constant surface tensions}
\begin{figure}
	\centering
	\includegraphics[width=\linewidth]{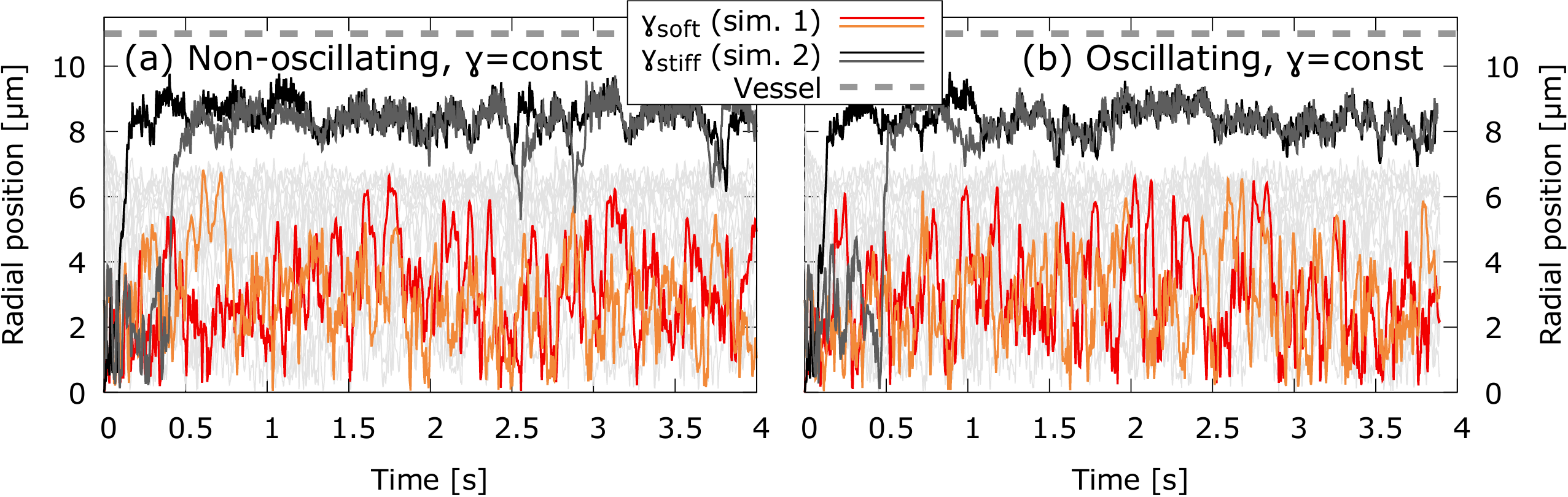}
	\caption{%
	Margination behavior of \emph{purely} soft and \emph{purely} stiff non-oscillating (a) and oscillating (b) microbubbles.
	I.e.\ $\gamma = \mathrm{const}$ in all cases.
	We depict the radial positions of the centroids of two bubbles with constant effective surface tensions in blood flow as a function of time.
	The surface tensions are set to $\gamma = \gamma_\mathrm{soft} = 0.5 \kappa_\mathrm{S}$ (red/orange) and $\gamma = \gamma_\mathrm{stiff} = 10 \kappa_\mathrm{S}$ (black/gray).
	For (b), bubbles oscillate with a frequency of $f = \SI{1}{kHz}$, leading to a variation of $R_\mathrm{eff}$ between $\SI{1.7}{\micro \meter}$ and $\SI{2.075}{\micro \meter}$.
	Curves for different $\gamma$ constitute independent simulations.
	The red blood cells, shown as light gray lines, 
	illustrate the cell-free layer between $\SI{7}{\micro\meter}$ and the wall.
	The vessel radius is $\SI{11}{\micro \meter}$, the hematocrit is fixed to $\SI{16}{\percent}$ and the maximal flow velocity is $u_\mathrm{max} \approx \SI{4.7}{mm/s}$. 
	The soft bubbles ($\gamma = \gamma_\mathrm{soft}$) remain in the center, whereas the stiff bubbles ($\gamma = \gamma_\mathrm{stiff}$) show margination.
	}
	\label{fig:ConstantGamma}
\end{figure}

In order to illustrate the general effect of margination, we first consider the case when the microbubbles are prepared once in the soft and once in the stiff state.
Figure~\ref{fig:ConstantGamma}~(a) shows two simulations without any volume oscillations.
The case $\gamma = \gamma_\mathrm{soft}$ corresponds to coated bubbles that are always in the soft state. 
Thus, they have a deformability comparable to the RBCs and remain in the center of the blood stream together with the erythrocytes.
On the other hand, setting $\gamma = \gamma_\mathrm{stiff}$ models pure bubbles that are much stiffer than the RBCs. Hence, they quickly marginate isotropically to the vessel wall.
Similar observations are made in figure~\ref{fig:ConstantGamma}~(b) for bubbles oscillating with a frequency of $f = \SI{1}{kHz}$ while keeping the effective surface tension constant.
These results demonstrate that the volume oscillations by themselves do not strongly affect particle migration for the presently chosen parameters.

\subsubsection{Lipid coated microbubbles with radius-dependent surface tension}
\begin{figure}
	\centering
	\includegraphics[width=\linewidth]{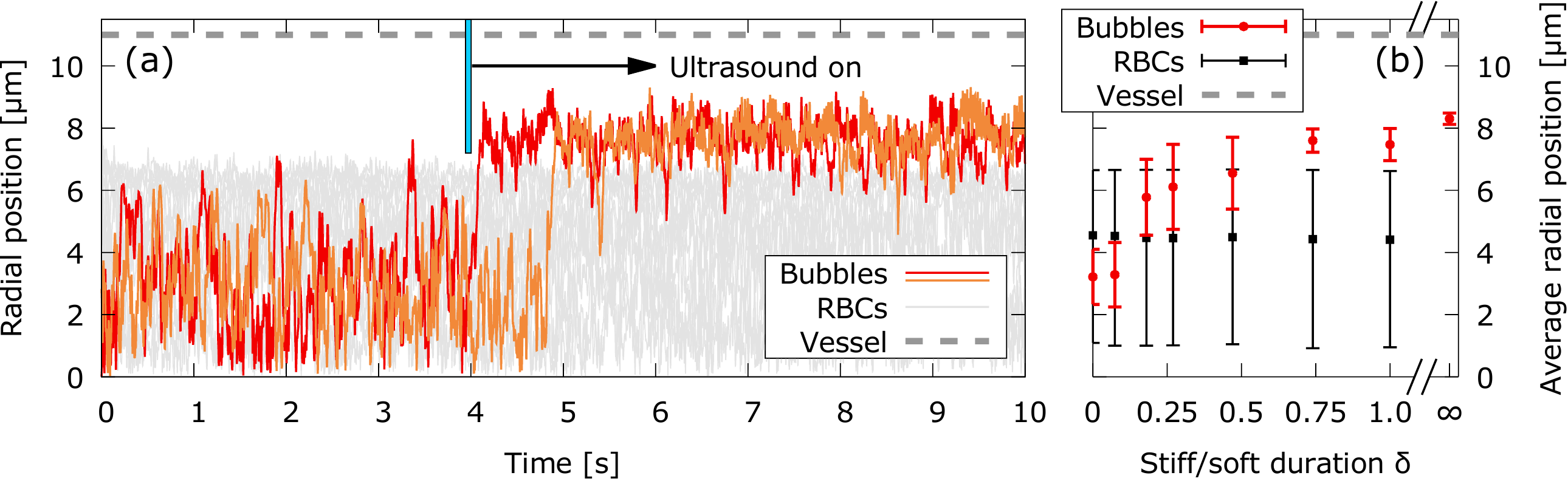}
	\caption{%
	Ultrasound-triggered margination.
	(a)~Radial positions of the centroids of two microbubbles coated with lipids, modeled according to Marmottant \textit{et al.}'s law \citep{Marmottant2005,Overvelde2010}.
	Since the ultrasound is off at the beginning, the bubbles are soft and thus remain in the vessel interior (effective radius $R_\mathrm{eff} = \SI{2}{\micro\meter}$). 
	When the oscillations are switched on at $\approx \SI{4}{s}$, ultrasound-triggered margination leads to rapid migration to the vessel wall.
	Here, $\delta = 1$, i.e.\ the bubbles are stiff for the first half of the ultrasound period and soft during the second one with their effective surface tension varying in the range $\gamma \in [0.5; 10] \kappa_\mathrm{S}$.
	The effective radii alternate between $\SI{1.7}{\micro\meter}$ and $\SI{2.23}{\micro\meter}$ in each period.
	See the supplementary material for a movie.
	(b)~Average radial positions of the oscillating bubbles and the RBCs for several different values of $\delta$. 
	The rightmost point corresponds to the limit $\delta \to \infty$ (i.e.\ always stiff) from fig.~\ref{fig:ConstantGamma}~(b).
	Error bars are determined as explained in the main text and the SI.}
	\label{fig:VariableGamma}
\end{figure}
To demonstrate ultrasound-triggered margination, we consider two lipid coated bubbles whose shells are modeled with an effective surface tension as described by equation~\eqref{eq:EffTens} and that are prepared in the soft state ($\delta=1$, $R_\mathrm{eff} = R_\mathrm{soft}$).
Figure~\ref{fig:VariableGamma}~(a) depicts the bubbles' radial trajectories from a simulation where initially no ultrasound is applied. 
The bubbles are preferably located in the RBC rich core, in agreement with figure~\ref{fig:ConstantGamma}~(a) and experimental observations \citep{Lindner2002}. This allows for secure travel through the vascular system.
Once the volume oscillations are activated after around $\SI{4.0}{s}$ to model an ultrasound source, the bubbles oscillate periodically in stiffness due to the lipid coating that is modeled according to equation~\eqref{eq:EffTens}.
This is in contrast to figure~\ref{fig:ConstantGamma} where the surface tension remains constant.
Most importantly, we observe fast migration towards the vessel wall within less than one second (see the SI for a movie). This time frame corresponds to a traveled distance of less than $\SI{4}{mm}$, highlighting the rapidity of the effect.
The cause of the fast margination is the lipid shell: As discussed in the previous section, the coating leads to a stiffening during the high-pressure state of the ultrasound signal and a corresponding softening during the low-pressure state \citep{Marmottant2005,Overvelde2010}.
As figure~\ref{fig:VariableGamma}~(a) clearly demonstrates the overall behavior is dominated by the stiff stage, as will be further analyzed below.

We continue to demonstrate the robustness of ultrasound-triggered margination by considering microbubbles that are very soft in equilibrium.
The bubbles then spend a much longer portion of the ultrasound period in the soft than in the stiff state ($\delta < 1$). 
Figure~\ref{fig:VariableGamma}~(b) depicts the results. The error bars are determined by considering first the minimal and maximal centroid position of all RBCs/bubbles as a function of time for $t > \SI{1}{s}$ or after definite margination, second a subsequent temporal average and third a weighted average over simulations with different starting configurations.
This is similar to the procedure by \citet{Muller2014} and is explained in more details in the supplementary information.
Figure~\ref{fig:VariableGamma}~(b) shows that the bubbles are still preferably located at the outside of the RBC rich core for $\delta < 1$, even for ratios as low as $\delta \approx 0.2$.
The margination is completely suppressed only at small values such as $\delta \approx 0.1$ where the soft time is around $10$ times longer than the stiff time.
The results from the SI for $P_\mathrm{A} = \SI{6}{kPa}$ show a transition at $\delta\approx 0.3$ indicating that the precise location of the transition depends on the details of the system setup but nevertheless happens for $\delta \ll 1$. Thus we can conclude that reliable margination is observed if the soft time is at most three times larger than the time in the stiff state ($\delta \gtrsim 0.3$).

The effect that small values of $\delta$ are sufficient to trigger ultrasound-triggered margination can be understood qualitatively: During the soft state, shearing by the flow and collisions with red blood cells cause deformations of the bubbles. 
Both are comparably slow processes.
During the subsequent stiff stage, however, a high surface tension forces the deformed object back to a spherical shape much more quickly.
More quantitatively, the typical relaxation time towards the spherical rest shape in the stiff state can be estimated as $\tau_\mathrm{stiff} = 2 R_\mathrm{stiff} \mu / \gamma_\mathrm{stiff} \approx \SI{0.1}{ms}$.
On the other hand, the time required by the flow to deform the bubble away from the spherical shape in the soft state can be estimated by assuming a simple Poiseuille profile with the center flow velocity $u_\mathrm{max} = \SI{4.7}{mm/s}$.
This leads to a shear rate of $s \approx \SI{544}{s^{-1}}$ if the bubble is positioned one diameter ($2 R_\mathrm{soft}$) away from the wall.
Hence, $\tau_\mathrm{deform} = 1/s \approx \SI{1.8}{ms}$, which is about one order of magnitude larger than the relaxation time scale $\tau_\mathrm{stiff}$ in the stiff state. This explains why the latter dominates the margination behavior.
We remark that the ratio $\tau_\mathrm{stiff}/\tau_\mathrm{deform}$ is formally equivalent to the capillary number $\mathrm{Ca} = 2 R_\mathrm{stiff} s \mu / \gamma_\mathrm{stiff}$ for the bubbles in the stiff state, i.e.\ $\mathrm{Ca}$ determines their behavior while being stiff.
The interpretation of the ratio $\tau_\mathrm{stiff}/\tau_\mathrm{deform}$, however, is fundamentally different here, as it makes a comparison between the two different states rather than making a statement about only one state.

\begin{figure}
	\centering
	\includegraphics[width=\linewidth]{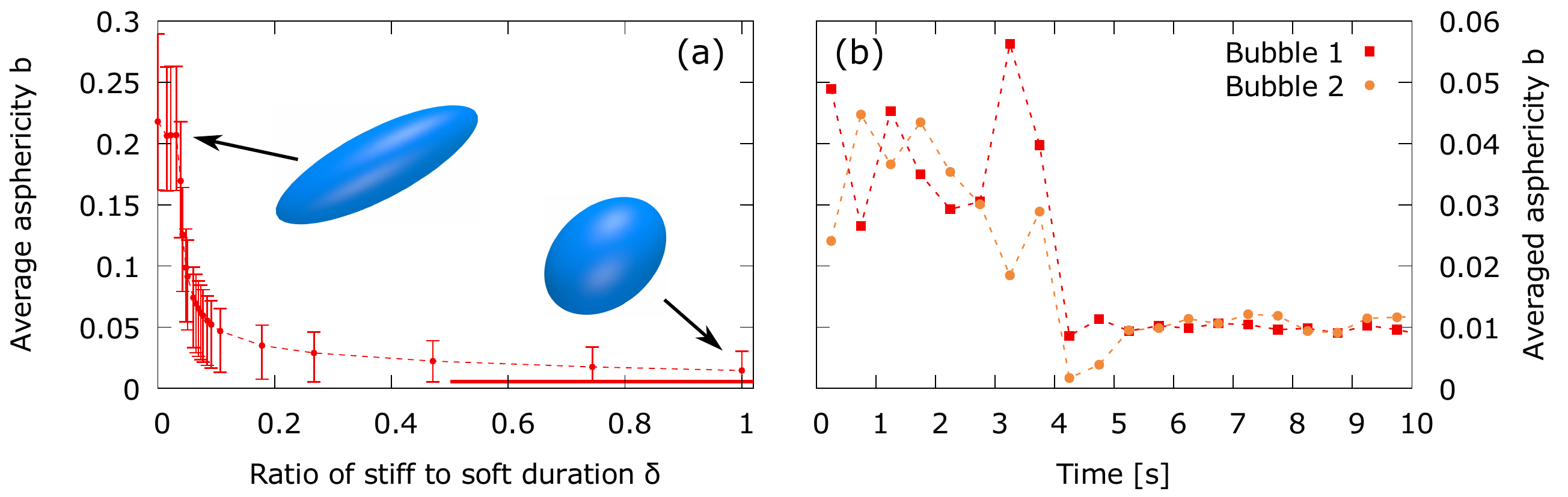}
	\caption{%
	Average asphericities of oscillating lipid coated microbubbles.
	(a) The left figure shows the result for a single microbubble in an infinite shear flow with shear rate $s = \SI{544}{s^{-1}}$ as a function of $\delta$.
	At $\delta \approx 0.05$ the deformation increases sharply while for $\delta > 0.1$ almost no deformation is seen.	
	The limiting value $\delta \rightarrow \infty$ is shown as a red solid line at the bottom. 
	The surface tension varies in $\gamma \in [0.5; 10] \kappa_\mathrm{S}$ and the oscillation frequency is $f = \SI{1}{kHz}$. 
	Error bars indicate the minimal and maximal asphericities during one ultrasound period.
	The two inset snapshots show the bubbles with their maximal deformation for $\delta = 0.032$ (left) and $\delta = 1$ (right).
	(b) Asphericity of the bubbles from figure~\ref{fig:VariableGamma}~(a) averaged over consecutive time intervals of $\SI{0.5}{s}$.
	}
	\label{fig:shear}
\end{figure}
The above argument can be explicitly confirmed by considering an oscillating microbubble in a simple linear shear flow in an infinite domain simulated with VCO-BIM ($V_\Gamma \rightarrow \infty$, compare section~\ref{sec:FinalBIEq}). 
As a measure of deformation, we extract the asphericity $b := \left[(\lambda_1 - \lambda_2)^2 + (\lambda_2 - \lambda_3)^2 + (\lambda_3 - \lambda_1)^2 \right] / (2 R_\mathrm{g}^4)$ from the shape, where $R_\mathrm{g}^2 := \lambda_1 + \lambda_2 + \lambda_3$ is the squared radius of gyration and $\lambda_1$, $\lambda_2$ and $\lambda_3$ are the eigenvalues of the gyration tensor \citep{Fedosov2011}.
For reference, the discocyte equilibrium shape of an RBC leads to $b \approx 0.15$.
Figure~\ref{fig:shear}~(a) shows that the bubble remains almost spherical for $\delta$'s as low as $0.1$, meaning that the bubble is stiff only during $\approx \SI{9}{\percent}$ of the ultrasound period. Only below a rather sharp threshold at $\delta \approx 0.05$ the bubble behaves akin to a truly soft object exhibiting strong deformation.
Note that this value matches well with the ratio $\tau_\mathrm{stiff} / \tau_\mathrm{deform} \approx 0.056$, and reasonably well with the bubbles' transition from the inner core to the outside in figure~\ref{fig:VariableGamma}~(b).
Furthermore, the value $b \approx 0.015$ at $\delta = 1$ approximately agrees with the asphericity observed for the simulation in figure~\ref{fig:VariableGamma}~(a) after the ultrasound was switched on, as depicted in figure~\ref{fig:shear}~(b).

\subsubsection{Further investigations}

\begin{figure}
	\centering
	\includegraphics[width=0.45\linewidth]{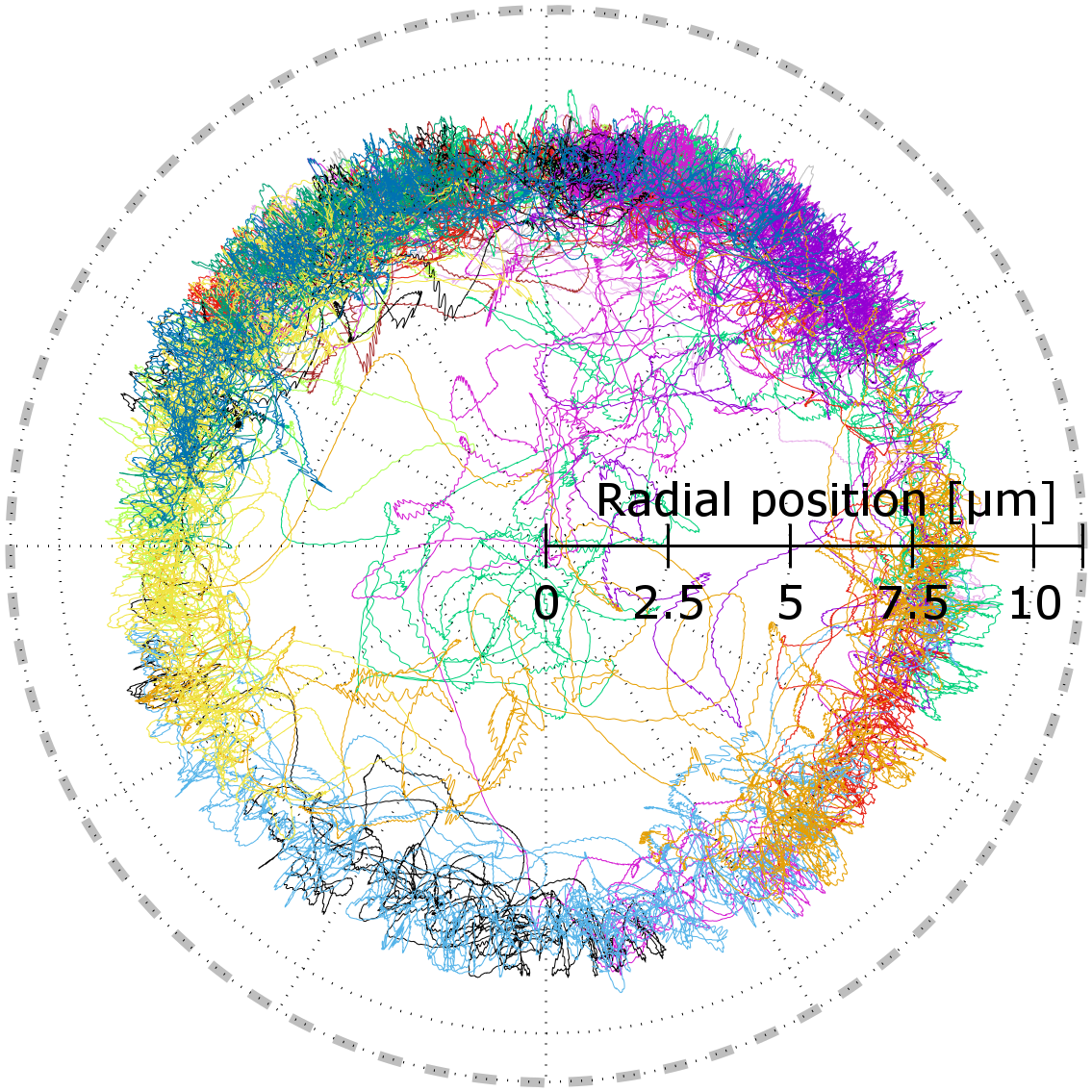}
	\caption{%
	Polar plot of several bubble trajectories (i.e.\ as viewed from the outlet).
	The figure shows the $\delta = 0.74$ and $\delta = 1$ simulations used for figure~\ref{fig:VariableGamma}~(b), representing different system realizations.
	Trajectories only shown for $t > \SI{1}{s}$ or after definite margination.
	Rare short-lived migration events to the inside occur.
	Each bubble in each simulation is shown in a different color.
	The outer gray dashed line depicts the vessel radius.
	}
	\label{fig:polar}
\end{figure}

Pure margination is an isotropic effect: There is no preferred initial migration direction nor a preferred position close to the wall.
We exemplify this in figure~\ref{fig:polar} where we show the trajectories of migrated bubbles from several simulations representing different system realizations.
Obviously, no specific accumulation point exists.
This is in contrast to migration induced by buoyancy or radiation forces at much higher frequencies \citep{Dayton1999,Rychak2007,Kilroy2014,Johnson2016}.
Note that the inclusion of radiation forces for the chosen parameters leaves the qualitative results unchanged, as described in detail in the SI.
The major influence on figure~\ref{fig:polar} comes therefore from the initial conditions and the length of the simulations.

\begin{figure}
	\centering
	\includegraphics[width=\linewidth]{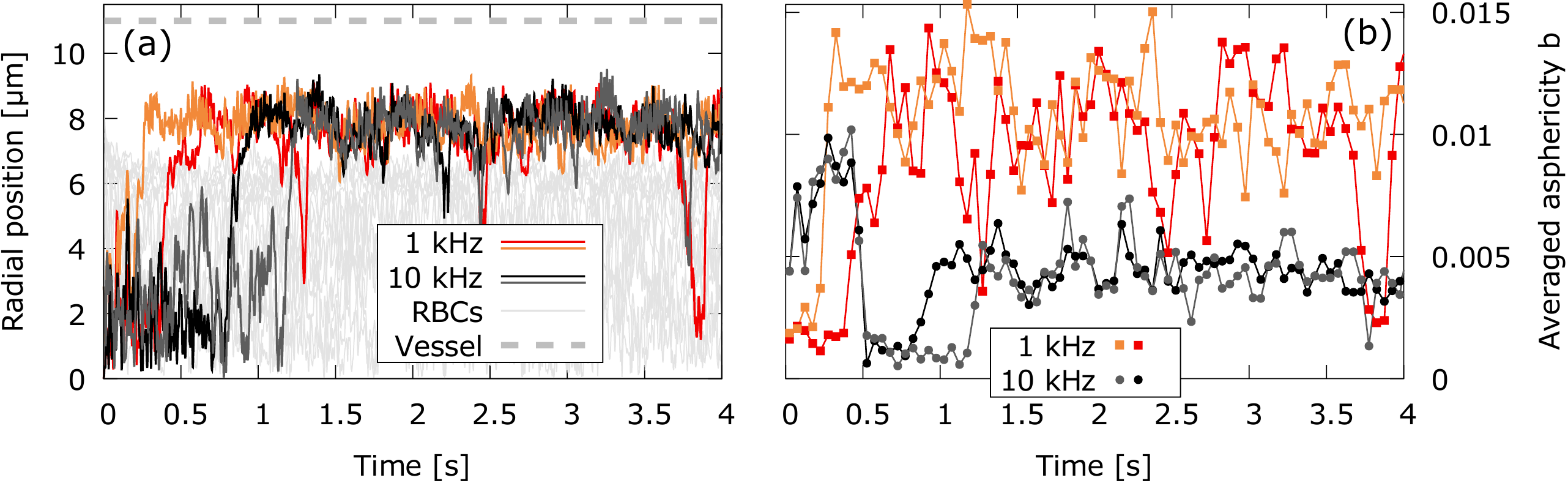}
	\caption{%
	Influence of frequency:
	Behavior of two oscillating lipid coated microbubbles for $\delta = 1$ and a hematocrit of \SI{16}{\percent}, once for a frequency of $f = \SI{1}{kHz}$ and once for $\SI{10}{kHz}$ (two distinct simulations).
	$u_\mathrm{max} \approx \SI{4.7}{mm/s}$.
	(a)~Radial positions of the centroids. The red blood cells are shown in light gray.
	(b)~Corresponding microbubble asphericities averaged over consecutive time intervals of $\SI{50}{ms}$.
	}
	\label{fig:10kHz}
\end{figure}

Figure~\ref{fig:10kHz}~(a) shows that increasing the frequency from $f = \SI{1}{kHz}$ to $\SI{10}{kHz}$ leaves the qualitative results for the radial position unchanged when radiation forces are neglected.
Most interestingly, however, the asphericity is roughly reduced by half in case of the faster oscillations (figure~\ref{fig:10kHz}~(b)).
The reason is that for $\SI{10}{kHz}$ less time within one period is available to deform the bubbles before the stiff state takes over, as suggested by the above time scale estimates.
This strongly indicates that higher frequencies reinforce the effect that small values of $\delta$'s are sufficient to obtain ultrasound-triggered margination.
Even more, this serves as a hint that the effect of UTM, which has been overlooked so far, might have provided a noticeable contribution to the effectiveness of microbubbles for targeted drug delivery observed in recent \textit{in-vivo} and clinical studies that used higher frequencies \citep{Lammertink2015,Kotopoulis2016}.

\begin{figure}
	\centering
	\includegraphics[width=\linewidth]{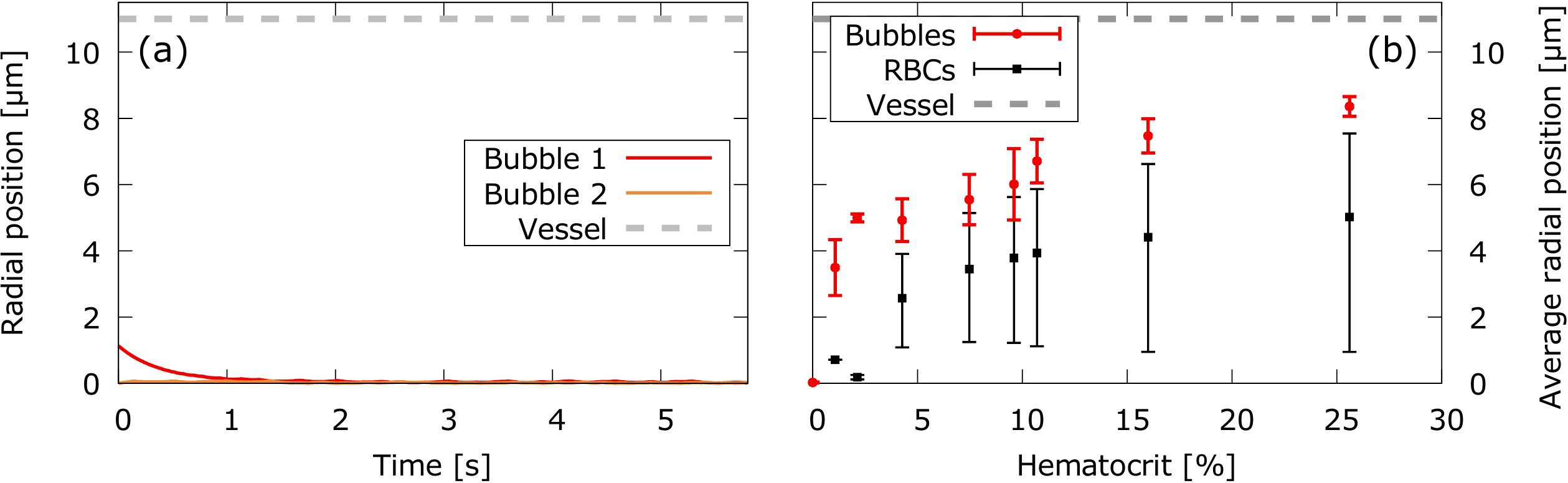}
	\caption{Influence of hematocrit: (a) Only bubbles ($\mathrm{Hematocrit} = 0$). (b) Average radial position for $\delta = 1$ as a function of hematocrit. Values and error bars extracted from several simulations as in figure~\ref{fig:VariableGamma}~(b).}
	\label{fig:onlyBubbles:noRadForce}
\end{figure}

Continuing, we demonstrate that UTM intrinsically hinges on the presence of the red blood cells.
If they are removed, the result in figure~\ref{fig:onlyBubbles:noRadForce}~(a) is obtained, showing clearly that oscillating lipid coated microbubbles move towards the center of the channel for $\delta = 1$.
This is in notable contrast to figure~\ref{fig:VariableGamma}~(a), where rapid margination for the same set of parameters is observed.
Hence, neglecting the influence of the red blood cells in \textit{in-vitro} experiments can easily lead to conclusions that no longer hold for the \textit{in-vivo} case.
On the other hand, with a finite hematocrit, margination is always observed in the sense that the RBCs are located at the interior and the bubbles form an outer layer (figure~\ref{fig:onlyBubbles:noRadForce}~(b)).
The position of this outer layer depends on the size of the RBC-rich inner region which grows when more RBCs are present \citep{Muller2014}, with the average radial position being reminiscent of a pitchfork bifurcation.
The reason is that the RBCs migrate to the center and push the bubbles to the outside because the latter are seen as stiffer on average as already described above.

\begin{figure}
	\centering
	\includegraphics[width=0.5\linewidth]{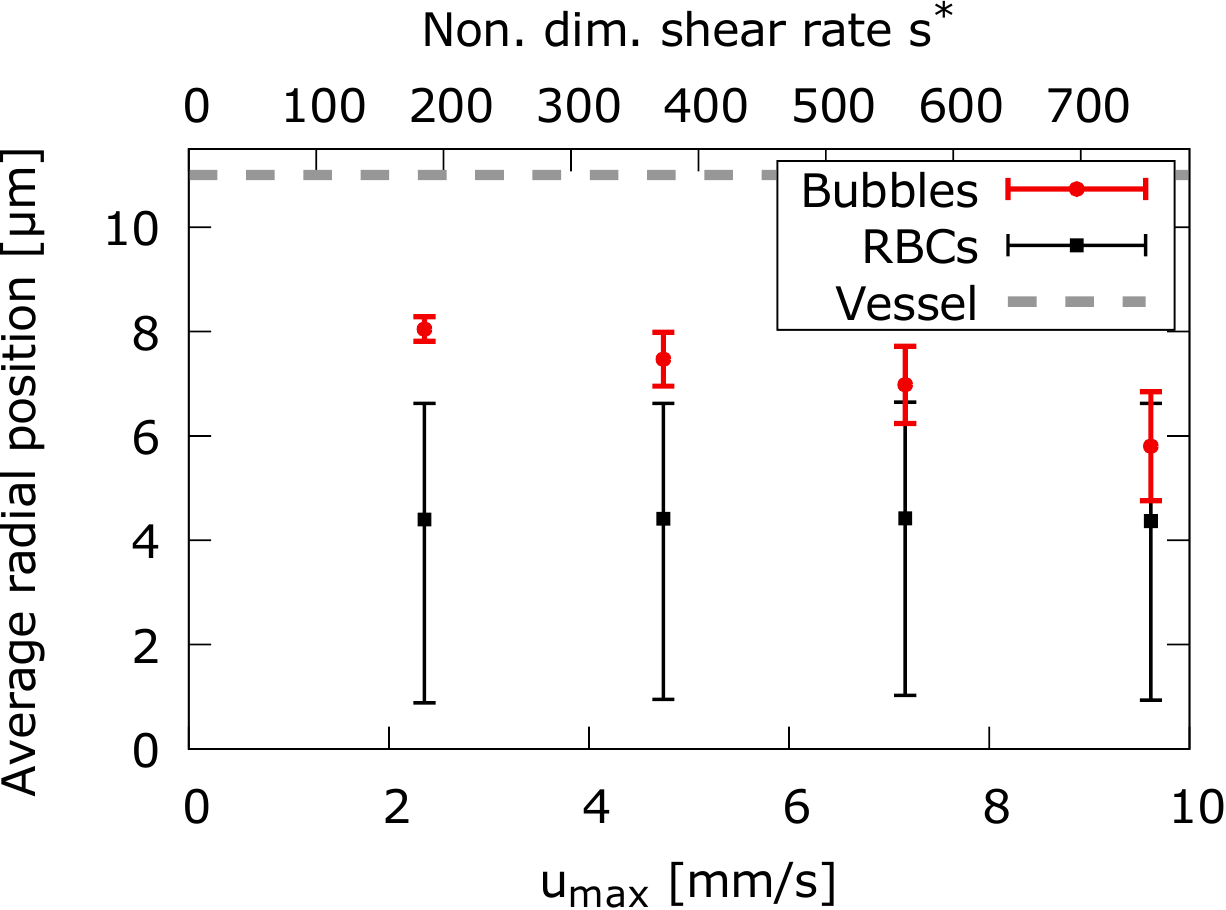}
	\caption{%
	Influence of the velocity for $\delta = 1$ and a hematocrit of \SI{16}{\percent}:
	Average radial position as a function of the flow velocity.
	Values and error bars extracted from several simulations as in figure~\ref{fig:VariableGamma}~(b).
	}
	\label{fig:velocity}
\end{figure}

Moreover, we depict in figure~\ref{fig:velocity} the influence of the flow velocity.
Margination still occurs in all cases, but higher velocities tend to decrease the radial position of the marginated bubbles.
The effect, however, is comparably small.
The upper horizontal axis displays a corresponding effective non-dimensional shear rate defined here by
\begin{equation}
	s^* := \frac{u_\mathrm{max}}{2 R_\mathrm{Vessel}} \frac{\mu D_\mathrm{RBC}^3}{\kappa_\mathrm{B}}
\end{equation}
with an effective RBC diameter $D_\mathrm{RBC} := \sqrt{A_\mathrm{RBC} / \pi}$ and the RBC surface $A_\mathrm{RBC} \approx \SI{137}{\micro\meter^2}$.
This definition is similar to the one by \cite{Muller2014} except that we use the maximal instead of the average flow velocity.
Hence, we find qualitative agreement with their results for spherical rigid particles \citep[compare fig.~3a in][]{Muller2014}: Margination is only a little affected by the shear rate if it is high enough.

Finally, we note that the effect of UTM does not change significantly if the effective surface tension in the soft state is decreased to $\gamma_\mathrm{soft} = 0.1 \kappa_\mathrm{S}$ or the stiff tension is increased to $\gamma_\mathrm{stiff} = 25 \kappa_\mathrm{S}$, if the size of the bubbles is halved to $R_0 = \SI{1}{\micro\meter}$, if the initial particle distribution is varied or if the simulation box length is doubled as shown in the SI. 
Moreover, the average of the nodal velocity of the vessel wall (sec.~\ref{sec:bloodFlowConst}) after a short initial startup phase is $\lesssim 0.08 \, u_\mathrm{max}$ and the nodes move less than $0.1 \, R_\mathrm{RBC}$, indicating that the spring wall sensibly replaces a completely stationary wall.
Nevertheless, we also show in the SI that the results do not change if the vessel wall is made $5$ times stiffer.

\section{Conclusion}
In the first part of our work we developed an extended boundary integral method to simulate volume-changing objects such as microbubbles in a 3D periodic domain (VCO-BIM). 
In contrast to all other commonly used \enquote{capsule-like} objects (vesicles, cells, drops), these bubbles contain a compressible gas with very low viscosity.
As a consequence their volume can change as a function of time.
This behavior leads to two additional terms in the boundary integral equations which arise from (i) integrals over the unit cell and (ii) ensuring uniqueness of the solution. 
We showed that the latter, which can be seen as part of a Wielandt deflation procedure, is optional for capsule-like objects with a finite inner viscosity but becomes a necessary ingredient for bubbles.
To this end, we proved that the resulting Fredholm integral equation has exactly one solution for an arbitrary number of bubbles and capsule-like entities with arbitrary viscosity ratios.
Although periodic boundary integral methods for cells and capsules have been amply used in the past, such a proof has so far not appeared in the literature.
The proof can be easily adapted for the case of other Green's functions, e.g., in infinite domains.

In the second part we used our method to show that lipid-coated microbubbles possess unique and highly desirable properties which are not found for other drug delivery agents. 
During transport from the injection site to the target organ, with no ultrasound present, the bubbles behave as soft objects akin to red blood cells, traveling near the center of blood vessels.
Application of a localized ultrasound at the target region then causes the bubbles to alternate between a soft and a stiff state. 
This leads to their isotropic margination towards the vessel wall within less than one second in the presence of red blood cells. 
Surprisingly, margination even happens when the time spent in the stiff state is more than three times smaller than the time in the soft state. 
We explain this observation by the fact that the effective surface tension (leading to a spherical shape during the stiff stage) acts on much shorter time scales than the surrounding flow (which deforms the bubble during the soft stage). 
This, together with the presented studies regarding frequency, hematocrit and flow velocity indicates that ultrasound-triggered margination is a robust effect.
Given that it leads to a uniform bubble distribution on the vessel wall while other targeting mechanisms such as radiation forces often cause large inhomogeneities, the here identified effect might open a promising route to design novel drug delivery systems in the future.

	\section*{Acknowledgments}
Funding from the Volkswagen Foundation, funding and support from the KONWIHR network and computing time granted by the Leibniz-Rechenzentrum on \mbox{SuperMUC} are gratefully acknowledged by A.~Guckenberger and S.~Gekle. A.~Guckenberger additionally thanks the Elitenetzwerk Bayern (ENB), Macromolecular Science, for their support.

	\appendix
\section{List of important symbols}\label{sec:ListOfSymbols}
For the reader's convenience, we provide in table~\ref{tab:Symbols} a short overview of the important symbols used in section~\ref{sec:Theory} together with their point of definition.
\begin{table}
	\begin{center}
		\def~{\hphantom{0}}
		\begin{tabular}{ll}
			Symbol  & Meaning \\[3pt]
			$\Obj_k$ & The $k$'th object (of any type), sec.~\ref{sec:SysComp}. \\
			$\Bub_k$ & The $k$'th bubble, sec.~\ref{sec:SysComp}. \\
			$\Caps_k$ & The $k$'th capsule-like entity (a capsule, vesicle, red blood cell, \ldots), sec.~\ref{sec:SysComp}. \\
			$\Wall_k$ & The $k$'th wall (walls do not have any inside), sec.~\ref{sec:SysComp}. \\
			$N_\Obj$ & The total number of objects, sec.~\ref{sec:SysComp}. \\
			$N_\Bub$, $N_\Caps$, $N_\Wall$ & The number of bubbles, capsules and walls, sec.~\ref{sec:SysComp}. \\
			$\Omega$ & The ambient fluid (the set outside of any object), sec.~\ref{sec:SysComp}. \\
			$\Gamma$ & The unit cell, sec.~\ref{sec:DefPeriod}. \\
			$\partial \Obj_k$ & The surface of object $\Obj_k$, sec.~\ref{sec:SysComp}. \\
			$V_{\Obj_k}$, $V_\Gamma$ & The volume of the $k$'th object and of the unit cell $\Gamma$, sec.~\ref{sec:DefPeriod}. \\
			$\vec{a}^{(1)}$, $\vec{a}^{(2)}$, $\vec{a}^{(3)}$ & The three base vectors spanning the unit cell $\Gamma$, sec.~\ref{sec:DefPeriod}. \\
			$\vec{X}^{(\vec{\alpha})}$ & A periodic grid vector in the real space with grid index $\vec{\alpha} \in \mathbb{Z}^3$, eq.~\eqref{eq:PerGridVec}. \\
			$\vec{b}^{(1)}$, $\vec{b}^{(2)}$, $\vec{b}^{(3)}$ & The reciprocal base vectors, sec.~\ref{sec:DefPeriod}. \\
			$\vec{k}^{(\vec{\kappa})}$ & A periodic grid vector in the reciprocal space with grid index $\vec{\kappa} \in \mathbb{Z}^3$, eq.~\eqref{eq:FourierDef}. \\
			$\vec{u}$ & The velocity, sec.~\ref{sec:SysComp}. \\
			$\vx$ & A generic point, often also the integration variable. \\
			$\vxz$ & The evaluation point, sec.~\ref{sec:BI}. \\
			$\vec{n}$ & The outside normalized normal vector, sec.~\ref{sec:SysComp}. \\
			$P$ & The pressure, eq.~\eqref{eq:Stokes}. \\
			$\mu$ & The viscosity of the ambient fluid $\Omega$, sec.~\ref{sec:SysComp}. \\
			$\sigma_{ij}$ & The stress tensor, eq.~\eqref{eq:Sigma}. \\
			$t$ & The time. \\
			$\lambda_{\Caps}$ & The viscosity ratio of capsule $\Caps$, sec.~\ref{sec:SysComp}. \\
			$\lambdabar_{\Obj}$ & The effective viscosity ratio of object $\Obj$, eq.~\eqref{eq:DefLambdabar}. \\
			$\vec{f}$, $\vec{f}^+$, $\vec{f}^-$ & The traction (general, outside and inside), eq.~\eqref{eq:Trac}. \\
			$\vec{\breve{f}}$ & The periodic part of the traction, eq.~\eqref{eq:f:Decomp}. \\
			$\triangle \vec{f}$ & The traction jump, eq.~\eqref{eq:TracJump}. \\
			$\vec{F}$ & The \enquote{unified traction} ($\fv^+$ or $\tfv$), sec.~\ref{sec:GenBI}. \\
			$(\mathcal{N}_{\partial \Obj_q} \vec{F})$ & The single-layer integral over the surface $\partial \Obj_q$, eq.~\eqref{eq:SL}. \\
			$(\mathcal{K}_{\partial \Obj_q} \vec{u})$ & The double-layer integral  over the surface $\partial \Obj_q$, eq.~\eqref{eq:DL}. \\
			$G_{ij}$ & The Green's function for the single-layer integral (Stokeslet), eq.~\eqref{eq:Green:G}. \\
			$T_{ijl}$ & The Green's function of the double-layer integral (Stresslet), eq.~\eqref{eq:Green:T}. \\
			$\breve{T}_{ijl}$ & The periodic part of $T_{ijl}$, eq.~\eqref{eq:Green:T}. \\
			$c_k$ & The prescribed velocity divergence of the bubbles, eqs.~\eqref{eq:Bubble:Div} and \eqref{eq:Bub:Flux:c}. \\
			$Q_{\Obj_k}$ & The flux out of or into the object $\Obj_k$, eq.~\eqref{eq:Bub:Flux:c}. \\
			$\chi^{(\Obj_k)}_j$, $\vec{\chi}^{(\Obj_k)}$ & The geometric centroid of object $\Obj_k$, eq.~\eqref{eq:Centroid}. \\
			$z^{(k)}_j$ & The Wielandt deflation prefactor, eq.~\eqref{eq:z:WithSurf}. \\
			$\braket{\bullet}_\Gamma$ & The volume average over the whole unit cell $\Gamma$, eq.~\eqref{eq:DefAverage}. \\
			$h_j$ & An eigensolution to the homogeneous version of the BI equation, eq.~\eqref{eq:FBI:NonUnique:Hom}. \\
			$a_j$ & Eigensolutions to the adjoint of the homogeneous equation, eq.~\eqref{eq:FBI:NonUnique:Adj}. \\
			$M_j[\vec{a}]$ & Most of the right-hand side of the adjoint equation, eq.~\eqref{eq:A:M}. \\
			$R_j$ & Collection of the terms missing in the homogeneous equation, eq.~\eqref{eq:CondR}. \\
		\end{tabular}
		\caption{List of important symbols used throughout the text.}
		\label{tab:Symbols}
	\end{center}
\end{table}

\section{Objects overlapping with the unit cell boundary}\label{sec:overlap}

Section~\ref{sec:Theory} assumed that all objects are completely located within the unit cell~$\Gamma$.
However, satisfying this condition is impossible for dense suspensions even for a single time step, and because flowing objects regularly leave and enter $\Gamma$ during the course of dynamic simulations.
Moreover, it is much more convenient to carry out integrations over the full surfaces of objects (i.e.\ including parts that lie partially outside of $\Gamma$), as otherwise the meshes would need to be split.
The possibility of doing this is often assumed in the literature.
Yet, due to the centroids that appear explicitly in the BI and FBI equations \eqref{eq:BI:Periodic} and \eqref{eq:FBI:Intro}, respectively, and the linear term in the stresslet~\eqref{eq:Green:T} it is not clear \textit{a priori} whether this is indeed possible since these parts contain striking non-periodicities.
Here we explicitly show that the equations nevertheless hold if applied correctly.

\subsection{The boundary integral equation with cut objects}
\begin{figure}
	\centering
	\includegraphics{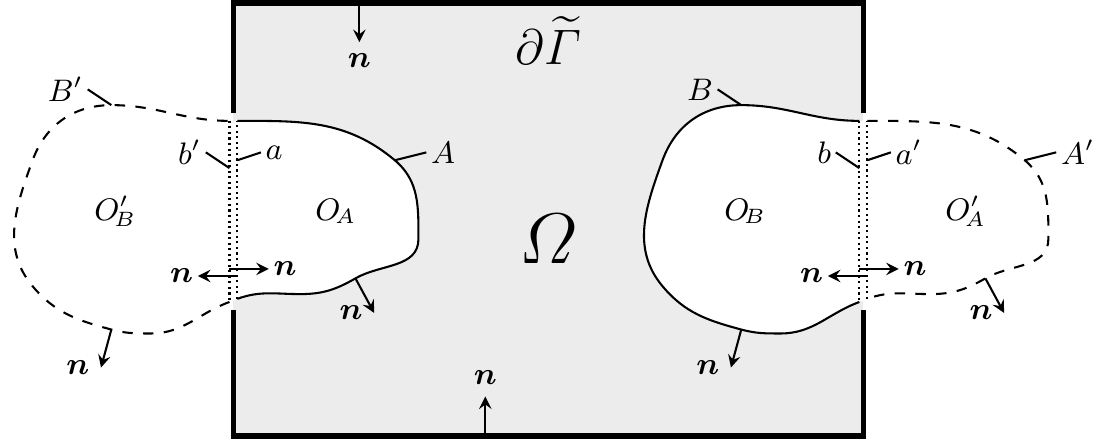}
	\caption{%
	Overlapping of objects with the unit cell boundary $\partial \Gamma = \partial \widetilde{\Gamma} \cup a' \cup b'$. 
	Primed quantities lie outside of the unit cell (except $a'$ and $b'$ which form portions of $\partial \Gamma$).
	A single object $\Obj_{\!B}^{\prime} \cup \Obj_{\!A}$ is shown.
	This object has a periodic image $\Obj_{\!B} \cup \Obj_{\!A}^{\prime}$.
	Furthermore, both are cut at the same place into two parts as they overlap with $\partial \Gamma$.
	The four closed sets are $\Obj_{\!B}^{\prime}$, $\Obj_{\!A}$, $\Obj_{\!B}$ and $\Obj_{\!A}^{\prime}$.
	The corresponding surfaces \emph{without} the cut faces are $B'$, $A$, $B$ and $A'$.
	For the left object, the cut faces (dotted) are $a$ and $b'$, where both lie on $\partial \Gamma$ and are identical except that their normal vectors point into opposite directions. 
	I.e.\ there is no gap between $a$ and $b'$. 
	Similar for its right image (but with $b$ and $a'$).
	Moreover, there is no real gap between \smash{$\partial \widetilde{\Gamma}$} and the objects.
	$\partial \Omega$ in this figure is given by \smash{$\partial \Omega = \partial \widetilde{\Gamma} \cup A \cup B$}; in the general case $\partial \Omega$ also contains the full surfaces of objects completely located within the unit cell.
	}
	\label{fig:PartialRemapping}
\end{figure}

We consider a system with overlapping objects as sketched in figure~\ref{fig:PartialRemapping}.
The outer boundary of $\Omega$ is thus not formed by $\partial \Gamma$ but rather by $\partial \widetilde{\Gamma}$, $A$ and $B$.
The usual derivation of the BI equation starts with integrating the reciprocal identity over $\Omega$ \citep[e.g.\@][ch.~2.3]{PozrikidisBook92}.
Subsequent use of the divergence theorem then explicitly introduces the boundaries of $\Omega$.
In section~\ref{sec:GenBI} the outer boundary was simply $\partial \Gamma$ as all objects were located within $\Gamma$, but here it is given by $\partial \widetilde{\Gamma} \cup A \cup B$.
This leads to
\begin{equation}\label{eq:PO:BIWithOmega}
	\begin{split}
		u_j(\vxz) =
		\ldots
		&-\frac{1}{8\pi\mu} (\mathcal{N}_{A \cup B} \vec{f}^+)_j(\vxz) 
		+ \frac{1}{8\pi} (\mathcal{K}_{A \cup B} \vec{u})_j(\vxz)
		\\
		&-\frac{1}{8\pi\mu} (\mathcal{N}_{\partial \widetilde{\Gamma}} \vec{f}^+)_j(\vxz) 
		+ \frac{1}{8\pi} (\mathcal{K}_{\partial \widetilde{\Gamma}} \vec{u})_j(\vxz)	
		\, , \quad \vxz \in \Omega \, , \quad j=1,2,3 \, .
	\end{split}
\end{equation}
We will use the notation introduced in figure~\ref{fig:PartialRemapping} from now on and only deal explicitly with one object and its periodic image (as also seen in this figure).
Additional objects that lie partially outside of the unit cell result in analogous terms. Objects that lie completely within $\Gamma$ do not require special treatment here.
Both types of omitted terms will be indicated by an ellipsis to shorten notation.

The first goal now is to recover integrals over the full unit cell surface $\partial \Gamma$ instead of only over $\partial \widetilde{\Gamma}$ in equation~\eqref{eq:PO:BIWithOmega}.
For this we exploit $\int_{a} + \int_{b'} = 0$ and $\int_{b} + \int_{a'} = 0$ for the single- or double-layer kernels. This holds since $a$ and $b'$ (and $b$ and $a'$) denote the same surface but with antiparallel normal vectors (again, see figure~\ref{fig:PartialRemapping}).
$a'$, $b'$ and $\partial \widetilde{\Gamma}$ can be combined to $\partial \Gamma$, and the remaining sets form the surfaces of the closed sets $\Obj_{\!A}$ and $\Obj_{\!B}$.
Thus, eq.~\eqref{eq:PO:BIWithOmega} becomes
\begin{equation}\label{eq:PO:BI:WithGammaDisturbance}
	\begin{split}
		u_j(\vxz) = 
		&-\frac{1}{8\pi\mu} (\mathcal{N}_{\partial \Obj_A \cup \partial \Obj_B} \vec{f}^+)_j(\vxz) 
		+ \frac{1}{8\pi} (\mathcal{K}_{\partial \Obj_A \cup \partial \Obj_B} \vec{u})_j(\vxz)
		+ \ldots\\
		&-\frac{1}{8\pi\mu} (\mathcal{N}_{\partial \Gamma} \vec{f}^+)_j(\vxz) 
		+ \frac{1}{8\pi} (\mathcal{K}_{\partial \Gamma} \vec{u})_j(\vxz)	
		\, , \quad \vxz \in \Omega \, , \quad j=1,2,3 \, .
	\end{split}
\end{equation}

If $\Obj_{\!A}$ and $\Obj_{\!B}$ are parts of a capsule filled with a Stokesian fluid, the reciprocal identity also holds for their insides.
Integrating it over their volumes, using the divergence theorem and adding the result to the above equation \citep[compare][pages~37 and 143]{PozrikidisBook92} recovers the form of equation~\eqref{eq:BI:GeneralWithGamma} that includes the viscosity ratios and the traction jumps, but now with separate terms for each object part:
\begin{equation}\label{eq:PO:BI:WithGamma}
	\begin{split}
		u_j(\vxz) = 
		&-\frac{1}{8\pi\mu} (\mathcal{N}_{\partial \Obj_A \cup \partial \Obj_B} \vec{F})_j(\vxz) 
		+ \frac{1 - \lambdabar_{\Obj_A \cup \Obj_B}}{8\pi} (\mathcal{K}_{\partial \Obj_A \cup \partial \Obj_B} \vec{u})_j(\vxz)
		+ \ldots\\
		&-\frac{1}{8\pi\mu} (\mathcal{N}_{\partial \Gamma} \vec{f})_j(\vxz) 
		+ \frac{1}{8\pi} (\mathcal{K}_{\partial \Gamma} \vec{u})_j(\vxz)	
		\, , \quad \vxz \in \Omega \, , \quad j=1,2,3 \, .
	\end{split}	
\end{equation}
$\lambdabar_{\Obj_A \cup \Obj_B} := \lambdabar_{\Obj_A} \equiv \lambdabar_{\Obj_B}$ denotes the effective viscosity ratio of the two object parts, and $\vec{f} := \vec{f}^+$.
Hence, the original equation~\eqref{eq:BI:GeneralWithGamma} remains valid if (i) only the surface parts located within $\Gamma$ are taken into account, and (ii) the surfaces of the cuts lying on $\partial \Gamma$ are added.

The next goal is to check if the unconnected set $\Obj_A \cup \Obj_B$ in eq.~\eqref{eq:PO:BI:WithGamma} can be replaced with the actual connected object $\Obj_A \cup \Obj_B'$ without changing the result.
We additionally have to see if the equation is invariant under translation of the evaluation point $\vxz$ by some periodic grid vector $\vec{X}^{(\vec{\alpha})}$.
This has to be done separately for the single- and double-layer integrals.
Moreover, we do this for some arbitrary point $\vxz \in \Gamma$ because both integral types are always well-defined and we need it for the FBI equation.

\subsection{The single-layer potentials}\label{sec:PO:SL}

First consider the single-layer integral for the \emph{outer} traction $\vec{f} := \vec{f}^+$ over $\partial \Obj_B'$ for some general evaluation point $\vxz \in \Gamma$ and an arbitrary grid vector $\vec{X}^{(\vec{\alpha}')}$ (cf.\ sec.~\ref{sec:DefPeriod}).
Outer tractions only appear for bubbles which always have closed surfaces (capsule-like objects and walls possess traction \emph{jumps} instead). 
Furthermore, the object $\Obj_B'$ is offset by construction from $\Obj_B$ by some particular grid vector $\vec{X}^{(\vec{\alpha})}$.
Hence, we compute
\begin{align}
	(\mathcal{N}_{\partial \Obj_B'} &\vec{f})_j(\vxz + \vec{X}^{(\vec{\alpha}')}) 
	= \oint_{\partial \Obj_B'} f_i(\vx) G_{ij}(\vx,\vxz + \vec{X}^{(\vec{\alpha}')}) \, \dd S(\vx) \nonumber \\
	&= \oint_{\partial \Obj_B} f_i(\vx + \vec{X}^{(\vec{\alpha})}) G_{ij}(\vx + \vec{X}^{(\vec{\alpha})},\vxz + \vec{X}^{(\vec{\alpha}')}) \, \dd S(\vx) \nonumber\\
	&= - \braket{\vec{\nabla} P}_{\Gamma} \cdot \vec{X}^{(\vec{\alpha})} \oint_{\partial \Obj_B} n_i(\vx) G_{ij}(\vx,\vxz) \, \dd S(\vx) + (\mathcal{N}_{\partial \Obj_B'} \vec{f})_j(\vxz) \nonumber\\
	&=(\mathcal{N}_{\partial \Obj_B} \vec{f})_j(\vxz) \, .
\end{align}
We made a simple substitution from the first to the second line.
From the second to the third line, we used the periodicity of the Green's function $G_{ij}$, eq.~\eqref{eq:G:Periodic}, as well as relation~\eqref{eq:f:Decomp}.
The last line follows because of equation~\eqref{eq:GnInt}.

If an object requires the traction jump $\triangle \vec{f}$ (capsules, walls) and is possibly open, an analogous result holds because $\triangle \vec{f} = \vec{f}^+ - \vec{f}^- = \breve{\vec{f}}^+ - \breve{\vec{f}}^-$ is periodic as the linear terms from eq.~\eqref{eq:f:Decomp} drop out.
Adding the $\Obj_A$ contribution, we have
\begin{equation}
	(\mathcal{N}_{\partial \Obj_A \cup \partial \Obj_B'} \vec{F})(\vxz + \vec{X}^{(\vec{\alpha}')}) = (\mathcal{N}_{\partial \Obj_A \cup \partial \Obj_B} \vec{F})(\vxz) \, , \quad \vxz \in \Gamma \, , \quad \forall \vec{\alpha}' \in \mathbb{Z}^3 \, .
\end{equation}
Thus, we find for some general object that the single-layer integrals are invariant under any possible periodic translations.
Moreover, the single-layer integral over $\partial \Gamma$ appearing in equation~\eqref{eq:PO:BI:WithGamma} simply vanishes for an arbitrary evaluation point $\vxz \in \mathbb{R}^3$ as in section~\ref{ch:UCellSurf:SL}.

\subsection{The double-layer potentials}\label{sec:PO:DL}

As for the single-layer potential, consider $\vxz \in \Gamma$ and an arbitrary grid vector $\vec{X}^{(\vec{\alpha}')}$.
Note that open objects (walls) do not require this integral (due to $\lambdabar_\Wall = 1$) and will therefore not be considered.
Then,
\begin{align}
	(\mathcal{K}_{\partial \Obj_B'} \vec{u})_j(\vxz + \vec{X}^{(\vec{\alpha}')})
	&= \oint_{\partial \Obj_B'} u_i(\vx) T_{ijl}(\vx, \vxz + \vec{X}^{(\vec{\alpha}')}) n_l(\vx) \, \dd S(\vx) \nonumber\\
	&= \oint_{\partial \Obj_B} u_i(\vx) T_{ijl}(\vx + \vec{X}^{(\vec{\alpha})}, \vxz + \vec{X}^{(\vec{\alpha}')}) n_l(\vx) \, \dd S(\vx) \nonumber\\
	&= -\frac{8\pi}{V_\Gamma} X^{(\vec{\alpha})}_j \oint_{\partial \Obj_B} u_i n_i \, \dd S + (\mathcal{K}_{\partial \Obj_B} \vec{u})_j(\vxz) \, ,
\end{align}
where a simple substitution was performed again, and the periodicity of the velocity (eq.~\eqref{eq:uFull:Periodic}), the normal vector and the second argument of the Stresslet from eq.~\eqref{eq:Green:T} was used.
The first argument of the Stresslet contributes the linear part.
After using the divergence theorem and adding the $\Obj_A$ term, we obtain for $\vxz \in \Gamma$
\begin{equation}\label{eq:PO:DLObj}
	(\mathcal{K}_{\partial \Obj_A \cup \partial \Obj_B} \vec{u})_j(\vxz) = \frac{8\pi}{V_\Gamma} X^{(\vec{\alpha})}_j \int_{\Obj_B} \vec{\nabla} \cdot \uv \, \dd x^3 + (\mathcal{K}_{\partial \Obj_A \cup \partial \Obj_B'} \vec{u})_j(\vxz + \vec{X}^{(\vec{\alpha}')}) \, .
\end{equation}
Hence, the double-layer integral is invariant under periodic offsets $\vec{X}^{(\vec{\alpha}')}$ regarding the evaluation point $\vxz$, but \emph{not} under periodic translations of some object when its flux is non-zero.
At first sight this would mean that bubbles cause major troubles and somewhat destroy the formalism for practical purposes as offsetting them leads to an additional term.

However, the BI equation additionally contains a double-layer integral over $\partial \Gamma$.
This integral also depends on the objects via the velocity.
We will see that the interplay between this integral and the double-layer integrals for the objects will recover the invariance.
Thus, consider
\begin{equation}
	(\mathcal{K}_{\partial\Gamma} \uv)_j(\vxz + \vec{X}^{(\vec{\alpha}')}) = -\frac{8\pi}{V_\Gamma} \oint_{\partial\Gamma} x_j u_i(\vx) n_i(\vx) \, \dd S(\vx) \, , \quad \vxz \in \Gamma \, .
\end{equation}
Note that the periodic contribution of the Stresslet vanishes as in section~\ref{ch:UCellSurf:DL}, and that periodic offsets of $\vxz$ leave the equation unchanged, compare expression~\eqref{eq:Green:T}.
Splitting up $\partial \Gamma$ and exploiting that $b$ and $a'$ (and $a'$ and $b$) denote the same surfaces but with antiparallel normal vectors, we have $\oint_{\partial\Gamma} = \int_{\partial \widetilde{\Gamma}} - \int_a - \int_b$ (with the same integrands as above).
Next, we add a zero by inserting $0 = \int_A + \int_B - \int_A - \int_B + \ldots,$ and then use $\oint_{\partial \Omega} = \int_{\partial \widetilde{\Gamma}} + \int_A + \int_B + \ldots,$ where the ellipsis contains analogous terms for other objects.
We also have $\int_A + \int_a = \oint_{\partial\Obj_A}$ and $\int_B + \int_b = \oint_{\partial\Obj_B}$.
With this we find
\begin{equation}
	\begin{split}
		(\mathcal{K}_{\partial\Gamma} \uv)_j&(\vxz + \vec{X}^{(\vec{\alpha}')}) =
		-\frac{8\pi}{V_\Gamma} \bigg[ 
		\oint_{\partial\Omega} x_j u_i n_i \, \dd S \\
		&- \oint_{\partial\Obj_A} x_j u_i n_i \, \dd S 
		- \oint_{\partial\Obj_B} x_j u_i n_i \, \dd S \bigg] + \ldots
		\, , \quad \vxz \in \Gamma \, .
	\end{split}
\end{equation}
Applying the divergence theorem as in section~\ref{ch:UCellSurf:DL} together with eq.~\eqref{eq:Incompress} yields
\begin{align}\label{eq:PO:DLUcell}
	\begin{aligned}
		(\mathcal{K}_{\partial\Gamma} \uv)_j(\vxz &+ \vec{X}^{(\vec{\alpha}')}) = 8 \pi \braket{u_j}_\Gamma \\
		&+ \frac{8\pi}{V_\Gamma} \Bigg[ \> \int\limits_{\Obj_A} x_j \vec{\nabla} \cdot \uv \, \dd x^3 + \int\limits_{\Obj_B} x_j \vec{\nabla} \cdot \uv \, \dd x^3 \Bigg] + \ldots \, ,
	\end{aligned}
	\\
	\vxz \in \Gamma \, , \quad j=1,2,3 \, . \nonumber
\end{align}
The appearance of the average flow is consistent with the results from section~\ref{ch:UCellSurf:DL}.
The two integrals vanish for capsules, i.e.\ just like in eq.~\eqref{eq:PO:DLObj} and as for the single-layer potential they do not cause any trouble.
Thus, we will now concentrate on the special case of bubbles.

To this end, consider the combined double-layer potentials for the bubbles and the unit cell: Define the double-layer parts from eq.~\eqref{eq:PO:BI:WithGamma} as
\begin{align}
	\mathrm{DL}_j(\vxz) := (\mathcal{K}_{\partial \Obj_A \cup \partial \Obj_B} \vec{u})_j(\vxz) + (\mathcal{K}_{\partial\Gamma} \uv)_j(\vxz) + \ldots \, , \\
	\vxz \in \Gamma \, , \quad j=1,2,3 \, , \nonumber
\end{align}
where $\lambdabar_{\Obj_A \cup \Obj_B} = 0$ for bubbles has been used.
By virtue of equations~\eqref{eq:PO:DLObj} and \eqref{eq:PO:DLUcell} we have
\begin{equation}
	\mathrm{DL}_j(\vxz + \vec{X}^{(\vec{\alpha}')}) = \mathrm{DL}_j(\vxz) \, , \quad \vxz \in \Gamma \, , \quad j=1,2,3
\end{equation}
and
\begin{equation}
	\begin{split}
		\mathrm{DL}_j(\vxz) = {}&{} (\mathcal{K}_{\partial \Obj_A \cup \partial \Obj_B'} \vec{u})_j(\vxz)
		+ \frac{8\pi}{V_\Gamma} \Bigg[ \>
			\int\limits_{\Obj_A} x_j \vec{\nabla} \cdot \uv \, \dd x^3 \\
			&+ \int\limits_{\Obj_B} \left(x_j + X^{(\vec{\alpha})}_j \right) \vec{\nabla} \cdot \uv(\vx) \, \dd x^3
		\Bigg] 
		+ \ldots \, , \quad \vxz \in \Gamma \, ,
	\end{split}
\end{equation}
where $\braket{\uv}_\Gamma$ is hidden in the ellipsis.
Using the periodicity of the velocity from eq.~\eqref{eq:uFull:Periodic}, we find that the last integral is identical to $\int_{\Obj_B'} x_j \vec{\nabla} \cdot \uv(\vx) \, \dd x^3$.
Thus,
\begin{align}
	\mathrm{DL}_j(\vxz) = (\mathcal{K}_{\partial \Obj_A \cup \partial \Obj_B'} \vec{u})_j(\vxz)
		+ \frac{8\pi}{V_\Gamma} \>\> \int\limits_{\mathclap{\Obj_A \cup \Obj_B'}} \> x_j \vec{\nabla} \cdot \uv(\vx) \, \dd x^3 + \ldots
		\, , \\
		\vxz \in \Gamma \, , \quad j=1,2,3 \, . \nonumber
\end{align}
or by means of equations~\eqref{eq:xDivU} and \eqref{eq:Bub:Flux:c}
\begin{align}
	\mathrm{DL}_j(\vxz) = (\mathcal{K}_{\partial \Obj_A \cup \partial \Obj_B'} \vec{u})_j(\vxz) + \frac{8\pi}{V_\Gamma} Q_{\Obj_A \cup \Obj_B'} \chi^{(\Obj_A \cup \Obj_B')}_j + \ldots \, , \\
	\vxz \in \Gamma \, , \quad j=1,2,3 \, . \nonumber
\end{align}
$\vec{\chi}^{(\Obj_A \cup \Obj_B')}$ is the centroid of the \emph{combined} object parts $\Obj_A$ and $\Obj_B'$ (i.e.\ the centroid of the non-split object), and $Q_{\Obj_A \cup \Obj_B'}$ the prescribed flux.

\subsection{Putting it all together}
Given the above results, the BI equation~\eqref{eq:PO:BI:WithGamma} for split bubbles is thus
\begin{equation}\label{eq:PO:BI:Remapped}
	\begin{split}
		u_j(\vxz) = 
		{}&-\frac{1}{8\pi\mu} (\mathcal{N}_{\partial \Obj_A \cup \partial \Obj_B'} \vec{F})_j(\vxz) 
		+ \frac{1}{8\pi} (\mathcal{K}_{\partial \Obj_A \cup \partial \Obj_B'} \vec{u})_j(\vxz)\\
		&+ \frac{8\pi}{V_\Gamma} Q_{\Obj_A \cup \Obj_B'} \chi^{(\Obj_A \cup \Obj_B')}_j
		+ \ldots
		\, , \qquad \vxz \in \Omega \, , \quad j=1,2,3 \, .
	\end{split}	
\end{equation}
If we did not move around the objects with the help of periodicity, equation~\eqref{eq:PO:DLUcell} would have led to the equivalent expression
\begin{equation}\label{eq:PO:BI:NotRemapped}
	\begin{split}
		u_j(\vxz) = 
		{}&-\frac{1}{8\pi\mu} (\mathcal{N}_{\partial \Obj_A \cup \partial \Obj_B} \vec{F})_j(\vxz) 
		+ \frac{1}{8\pi} (\mathcal{K}_{\partial \Obj_A \cup \partial \Obj_B} \vec{u})_j(\vxz)\\
		&+ \frac{8\pi}{V_\Gamma} \left[
			Q_{\Obj_A} \chi^{(\Obj_A)}_j + Q_{\Obj_B} \chi^{(\Obj_B)}_j
			\right]
		+ \ldots
		\, , \qquad \vxz \in \Omega \, , \quad j=1,2,3 \, .
	\end{split}	
\end{equation}
Comparing these two equations, we can draw the following conclusion regarding the original BI equation~\eqref{eq:BI:Periodic}:
If some bubble overlaps with the unit cell's boundary, we can either split it up and use the two unconnected parts ($\Obj_A$ and $\Obj_B$) on the opposite sides of the unit cell separately, including different centroids (equation~\eqref{eq:PO:BI:NotRemapped}).
Or, more conveniently and intuitively, we can simply integrate over the surface of the whole connected bubble $\Obj_A \cup \Obj_B'$ including the parts that lie outside of $\Gamma$ and use its actual centroid (equation~\eqref{eq:PO:BI:Remapped}).
This is highly desirable for the numerical implementation because we only have to deal with whole objects and no splitting of the meshes is required.
It also shows that the choice of the unit cell's position in the 3D Cartesian coordinate system does not matter.

We further note that the non-zero contributions from the double-layer integrals over $\partial \Gamma$ containing the centroids are crucial to obtain invariance for bubbles, and hence the above results are non-trivial.
If they were missing (by assuming $\mathcal{K}_{\partial\Gamma} \vec{u} = 0$), eq.~\eqref{eq:PO:DLObj} would have introduced an additional position dependent term.
This would lead to changes in the flow field if a bubble is moved by a periodic grid vector\,--\,which is clearly unphysical.

For objects other than bubbles, the fluxes are missing and additional viscosity ratios appear.
Nevertheless, the above statement (that we can simply choose the whole objects) remains true because the individual single- and double-layer integrals are invariant under periodic translations for objects with zero flux (compare equations \eqref{eq:PO:DLObj}, \eqref{eq:PO:DLUcell} and section~\ref{sec:PO:SL}).

Analogous conclusions can be drawn for the FBI equation~\eqref{eq:FBI:Intro} which in the end is evaluated by our numerical code:
First, sections~\ref{sec:PO:SL} and \ref{sec:PO:DL} considered a general evaluation point $\vxz \in \Gamma$ and thus remain valid if $\vxz$ is located on the surface of some object.
We additionally saw that the integrals are invariant if $\vxz$ is moved by some periodic grid vector $\vec{X}^{(\vec{\alpha}')}$, allowing $\vxz$ to be on the parts of surfaces that lie outside of~$\Gamma$.
Second, the imposed flow as well as the Wielandt deflation term are periodic due to equation~\eqref{eq:z:WithSurf}.
Third, the proof from section~\ref{sec:Proof} is largely independent of the position of the objects. 
Where integrals over $\partial \Gamma$ occur (e.g.\ in the energy conservation statements), they can be reconstructed from $\partial \widetilde{\Gamma}$ in the same way as was done for equation~\eqref{eq:PO:BI:WithGammaDisturbance}, leaving the procedure unchanged.
Replacing the object parts with the whole objects therefore does not affect the proof in section~\ref{sec:Fredholm}.

	\bibliographystyle{apsrev4-1_mod}
	\bibliography{literature}
	


\section*{Supplementary material}

affect the overall results qualitatively (i.e. margination on average is still observed). Furthermore, halving the bubbles' equilibrium radius or making the vessel wall stiffer does not lead to significant changes either (fig. S8). Note that depending on the exact history, short lived migrations toward the vessel center can occur sometimes, which are nevertheless again followed by rapid movement to the vessel walls.

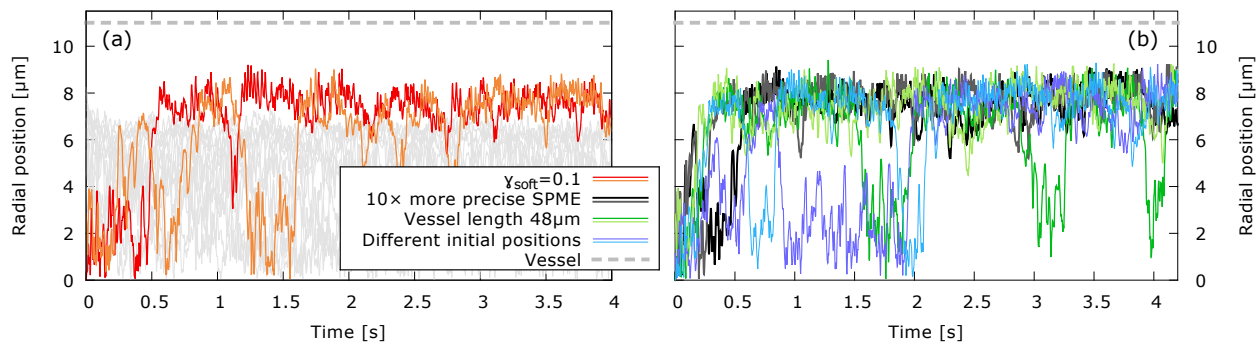

Figure S7: Radial positions of the centroids of two oscillating lipid coated microbubbles ($f = 1\,\text{kHz}$, $P_A = 45\,\text{kPa}$) for $\delta = 1$ with red blood cells. Radiation forces *not* included. (a) The surface tension in the soft state is reduced to $\gamma_{\text{soft}} = 0.1\kappa_S$, retaining the effect of margination without qualitative changes (red/orange). The gray lines indicate the red blood cells. (b) Three different simulations for $\gamma_{\text{soft}} = 0.5\kappa_S$: One with more precise SPME parameters (cutoff errors below $\lesssim 0.001\,\%$, i.e. one order of magnitude smaller than usual; also see section 2.5 in the main manuscript; black/gray lines), one with the larger system from figure 2 (a) from the main text (vessel length 48 µm and hematocrit 16 %; dark/light green), and one with different initial positions for all particles compared to the remaining simulations (RBCs initialized in regular arrays; purple/blue). Red blood cells omitted for clarity.

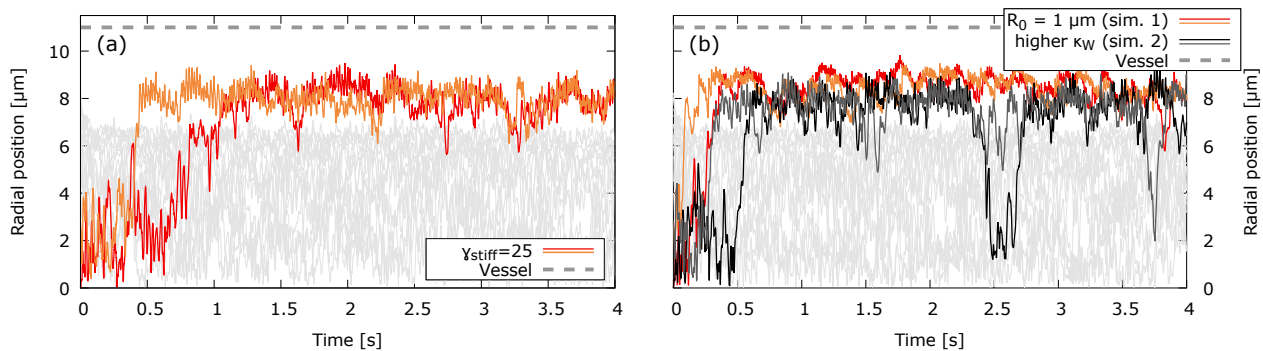

Figure S8: Behavior of two oscillating lipid coated microbubbles for $\delta = 1$, $f = 1\,\text{kHz}$ and $P_A = 6\,\text{kPa}$ with red blood cells. The primary radiation force is included ($|\boldsymbol{F}_{\text{rad}}| \approx 1.5 \times 10^{-17}\,\text{N}$). (a) The surface tension in the stiff state is increased to $\gamma_{\text{stiff}} = 25\kappa_S$. (b) Two simulations, once for bubbles of half the usual size ($R_0 = 1\,\mu\text{m}$) and once for a five times stiffer vessel wall ($\kappa_{\text{Wall}} = 31.25 \times 10^6\,\text{N/m}^3$). The qualitative results remain unchanged in all cases.

## S4 Code verification

We performed extensive testing of our code to ensure correctness of both the chosen algorithms as well as of the implementation itself. For example, we compared the results of the single- and double-layer integrals for both the infinite and the periodic Green's functions with analytically known values (similar to ref. [8]). The red blood cell model was validated by considering, amongst others, the deformation of a capsule in an infinite shear flow, as published in reference [7] and further explored below, as well as by comparing with analytical calculations for a diffusing particle near elastic membranes [9–14]. We also used the code to compute the shapes assumed by red blood cells in microchannels [15].

For further verifications we consider a single bubble in an extensional flow in figure S9. More precisely, the flow is solved with VCO-BIM in an infinite domain ($V_\Gamma \to \infty$) as explained in the main text with the imposed

flow set to $\boldsymbol{u}^\infty(\boldsymbol{x}) = s(2x, -y, -z)$, where $s$ is the shear rate (for infinite domains, more general imposed flows $\boldsymbol{u}^\infty(\boldsymbol{x})$ are possible instead of only a constant flow $\langle\boldsymbol{u}\rangle_\Gamma$ [16]). The bubble starts with 5120 triangles and is refined as needed [17]. Moreover, the mean curvature is computed via Method C from ref. [7] and the mesh stabilization routine from equation (2.72) from the main text is employed. The proper dimensionless parameter is the capillary number $\text{Ca} = s\mu R/\gamma$, where $\mu$ is the dynamic viscosity of the ambient fluid, $R$ the initial bubble radius and $\gamma$ the surface tension. For comparison with the literature, we extract the Taylor deformation parameter $D = (a-c)/(a+c)$. The length of the largest half-axis $a$ and of the smallest half-axis $c$ of the deformed object are computed from an ellipsoid with the same inertia tensor [18, 19]. Figure S9 (a) shows a cut through the bubble in the $z = 0$ plane in the stationary state and compares it with the shapes found by Youngren and Acrivos using an axisymmetric boundary integral method [20]. Furthermore, figure S9 (b) depicts the stationary value for the deformation parameter $D$ as a function of the capillary number. We compare it with the numerical results of Youngren and Acrivos [20] and with the analytical $\mathcal{O}(\text{Ca}^2)$ theory of Barthès-Biesel and Acrivos [21]. In both cases very good agreement is observed. We also note that the deformation in the $\text{Ca} = 0.1$ case corresponds to an asphericity of around 0.295, which is larger than any values observed for the full setup from the main text. Hence we conclude that we obtain correct behavior within the relevant deformation range.

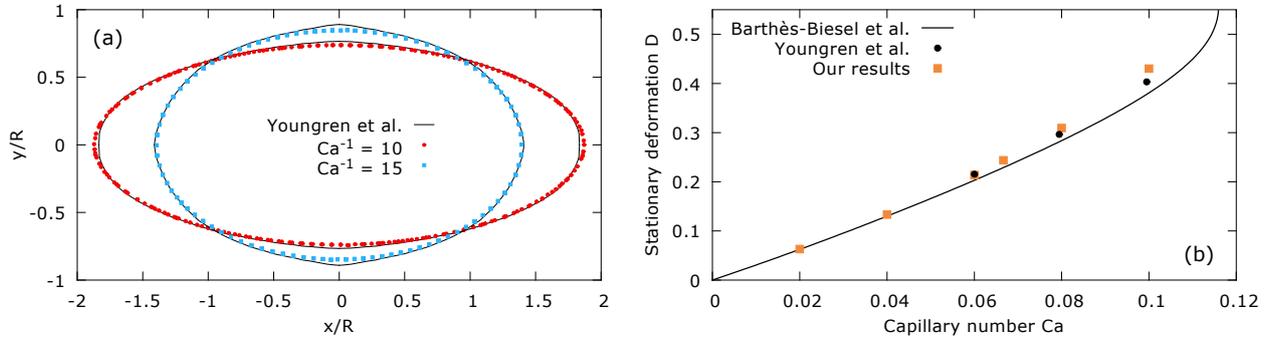

Figure S9: Verification of the bubble in an infinite extensional flow. (a) Cut through the stationary 3D shape of the bubble in the $z = 0$ plane for two different capillary numbers. The black lines depict the results obtained by Youngren and Acrivos using an axisymmetric boundary integral method [20]. (b) Stationary Taylor deformation parameter $D$ as a function of the capillary number. The black dots depict the result by Youngren and Acrivos [20], the orange squares our results. Furthermore, we show the analytical theory of Barthès-Biesel and Acrivos as a black line [21]. Note that it quickly diverges contrary to numerical evidence [20]. Hence, its validity appears to be limited to $\text{Ca} \lesssim 0.06$ where excellent agreement is found.

In reference [7] we treated the case of a capsule in an infinite shear flow, and found very good matching with the literature (figures 12 and 17 therein). For the *periodic* system we show a similar result in figure S10. Namely, we place two flat walls with a distance $h = 19R$ in a cubic unit cell with side lengths $20R$ together with an initially spherical capsule of radius $R$. We implement the shear flow by prescribing the velocities $\boldsymbol{u} = (\pm sh/2, 0, 0)$ at the walls, whereas the top (bottom) sign corresponds to the top (bottom) wall and $s$ is the shear rate. The capsule is endowed not only with some shear elasticity modeled according to the neo-Hookean law (shear modulus $\kappa_S$ [7]), but also with some bending rigidity following the Canham-Helfrich law [22–24] (bending modulus $\kappa_B$, Method C from ref. [7], flat reference state). We set the inner viscosity to be identical to the dynamic viscosity $\mu$ of the ambient fluid. Hence, two dimensionless parameters are relevant: The capillary number $\text{Ca} = s\mu R/\kappa_S$ and the reduced bending modulus $\hat{\kappa}_B = \kappa_B/(R^2\kappa_S)$. As before, we extract the Taylor deformation parameter $D$. See reference [7] for further details and section 3.2 in the main text for the remaining parameters. Varying the distance between the walls does not change the results significantly compared to $h = 19R$. We therefore effectively mimic an infinite system and comparisons with results from unbounded flows are appropriate. However, the keypoint here is that we use the very same SPME code that is also employed for the simulations in the main paper, and thereby further validate the implementation. Numerical parameters include 1280 triangles for the capsule (which is well converged [7] and roughly corresponds to the maximal resolution used for the red blood cells in the margination simulations) and 800 triangles per wall. The SPME error is $\lesssim 0.01\%$ (similar to the main simulations). Obviously, figure S10 shows that our results compare very favorably with data extracted from Tsubota [25], proving that our code works as intended.

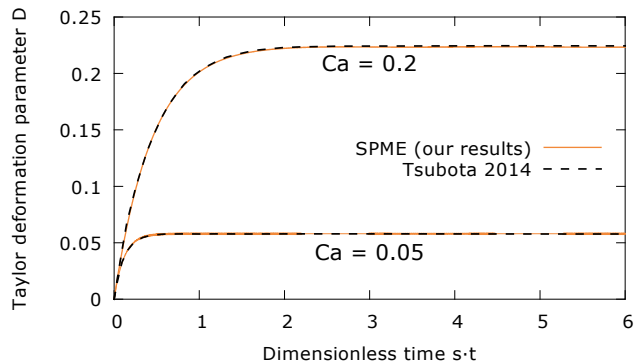

Figure S10: Deformation of a single spherical capsule (endowed with shear elasticity and bending rigidity) in a periodic system for two different capillary numbers Ca, compared to results from reference [25] ("model H" therein). We set $\hat{\kappa}_B = 2/15$. Since the wall distance $h$ is chosen sufficiently large, the data matches very well although the values from the literature were obtained for a capsule in an infinite (rather than a periodic) system.

## S5 Dynamic mesh refinement

As outlined in the main text, we employ Rivara's longest-edge bisection algorithm [17] in order to refine the triangular meshes locally when objects come close to each other and at high curvature regions. Lower resolutions are sufficient for the other regions, i.e. we coarsen the previously refined areas again in this case. Example snapshots are shown in figure S11.

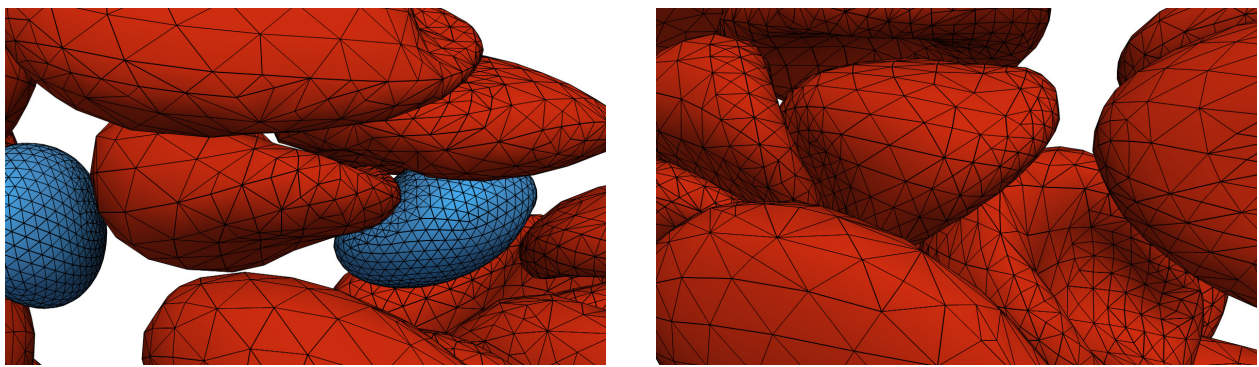

Figure S11: Example snapshots from the simulations, highlighting the dynamic mesh refinement due to close contact and high curvature regions.



\end{document}